%% file: main.tex
\documentclass[twocolumn]{aastex63}
\usepackage{savesym}            % needed to avoid conflict with siunitx 
\savesymbol{tablenum}
\usepackage{siunitx}
\restoresymbol{SIX}{tablenum}
\usepackage{siunitx}
\usepackage{xspace}
\usepackage{xcolor}
\usepackage{microtype}  % much better typesetting and linebreaks
\usepackage[nolist,nohyperlinks]{acronym}
\usepackage{placeins}   % to avoid new sections starting before previous figures are inserted (with \FloatBarrier command)
\usepackage{float}  % option for more aggressive figure placement with H

\begin{acronym}
    \acro{ao}[AO]   {Adaptive Optics}
    \acro{alma}[ALMA] {Atacama Large Millimeter Array}
    \acro  {psf}[PSF]   {Point Spread Function}
    \acro  {wispit}[WISPIT]   {WIde Separation Planets In Time}
    \acro  {yses}[YSES]   {Young Suns Exoplanet Survey}
    \acro  {sphere}[SPHERE]   {Spectro-Polarimetric High-contrast Exoplanet REsearch}
    \acro  {irdis}[IRDIS]   {Infrared Dual-band Imager and Spectrograph}
    \acro  {dpi}[DPI]   {Dual Polarization Imaging}
    \acro  {adi}[ADI]   {Angular Differential Imaging}
    \acro  {rdi}[RDI]   {Reference Differential Imaging}
    \acro  {pdi}[PDI]   {Polarimetric Differential Imaging}
    \acro  {pca}[PCA]   {Principal Component Analysis}
    \acro  {cpd}[CPD]   {Circumplanetary Disk}
    \acro  {sed}[SED]   {Spectral Energy Distribution}
    \acro  {ofti}[OFTI]   {Orbits For The Impatient}
    \acro  {fwhm}[FWHM]   {Full Width at Half Maximum}
    \acro{mse}[MSE]{Mean Squared Error}
    \acro{lcc}[LCC]{Lower Centaurus Crux}
    \acro{scocen}[Sco-Cen]{Scorpius-Centaurus}
    \acro{eagles}[EAGLES]{Estimating AGes from Lithium Equivalent widthS}
    \acro{vosa}[VOSA]{Virtual Observatory SED Analyzer}
    \acro{irdap}[IRDAP]{IRDIS Data reduction for Accurate Polarimetry}
\end{acronym}

\newcommand{\wis}{WISPIT\xspace} % noice!
\newcommand{\wisa}{WISPIT~2\xspace}
\newcommand{\wisb}{WISPIT~2b\xspace}

\newcommand{\mj}{\ensuremath{\mathrm{M_{Jup}}}\xspace}

\newcommand{\pmra}{$\mu_{\alpha *}$}
\newcommand{\pmdec}{$\mu_{\delta}$}
\newcommand{\masyr}{mas\,yr$^{-1}$}

\received{July 3, 2025}
\revised{July 28, 2025}
\accepted{\today}
\submitjournal{ApJL}

\shorttitle{\wisa}
\shortauthors{van Capelleveen et al.}
\begin{document}
%TC:ignore
\title{WIde Separation Planets In Time (WISPIT): \\ A gap-clearing planet in a multi-ringed disk around the young solar-type star WISPIT 2}

\correspondingauthor{Richelle van Capelleveen}
\email{capelleveen@strw.leidenuniv.nl}

% author list is generated from the Google spreadsheet of authors and then running
% python make_authorlist_aastex.py > authorlist.tex
% MAK 2025/06/21

\input{authorlist}



\begin{abstract}

In the past decades several thousand exoplanet systems have been discovered around evolved, main-sequence stars, revealing a wide diversity in their architectures.
To understand how the planet formation process can lead to vastly different outcomes in system architecture we have to study the starting conditions of planet formation within the disks around young stars.

In this study we are presenting high resolution direct imaging observations with VLT/SPHERE of the young ($\sim$5\,Myr), nearby ($\sim$133\,pc), solar-analog designated as \wisa ($=$\,TYC~5709-354-1).
These observations were taken as part of our survey program that explores the formation and orbital evolution of wide-separation gas giants.
WISPIT\,2 was observed in four independent epochs using polarized light and total intensity observations.
They reveal for the first time an extended (380\,au) disk in scattered light with a multi-ringed sub-structure.
We directly detect a young proto-planet \wisb, embedded in a disk gap and show that it is co-moving with its host star.
Multiple SPHERE epochs demonstrate that it shows orbital motion consistent with Keplerian motion in the observed disk gap.
Our $H$ and $K_s$-band photometric data are consistent with thermal emission from a young planet.
By comparison with planet evolutionary models, we find a mass of the planet of $4.9^{+0.9}_{-0.6}$\,\mj.
This mass is also consistent with the width of the observed disk gap, retrieved from hydrodynamic models.

\wisb is the first unambiguous planet detection in a multi-ringed disk, making the \wisa system the ideal laboratory to study planet-disk interaction and subsequent evolution.

\end{abstract}


%TC:endignore

%\newpage
\section{Introduction}
\label{section: introduction}
It has only been three decades since the first exoplanet detection, but tremendous progress has been made since---to date there are nearly 6000 confirmed exoplanets.
These planets span a wide range of masses, are found at separations from less than an au \citep[e.g.][]{Dawson2010,Goffo2023} to several hundreds of au from their host stars \citep[e.g.][]{Janson2021,Zhang2021}, and exhibit diverse atmospheric chemistries \citep[e.g. ][]{MacDonald2017, Rustamkulov2023, Gapp2025}, with some planets even hosting circumplanetary disks \citep[e.g.][]{Benisty2021,Close2025a}.
They have been found around a variety of stellar types, including stellar multiples \citep[e.g.][]{Sigurdsson2003,Dupuy2018,Kostov2020}, though the majority have been detected around single stars. 
This diversity raises a fundamental question: are planetary properties inherited from their natal disks, or shaped by later evolutionary processes?
%
%altervative sentence
%This wide variety of exoplanets raises the question as to whether they inherit their properties from their birth disks, or whether the variety is (partially) introduced by factors in later stages of their evolution.
Addressing this question requires a detailed understanding of the environments in which planets form---their protoplanetary disks.

As the disk and planet evolve simultaneously, the disk affects the planet and the planet in turn affects the disk. 
This is evident in the formation of substructures and in the distribution of gas and dust.
%
%Investigating these effects requires high-resolution imaging of gas and dust, which was historically limited by the angular resolution of available instruments. 
%
%The first surveys to systematically observe protoplanetary disks targeted Herbig Ae/Be and T Tauri stars with Subaru/HiCIAO {\color{red}e.g. REF SEEDS program, examples} and VLT/NACO {\color{red}e.g. examples}. 
%
%These early efforts focused on bright disks around bright stars due to the constraints of \ac{ao} systems available at the time. 
%
%This all changed with the advent of extreme \ac{ao}-assisted facilities, which significantly expanded not only the number of observed disks but also the range of masses and ages of the observed targets {\color{red}e.g. REF}.
%
Combined observations from high-contrast imaging---sensitive to thermal emission and scattered light from (sub)micron-sized dust---and the \ac{alma}, which traces gas and millimeter dust, have revealed a wide variety of such substructures. 
These findings have provided critical inputs for theoretical models of disk dynamical evolution and planet-disk interactions \citep[see review][and citations therein]{Bae2023}.

The next logical step towards advancing our understanding of planet-disk interactions in early planet formation is to test these models against observations of planet-forming disks with embedded protoplanets. 
This remains challenging, however, as to date only one system---PDS~70---has been unambiguously confirmed to host embedded protoplanets \citep{Keppler2018,Haffert2019}.
While several notable candidate systems exist \citep[e.g.][]{Gratton2019,Currie2022}, confirmation is hindered by the difficulty of disentangling planet signal from disk signal \citep[e.g.][]{Currie2019,Folette2017}.
%
%NOTE TO SELF, COPY BELOW FROM COMPARISON SECTION
%Besides the PDS\,70 system, there are currently two strong candidates for directly detected, embedded protoplanets in the AB\,Aur system (\citealt{Currie2022}) as well as most recently in the HD\,169142 system (\citealt{Gratton2019, Hammond2023}).
In the case of direct detections, the challenge lies in determining whether the detected emission originates from a planet or from disk structures.
For indirect methods, such as detections based on kinematic signatures, the difficulty lies in distinguishing between deviations from Keplerian velocity caused by an embedded planet and those resulting from intrinsic disk dynamics in the absence of a planet \citep{Teague2025}.
These difficulties are compounded by the technical complexity of detecting low-mass planets, especially through accretion signals \citep{Benisty2023,Close2025a}. 
This scarcity of testbeds, both in number and diversity, leaves key questions about planet formation unresolved. 

One such question concerns the formation of wide-separation giant planets orbiting at semi-major axes larger than 50\,au. 
It remains unclear whether these are formed in situ through gravitational instability, either through interstellar cloud fragmentation or circumstellar disk fragmentation \citep{Boss1997,Kroupa1995}, or whether they were formed closer to the star through accretion processes \citep{Pollack1996} and migrated outward later through scattering events.
Along with the broader goal of discovering planets and determining their occurrence rates around stars similar to our Sun, this gave rise to the \acl{yses} \citep[\acs{yses}; ][and van Capelleveen in prep.]{Bohn2021}, a VLT/SPHERE direct imaging survey targeting 70 young ($14\pm3$ Myr), solar-mass stars in the \ac{lcc} subgroup of the \ac{scocen} OB association.
\acused{yses}
Building on the success of \ac{yses}, the \acl{wispit} (\acs{wispit}; van Capelleveen, submitted; van Capelleveen, in prep) survey extends this sample to younger ages---the median age is 8.5 Myr---and to other regions of the sky.
\acused{wispit}
This ongoing survey comprises a total of 178 young suns, making it the closest and largest selection of young solar-mass stars.

\begin{figure}
\center
\includegraphics[width=\hsize]{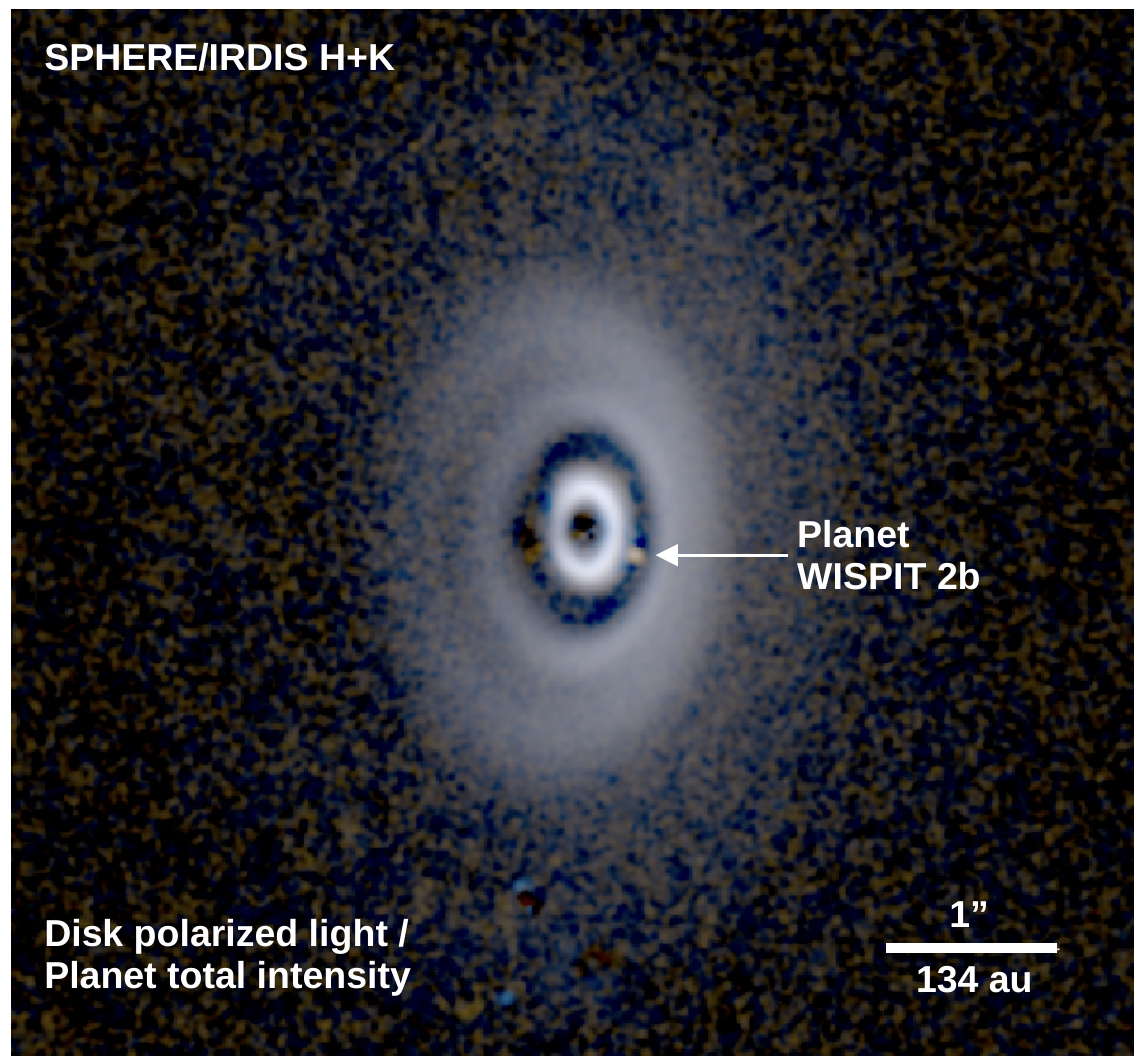} 
\caption{SPHERE/IRDIS multi-band image of the \wisa system.
The H-band $Q_\phi$ image was added as blue channel and the median combination of $H$-band and $K_s$-band $Q_\phi$ images was added as green channel.
The red channel is a combination of $K_s$-band $Q_\phi$ image and $K_s$-band cADI image in which we masked all but the gap containing the thermal emission from \wisb. 
For more details see Appendix~\ref{app:composit_fig}.
} 
\label{fig:sphere_rgb}
\end{figure}

The best way to test (wide separation) planet formation and planet-disk interaction theories is finding unambiguous planet signals embedded in disks around young stars.
In this work we present such a discovery: a robust detection of a planetary companion embedded in a ringed disk around \wisa (=TYC~5709-354-1) - see Figure~\ref{fig:sphere_rgb}. 
In Section~\ref{sec:wispit2stars} we present the stellar properties of \wisa, followed by our observations and data processing in Section~\ref{sec:obsandprod}.
% Note to co-authors, it will be in simbad as \wisa so just giving its current main ID
%
In Section~\ref{sec:disk} we detail the morphology and analysis of the scattered light from the multi-ringed disk.
The characterization of multiple epochs of the planet along with its orbital properties are detailed in Section~\ref{sec:planet}, its subsequent interaction with the disk follows in Section~\ref{sec:planetdiskinteraction}. 
The planetary interpretation of \wisb is additionally strengthened by its detection in H$\alpha$ observations (Close et al. submitted; companion letter~2).
Our discussion and conclusions are presented in Section~\ref{sec:conc}.
%--------------------------------------------------------------------
\section{Stellar properties of \wisa}
\label{sec:wispit2stars}
\wisa was flagged as a pre-main sequence star candidate by \citet{Zari2018}, and is fairly proximate \citep[$d=133.35^{+0.37}_{-0.38}$~pc; ][]{BailerJones2021} which is how it came to be included in the \wis survey.
Its combined stellar properties are listed in Table~\ref{tab:star}. 
\wisa is located in the outskirts of the Scorpius-Centaurus OB association but has not been assigned membership to any of its main subgroups, and its astrometry is not consistent with membership in any of them. 
We discuss its co-moving co-distant stars and membership to recently identified young stellar associations in Appendix~\ref{app:membership}, and found that it is likely part of the Theia~53 \citep{Kounkel2019,Kerr2021,Hunt2024} group. 
Based on lithium equivalent width measurements of three stars assigned to this group, we used \ac{eagles} and \ac{eagles} v2.0 to estimate the age of this cluster, resulting in age constraints of $<13$~Myr and $11.1^{+5.9}_{-8.1}$~Myr, respectively (see Appendix~\ref{app:age} for more details).
%%%%% added paragraph break

As this star has not previously been characterized in detail, we modeled its \ac{sed} to constrain its physical properties.
We performed a $\chi^2$ fit of 17 photometric points from TYCHO \citep{Hog2000}, Pan-STARRS \citep{Chambers2016}, Gaia DR3 \citep{Gaia_DR3}, DENIS \citep{Epchtein1999} and 2MASS \citep{Cutri2003} to synthetic models from BT-Settl CIFIST \citep{Allard2013} using the \ac{vosa} \citep{Bayo2008}.
Photometric data from WISE \citep{Cutri2012} were not included in the fit because \ac{vosa} identified photometric excess starting at band W1 (3.4\micron).
The extinction was constrained to the 1$\sigma$-range derived in Appendix~\ref{app:reddening}, $A_V=0.136\pm0.087$~mag.
While this does not account for possible extinction from the disk itself, we expect such effects to be minimal due to the disk's $\sim45^\circ$ inclination (see Section~\ref{sec:disk}) and a cleared inner gap, making the interstellar extinction a reasonable approximation.
The other model parameters for the fit were constrained to be in ranges $3000\leq \mathrm{T_{eff}} \leq 6000$, and $3.5\leq\log{g}\leq4.5$.
%%%%% start edit
We adopted the best-fit model consistent with \citet{Prato2023}, and used the resulting bolometric luminosity of $L_{\rm bol} = 0.699 \pm 0.021\,\mathrm{L_\odot}$ and temperature of $T_{\rm eff} = 4400 \pm 50$\,K to retrieve a mass of $1.08^{+0.06}_{-0.17}\,\mathrm{M_\odot}$ and age of $5.1^{+2.4}_{-1.3}$~Myr by comparing to BHAC15 \citep{Baraffe2015} isochrones. The uncertainties are adjusted to account for systemic errors due to choice of stellar evolution model (see Appendix~\ref{app:age}).
While on the lower end, the resulting age is consistent with that of the group to which \wisa likely belongs.
%%%%% end edit
\textit{From this, we conclude that \wisa is a young ($\sim$5 Myr) solar-mass star.}

\begin{figure}[t]
    \centering
    \includegraphics[width=\hsize]{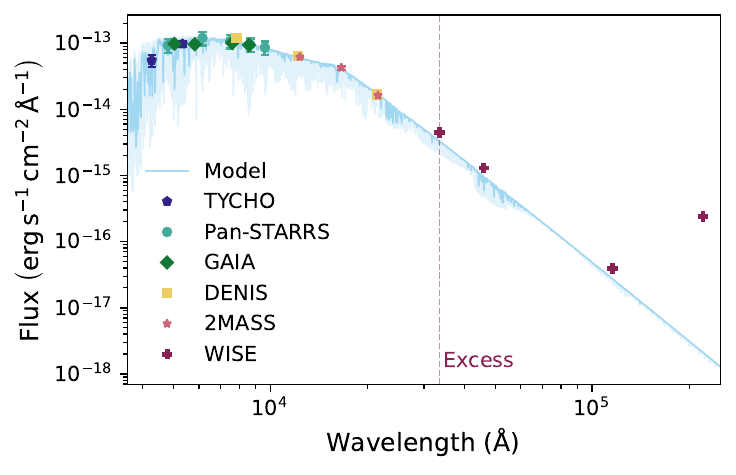}
    \caption{\ac{sed} of \wisa.
    Photometric data from various sources are shown with colored markers.
    The blue curve represents the best-fit BT-Settle-CIFIST model ($\chi^2=7.94$), with the low resolution (high opacity) version overlayed on the high resolution (low opacity) model.}
    \label{fig:sed}
    \end{figure}

The parameters obtained from the \ac{sed} fit, as well as the mass and age derived from the isochrone fit, are provided in Table~\ref{tab:star}.
Figure~\ref{fig:sed} presents the the best-fit synthetic spectrum overlaid on the photometry, which was dereddened by \ac{vosa} using the extinction law by \citet{Fitzpatrick1999} improved by \citet{Indebetouw2005} in the infrared.
This \ac{sed} already reveals an increased infrared excess at W4 (22\micron), hinting at the presence of a disk.

\begin{table}[htb!]
\centering
\caption{Stellar parameters of \wisa}
\label{tab:star}
\def\arraystretch{1.2}
\setlength{\tabcolsep}{13pt}
\begin{tabular}{@{}lll@{}}
\hline\hline
Parameter & Value & Ref.\\ % table heading
\hline
Gaia DR3 & 4207586980945067648 & (1)\\
2MASS & J19231702-0740550   & (2)\\
WISE & J192317.04-074055.2 & (3)\\
TIC & 98898373           & (4)\\
TYC & 5709-354-1  & (5)\\
\hline
RA* $\alpha$  [deg] & 290.82100005622 & (1)\\
Dec* $\delta$ [deg] & -07.68208608363 & (1)\\
Parallax $\varpi$ [mas] & $7.4649\pm0.0214$ & (1)\\
Distance $d$ [pc] & $133.35^{+0.37}_{-0.38}$ & (6)\\
pmra $\mu_{\alpha}$ [mas/yr] & $6.308\pm0.024$ & (1)\\
pmdec $\mu_{\delta}$ [mas/yr] & $-27.138\pm0.018$ & (1)\\
$v_r$ [km/s] & $-16.23\pm14.58$ & (1)\\
\hline
$G$ [mag] & $11.186381\pm0.003952$ & (1)\\
$B_p - R_p$ & $1.377451$ 	& (1)\\
$B_p-G$ & $0.550725$ & (1)\\
$G-R_p$ & $0.826726$ & (1)\\
$J$ [mag]   & $9.310\pm0.044$  & (2)\\
$H$ [mag]   & $8.591\pm0.078$ & (2)\\
$K_s$ [mag] & $8.577\pm0.021$  & (2)\\
$W1$ [mag]  & $8.475\pm0.028$ & (3)\\
$W2$ [mag]  & $8.444\pm0.027$ & (3)\\
$W3$ [mag]  & $8.046\pm0.027$ & (3)\\
$W4$ [mag]  & $3.300\pm0.027$ & (3)\\
$FUV$ & $20.373\pm0.217$ & (7)\\
$NUV$ & $18.168\pm0.052$ & (7)\\
$\mathrm{P_{rot}}$ [day] & 4.7004 & (8)\\
\hline
$A_V$ [mag] & $0.171 \pm 0.050$ & (9)\\
$T_{\rm eff}$ [K] & $4400 \pm 50$ & (9)\\
$\log g$ [dex] & $4.00 \pm 0.25$ & (9)\\
$L_{\rm bol}$ [$\mathrm{L_\odot}$] & $0.699 \pm 0.021$ & (9)\\
$R$ [$\mathrm{R_\odot}$] & $1.418 \pm 0.004$ & (9)\\
\hline
Age [Myr] & $5.1^{+2.4}_{-1.3}$ & (9)\\
Mass [$\mathrm{M_\odot}$] & $1.08^{+0.06}_{-0.17}$ & (9)\\
\hline
\end{tabular}
\tablecomments{*=ICRS, epoch J2016.0}
References:
(1) \citet{Gaia_DR3},
(2) \citet{Cutri2003},
(3) \citet{Cutri2012},
(4) \citet{Stassun2019},
(5) \citet{Hog2000},
(6) \citet{BailerJones2021},
(7) \citet{Bianchi2011},
(8) \citet{Watson2006},
(9) This work.
\end{table}

% https://gea.esac.esa.int/archive/documentation/GDR3/Data_analysis/chap_cu8par/sec_cu8par_apsis/ssec_cu8par_apsis_flame.html

% FLAME requires atmospheric parameters (Teff, logg, [M/H]) to derive a bolometric correction BCG, see Section 11.2.3. Then, using extinction AG, magnitude and an estimate of the distance, L is derived from the standard equation

%--------------------------------------------------------------------

\section{Observations and data processing}
\label{sec:obsandprod}

\begin{table*}
\caption{SPHERE/IRDIS observations of \wisa.}
\label{tab:obs_setup}
\centering
\def\arraystretch{1.2}
\setlength{\tabcolsep}{9pt}
\begin{tabular*}{\textwidth}{@{}llllllll@{}}
\hline\hline
Observation date & Filter & Coronagraph & NEXP$\times$NDIT$\times$DIT\ & $\omega$ & $ X$ & $\tau_0$ \\
(yyyy-mm-dd) & & & (1$\times$1$\times$s)  & (\arcsec) & & (ms)\\
\hline
2023-10-19 & $H$ &N\_ALC\_YJH\_S& 4$\times$2$\times$32  & $0.545\pm0.005$ & $1.284\pm0.005$ & $4.1\pm0.3$\\
2024-10-04 & $H$ &N\_ALC\_YJH\_S& 4$\times$2$\times$32  & $0.990\pm0.097$ & $1.397\pm0.007$ & $4.5\pm0.3$ \\
2025-03-21 & $H$ &N\_ALC\_YJH\_S& 48$\times$1$\times$64  & $0.390\pm0.048$ & $1.451\pm0.063$ & $5.7\pm1.3$ \\
2025-04-26 & $K_s$ &N\_ALC\_Ks& 48$\times$1$\times$64  & $0.275\pm0.037$ & $1.146\pm0.067$ & $7.4\pm1.5$ \\
\hline
\end{tabular*}
\tablecomments{Observation setup and conditions for all \wisa observations.
All filters are SPHERE broadband filters.
The total integration time is the product of the number of exposures (NEXP), the number of subintegrations per exposure (NDIT) and the detector integration time (DIT).
The seeing is denoted by $\omega$, the airmass by $X$ and the coherence time by $\tau_0$.}
\end{table*}

\subsection{Description of observations}
\label{sec:observations}
All observations were performed at the VLT with the \ac{sphere} instrument \citep{Beuzit2019} using the \acl{irdis} \citep[\acs{irdis}; ][]{Dohlen2008} camera.
\acused{irdis}
We present four different observational epochs taken on 2023-10-19, 2024-10-04, 2025-03-21 and 2025-04-26. 
Details on the exposure settings and weather conditions for all epochs are given in Table~\ref{tab:obs_setup}.
In the following we give a brief summary.

The first two observational epochs were taken as part of the main WISPIT survey program (van Capelleveen et al., in prep.).
Both of these observation epochs were taken in the $H$-band in the classical imaging mode with pupil stabilization.
These two epochs have a total exposure time of 4.3\,min each.
%
%The observation epoch on 2025-03-21 was taken with the \ac{dpi} mode \citep{deBoer2020,vanHolstein2020} of the \ac{irdis}, also in the broad-band $H$ filter.
The observation epoch on 2025-03-21 was taken in the broadband $H$-filter with \ac{irdis} \ac{dpi} mode \citep{deBoer2020,vanHolstein2020}.
A total of 12 polarimetric cycles were recorded, each consisting of the usual four images taken at different half wave plate positions.
This led to a total exposure time of 51.2\,min.
While the observation sequence was taken in pupil stabilized mode to maximize polarimetric efficiency, the overall parallactic field rotation was small (2.3$^\circ$).
The data set taken on 2025-04-26 was taken in an identical manner using the broadband $K_s$-filter and additionally using the ``star-hopping'' technique \citep{Wahhaj2021}.
This technique alternates every $\sim10\,$min between science target and reference star.
Similar to the previous observation we obtained 12 polarimetric cycles on the science target for a total exposure time of 51.2\,min.
For the reference star we recorded 20 frames interspersed with the science target and with the same individual frame exposure time.
The observation sequence was carried out in pupil stabilized mode to allow for multiple differential imaging processing approaches as well as to maximize polarimetric efficiency.
A total parallactic angle rotation of 27.5$^\circ$ was recorded. 

\subsection{Data reduction}
\label{sec:data_reduction}
%\subsubsection{Total intensity imaging}

We used all four \ac{sphere} observation epochs to produce total intensity images of the \wisa system.
Total intensity observations are sensitive to scattered light from the circumstellar dust, as well as the thermal emission of embedded planets.
Depending on the observation epoch, we used several differential imaging post-processing techniques to remove the stellar speckle field from the images.
We summarize these for all observation epochs along with the standard data reduction.

All \ac{sphere} total intensity observations in broadband filters were (pre-)processed with {\tt PynPoint} \citep{Amara2012,Stolker2019}. 
The reduction workflow includes bad-pixel correction, flat-fielding, sky subtraction, and anamorphic distortion correction. 
The anamorphic distortion was corrected by scaling the $y$-axis by a factor of $1.0062\pm0.0002$, following the procedure outlined in the SPHERE documentation\footnote{SPHERE manuals: \url{https://www.eso.org/sci/facilities/paranal/instruments/sphere/doc.html}} and described by \citet{Maire2016}. 
Images were aligned to the sky's parallactic angle and corrected for the pupil offset of $135.99\pm0.11\deg$, and an additional rotation of  $1.76\pm0.04\deg$ was applied to correct for the true North offset \citep{Maire2021}. 
The pixel scales for astrometric calibration are $12.246\pm0.009\,\si{mas.yr^{-1}}$ in $H$-band and $12.266\pm0.009\,\si{mas.yr^{-1}}$ in $K_s$-band based on the five-year analysis of \ac{sphere} astrometric calibration data presented in \citet{Maire2021}.
%

%\begin{itemize}
%    \item Few sentences on ADI / PCA for the K band observation with enough field rotation
%\end{itemize}

\begin{figure*}
\center
\includegraphics[width=0.98\textwidth]{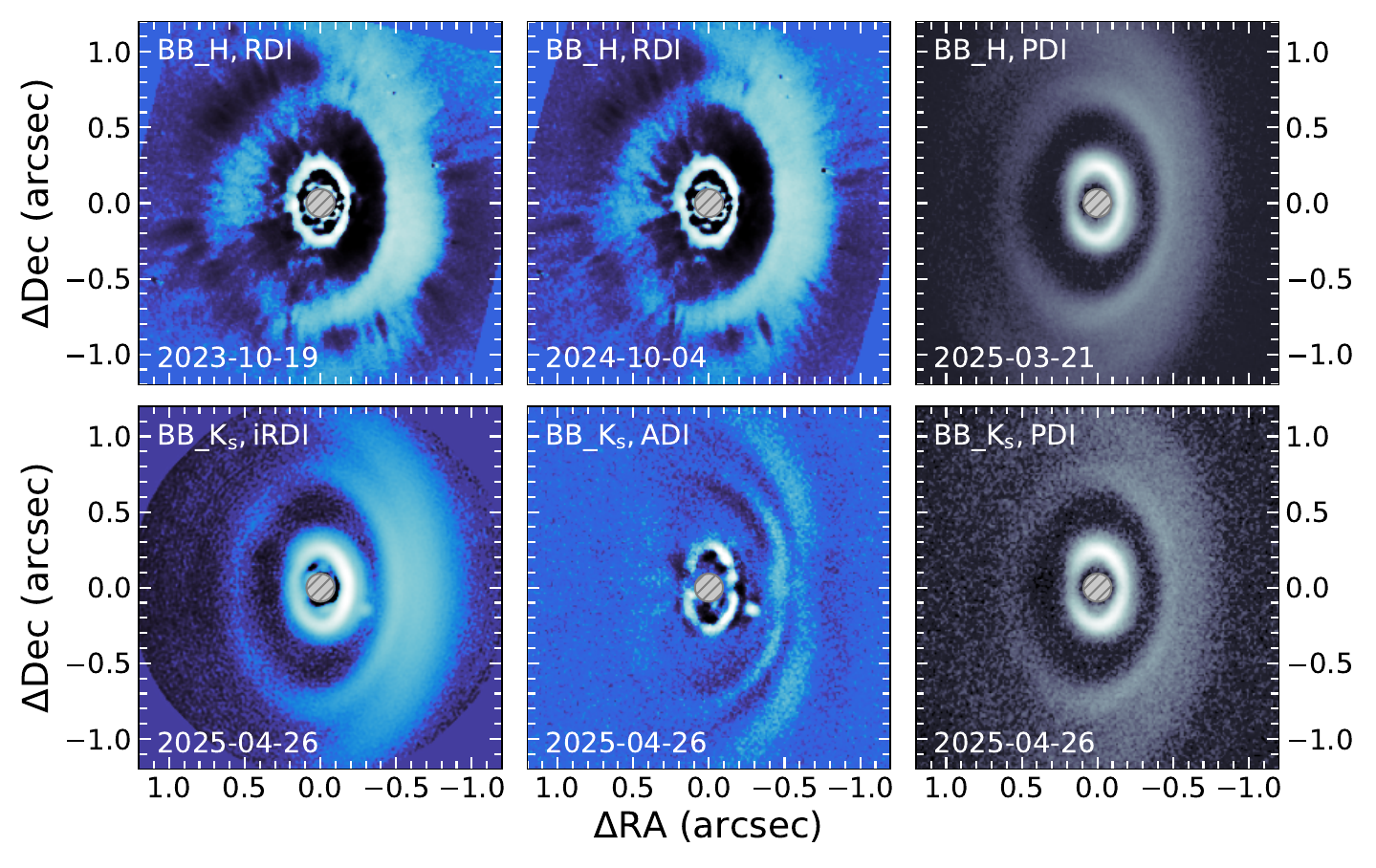} 
\caption{SPHERE/IRDIS observations of the \wisa system.
The gray, hashed disk in the image center indicates the size of the coronagraphic mask.
The differential imaging method and observed waveband for each image are indicated in the top left corner.
Blue-hued images reduced with the \ac{adi} or \ac{rdi} (50 principal components) methods are showing total intensity, sensitive to disk scattered light and thermal emission from embedded planets.
The gray-hued images are $Q_\phi$ images (reduced with the \ac{pdi} method) showing linearly polarized scattered light, not sensitive to thermal emission.
} 
\label{fig:obs-results}
\end{figure*}

\begin{figure*}[]
    \centering
    \includegraphics[width=0.999\textwidth]{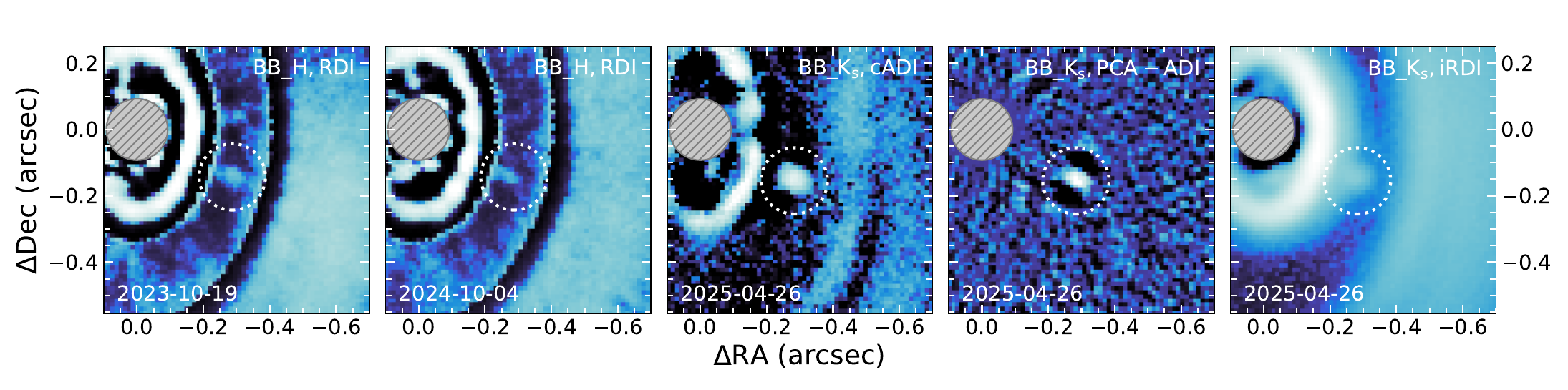}
    \caption{All detections of the embedded planet \wisb in the various observation epochs and filters.
    We indicate the embedded planet position with a white, dotted circle.
    The coronagraph position is indicated with a gray, hashed mask.
    For the 2025 $K_s$-band data we show that the embedded planet is recovered with  classical and \ac{pca}-based \ac{adi} as well as in iRDI.}
    \label{fig:planet-images}
    \end{figure*}

Due to the short integration time, the two initial $H$-band observations taken in 2023 and 2024 only have $\sim$1 degree of field rotation, which makes them unsuitable for reduction with \ac{adi}.
Instead we used \acl{rdi} \citep[\acs{rdi}; ][]{Smith1984,Lafreniere2009,Lagrange2009} with \acl{pca} \citep[\acs{pca}; ][]{Amara2012,Soummer2012}.
\acused{rdi}
\acused{pca}
We leveraged observations from \ac{yses} to create a reference library. 
We built on the reference library used in \citet{Bohn2021} and excluded \ac{yses} observations affected by image misalignment, binarity, the presence of contaminating sources in the field of view, or poor image quality, resulting in a library of 61 observations comprising 340 frames. 
The observations in this library are listed in Table~\ref{tab:reference_library} in Appendix~\ref{app:reflib}.
From this set, for each \wisa observation, the 275 frames with the highest correlation based on \ac{mse} were selected for the reference library, a strategy proven to be effective in increasing \ac{rdi} performance \citep[e.g.][]{Ruane2019,Xie2022,Sanghi2024}. 
We produced two different sets of reductions for each of these two epochs.
Initially we performed \ac{rdi} with 50 principal components on the entire image.
The results are shown in the first two panels of Figure~\ref{fig:obs-results} revealing extended circumstellar disk structure.
To increase sensitivity and minimize over-subtraction in the potential planet-hosting gap between the surrounding bright rings, we isolated the gap with an elliptical mask and applied the reduction routine separately to the gap and the rest of the image.
The results of this approach, using 30 principal components for the 2023 epoch and 40 for the 2024 epoch, are displayed in Figure~\ref{fig:planet-images}.

As there was very little parallactic field rotation in the following $H$-band epoch taken in 2025, and no reference library was available for the \ac{sphere} polarimetric imaging mode, we did not perform a total intensity differential imaging reduction for this data set.

The 2025 $K_s$-band observation sequence had significant parallactic angle rotation and thus allowed for \ac{adi} to be performed, additionally the recorded interspersed reference star images allowed also for dedicated \ac{rdi}. 
%
%For the \ac{adi} we used two approaches.
%
Initially we performed classical ADI (cADI) as outlined by \citet{Marois2006}.
The result is shown in Figure~\ref{fig:obs-results}.
%
%To increase sensitivity for potential embedded planets we then also performed \ac{pca}-based \ac{adi} (PCA-ADI) with {\tt PynPoint} as outlined in \citet{Amara2012}.
%
%We subtracted a total of {\color{red} XXXX} principal components, which yielded an image which retained some disk signal, and which is shown in Figure~\ref{fig:obs-results}.
%
We then performed more aggressive processing using \ac{pca}-based \ac{adi} (PCA-ADI) with {\tt PynPoint} as outlined in \citet{Amara2012}, focusing on the embedded planet.
We used 5 principal components for this reduction, the result of which is shown in Figure~\ref{fig:planet-images}.
Finally we performed \ac{pca}-based \ac{rdi} on the data set using the dedicated reference star observations.
To prevent over-subtraction due to the bright circumstellar disk structures, we followed the approach outlined in \citet{Ginski2021} for iterative reference differential imaging (iRDI)\footnote{We note that this approach is not suitable for the initial shorter $H$-band observation epochs as it requires a dedicated reference star sequence and does not work well with a library of observations.}.
The result of this approach is shown in Figures~\ref{fig:obs-results} and \ref{fig:planet-images}. 

%\begin{itemize}
%    \item reduced with RDI
%    \item library from YSES survey, manually removed misaligned images, binary stars, observations with sources in field or very low quality observations
%    \item further selection on most correlated frames with median observation based on MSE
%    \item elliptical mask to prevent over subtraction due to the bright inner ring (ring 3) and ring 2
%\end{itemize}

%\subsubsection{Polarimetric imaging}

Both the $H$ and $K_s$-band 2025 observations were performed in the \ac{irdis} \ac{dpi} mode, making them suitable for \ac{pdi} \citep{Kuhn2001}.
While \ac{pdi} is one of the best techniques to reveal (polarized) scattered light from circumstellar material \citep[see e.g. the discussion in ][]{Benisty2023}, it is not sensitive to unpolarized thermal emission of embedded planets.
\ac{pdi} was performed using the \acl{irdap} \citep[\acs{irdap}; ][]{vanHolstein2020} pipeline.
\acused{irdap}
As a result \ac{irdap} produces the Stokes $Q$ and $U$ images that contain the linearly polarized signal from the circumstellar disk, while having removed the unpolarized signal from the central star.
Using an instrument model as well as dedicated measurements within the final $Q$ and $U$ images, \ac{irdap} also removes residual stellar polarization which can be induced either by the telescope and instrument or by interstellar or local dust around the star.
For the polarimetric observations we generally show the $Q_\phi$ images which contain the expected azimuthally polarized scattered light signal from single scattering events as positive signal \citep[see e.g. ][ for a description of the $Q_\phi$ formalism]{Monnier2019}.
The polarized light $H$ and $K_s$-band images are both shown in Figure~\ref{fig:obs-results}.
Additionally we show the associated Stokes $Q$ and $U$ images in Appendix~\ref{app:stokes}. 

%--------------------------------------------------------------------

\section{An extended multi-ringed disk seen in scattered light}
\label{sec:disk}

\begin{figure}[htb!]
    \centering
    \includegraphics[width=0.49\textwidth]{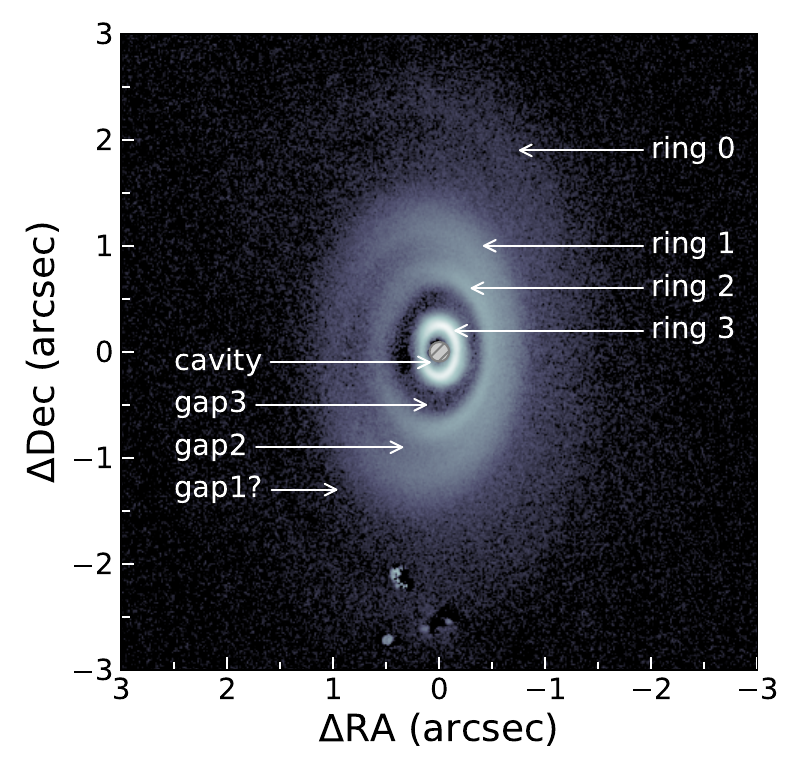}
    \caption{Polarized light $Q_\phi$ image of the \wisa system taken in the $H$-band.
    We indicate the various sub-structures that we are detecting within the scattered light signal of the planet-forming disk.
    Clusters of bad pixels from the detector are seen near the lower edge of the image.}
    \label{fig:disk-annotated}
    \end{figure}

Our \ac{sphere} observations resolve for the first time an extended circumstellar disk surrounding \wisa.
%%%%% start edit
Using the polarimetric $H$-band image as a reference we find detectable signal out to 2.8\,arcsec (380\,au) from the star along the north-south direction, which appears to coincide with the major axis of the disk (forward-scattering, near-side to the west).
%%%%% end edit
Within this region the disk appears highly structured with a set of four concentric rings separated by gaps of different sizes and contrast.
We indicate the individual rings and gaps in Figure~\ref{fig:disk-annotated}.
We used an outside-in labeling strategy for the various structures, as future observations at higher angular resolutions may well detect additional structures further in.
We detect no clear scattered light signal inside ring 3 and down to the coronagraphic mask, which may indicate that there is indeed a cavity in the disk at this position (also seen at $z'$ band in Close et al. submitted; companion letter~2).
By far the most prominent gap in the disk is located between ring 2 and ring 3.

To obtain an overall picture of the disk morphology we fitted simple geometrical models to the individual rings.
Features were extracted using a semi-automatic algorithm which involves edge detection and subsequent ellipse fitting based on the points extracted.
For the edge detection we used radial cuts from the stellar position to the outer disk regions.
For the well defined rings 1 and 2 we found the ring position by fitting one-dimensional Gaussian profiles to the radial cuts.
For the innermost ring (ring 3) this method proved problematic due to its somewhat extended flux toward the West, which may be related to the bottom side of the visible disk as we discuss in Appendix~\ref{app:inner disk bottom}.
In this case, we used the simple maximum of the disk flux along each radial profile instead.
For the outermost ring (ring 0), the signal was too faint for Gaussian fitting, thus in this case we also used the largest value in the profile (including some thresholding to exclude spurious data points).
%

%For rings 1, 2 and 3, Gaussian fitting was used.
%
In addition to the visible rings, we also measured the position and width of the gap between rings 2 and 3 by applying inverse Gaussian fitting, where the algorithm targets the intensity minima along the pixel line rather than the peaks.
%
%A summary of the adopted measurement convention and the results of each of the positional fits are listed in Appendix~\ref{app:geometric_fitting_table}.
The results of each of the positional fits are listed in Appendix~\ref{app:geometric_fitting_table}.
The approximate radial locations for ring 0 -- 3 are at 316\,au, 164\,au, 97\,au and 38\,au.
The center of the gap is located at approximately 69\,au and the gap has a width of 59\,au in the $H$-band image, based on the \ac{fwhm} of the inverse Gaussian used for fitting.
%
% We find an average inclination of $43.990 \pm 0.866^\circ$ and position angle of $88.686 \pm 1.12^\circ$ for the $H$-band and for the $K_s$-band, $45.863 \pm 1.269^\circ$, $89.308 \pm 2.327^\circ$ respectively.
We find an average inclination of $43.99 \pm 0.87^\circ$ and position angle of $358.7 \pm 1.1^\circ$ in the $H$-band, and $45.86 \pm 1.27^\circ$ and $359.3 \pm 2.3^\circ$ in the $K_s$-band, respectively (see Fig.~\ref{app:fig:disk_deprojection} in Appendix~\ref{app:geometric_fitting_table} for convention).
%\footnote{The position angle follows the convention that it is measured from the North in direction of East (counter-clockwise) towards the disk major axis on the disk near side.}.

    \begin{figure*}[htb!]
    \centering
    \includegraphics[width=0.99\textwidth]{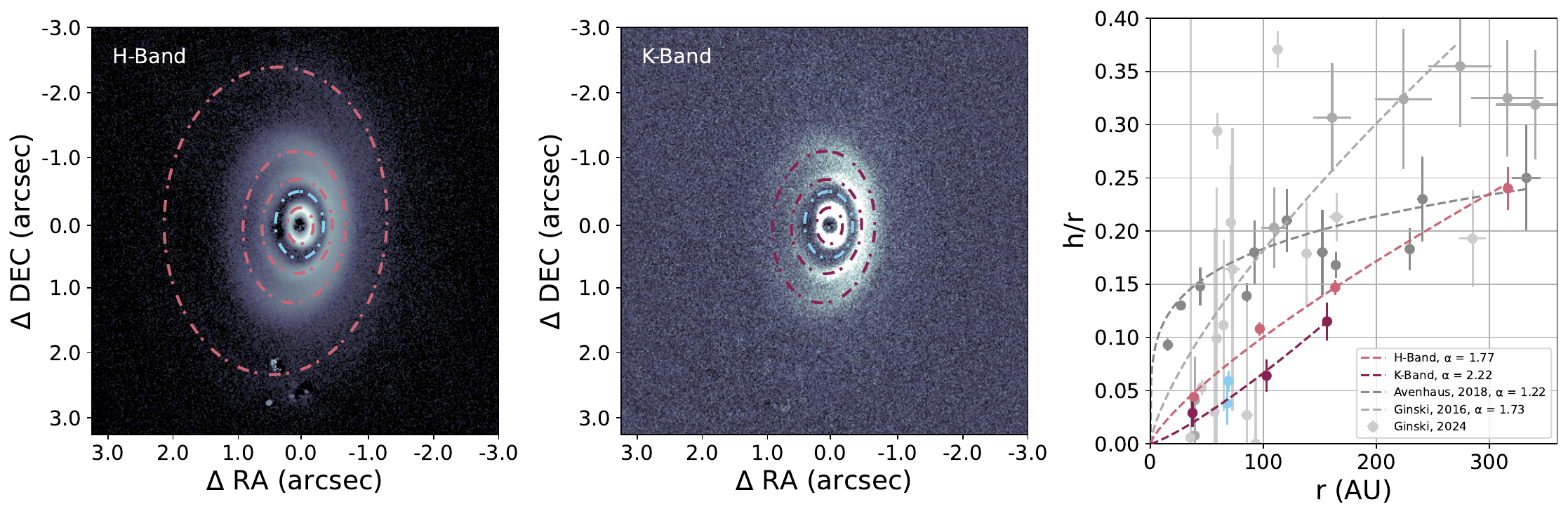}
    \caption{Left and Middle: Geometric fitting of the disk in both $H$-band and $K_s$-band images.
    Right: The aspect ratio $(h/r)$ vs. radius (au) of both bands.
    For comparison we include the literature measurements of \citet{Ginski2016,Avenhaus2018,Ginski2024} for a total of 17 disks (some of which also have a multiple ringed sub-structure) as gray data points.}
    \label{fig:geometric_fitting}
    \end{figure*}

Following \citet{deBoer2016} we used the offset of the ellipse center from the stellar position along the minor axis to measure the height of the disk scattering surface above the disk mid-plane.
This assumes that the identified rings are not significantly eccentric, as an inherent eccentricity can also lead to a center offset along the minor disk axis independent of the disk vertical structure.
Figure~\ref{fig:geometric_fitting} shows the aspect ratio (the disk height divided by the radial separation) $h/r$ vs. $r$, which can be described by a power law, 

\begin{equation}
\label{eqn:h/r}
\frac{h}{r} = \frac{h_0}{r_0}\left(\frac{r}{r_0}\right)^{\alpha-1},
\end{equation}

where $h_0$ describes the height at radius $r_0$, and $\alpha$ is the flaring index.
We fit this power law to the data using ring 1 as the reference point ($h_0$ = 24.0\,au, $r_0$ = 163.6\,au), yielding flaring indices of 1.77 in the $H$-band and 2.22 in the $K_s$-band.
Previous determinations of the flaring index include 1.22 from \citet{Avenhaus2018} for a joint fit of a set of 5 T~Tauri stars and 1.73 from \citet{Ginski2016} for the extreme case of the Herbig star HD\,97048.
Our $H$-band result closely matches the latter, while the $K_s$-band flaring index is significantly higher.
This high value is driven by the strong discrepancy between the relatively flat inner disk (ring 3) and the strongly flared outer disk (rings 1 and 2).
As is visible in Figure~\ref{fig:geometric_fitting} (right panel), the inner disk (ring 3) in the \wisa system is indeed among the gas-rich disks with the lowest aspect ratio reported in the literature to date \citep{Avenhaus2018, Ginski2016, Ginski2024}.
While the disk vertical profile is then subsequently rising steeply (as indicated by the large flaring index we find), the disk vertical extent remains on the low end compared to the full literature population with the outermost well defined ring 1 having only an aspect ratio of $0.15\pm0.01$.
We also note that the overall aspect ratio that we recover is smaller at all radial locations for the $K_s$-band than for the $H$-band.
This is an expected behavior related to lower dust opacities at longer wavelengths, which look deeper into the disk at longer wavelengths.

%--------------------------------------------------------------------
\vspace{0.5cm}
\section{Thermal emission from an embedded planet}
\label{sec:planet}

\subsection{Astrometric analysis}
\label{sec:astr_planet}
We extracted the astrometry and photometry of the observation in $K_s$ band following the methods of \citet{Stolker2020b}.
We used the \texttt{SimplexMinimizationModule} of \texttt{PynPoint} to first obtain an approximate position and flux of the companion by minimizing the flux residuals, evaluated in a 5 pixel aperture around the injection position, with a downhill simplex method.
We then performed Bayesian inference using Markov chain Monte Carlo \citep[MCMC; ][]{MacKay03} to sample from the posterior distribution, using the \texttt{MCMCsamplingModule}.
The systemic uncertainties of the injection and minimization approach are derived with \texttt{SystematicErrorModule}, which injects positive artificial companions with the same magnitude contrast and at the same radial separation as \wisb at positions equidistantly distributed in polar space, and then extracts the astrometry and photometry of these artificial companions with the same method as described above.
The error on the pixel position of the companion is the combination of the standard deviation across the injected positions and the astrometric uncertainty of the fit of the companion.

The planet signal is significantly less prominent in the $H$-band observations than in the $K_s$-band observations due to shorter exposures and negligible field rotation.
As a result, extracting the astrometry from the $H$-band snapshots requires a different approach.
In these observations, the region near the companion is dominated by speckle noise, partially introduced by the \ac{rdi}/\ac{pca} subtraction. 
To robustly extract the position of the companion, we performed a 2D Gaussian fit to each of the 8 individual frames, using the residuals from all \ac{rdi}/\ac{pca} reductions resulting from subtraction of 0 to 100 principal components in steps of 2, with bounds applied to the fit parameters.
For each of the 8 frames, this yielded 50 pixel position measurements, with uncertainties from the covariance matrix of the Gaussian fit.
Details of the fitting routine and bounds are provided in Appendix \ref{app:gauss_fit_astrometry}.
For each frame, we filtered out unsuccessful fits---specifically, fits that failed to converge, reached imposed bounds, or had positional uncertainties exceeding 100 pixels (more than half the size of the cropped frame).
The remaining fits were averaged per frame, with uncertainties derived from both the spread in measurements and the uncertainties of the individual fits.
The final astrometric measurement was obtained by computing the weighted mean and standard deviation across the 8 independent frames.

In conversion from pixel positions to separation (arcsec) and position angle (degrees) of both $H$-band and $K_s$-band observations, uncertainties in pixel scale, true North correction, and pupil offset (as outlined in Section \ref{sec:data_reduction}), as well as the centering precision of 2.5 mas of the star behind the coronagraph, were included in the error budget.
The resulting astrometric measurements are reported in Table~\ref{tab:astrometry}.

\begin{figure}[htb!]
    \centering
    \vspace{0.25cm}
    \includegraphics[width=\hsize]{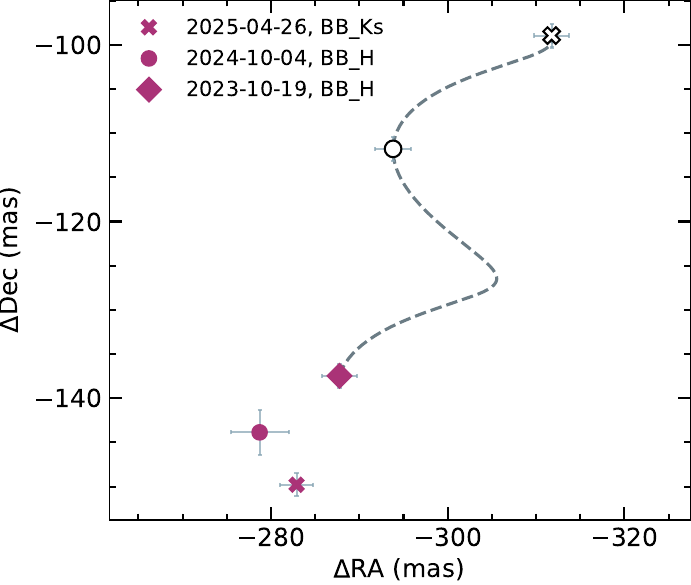}
    \caption{Proper motion analysis of \wisb.
    Each epoch is represented by a unique marker shape: diamond (2023), circle (2024) and cross (2025).
    The colored version of each marker denotes the measured position of the companion.
    The unfilled (black outline, white center) version of each marker shows the expected position of the source if it were a stationary background object.
    The dashed curve illustrates the parallactic motion of such a background object from first to last epoch.}
    \label{fig:ppm_plot}
    \end{figure}

The predicted parallactic motion of a stationary background object was calculated using the Gaia DR3 distance to \wisa (see Table~\ref{tab:star}).
Figure~\ref{fig:ppm_plot} plots the measured companion positions for all three epochs relative to these background tracks, confirming that the companion is inconsistent with a background source.

% MAK commented out Orbital fit out b/c it's a single section
% \subsubsection{Orbital fit}
Given our astrometric measurements of the planet, we tested whether the change in position relative to the central star (apparent in Figure~\ref{fig:ppm_plot}), is consistent with the expected orbital motion of a low mass planet.
However, the astrometric uncertainties for the $H$-band epoch on 2024-10-04 are notably larger than those of the other two epochs, as also seen in Figure~\ref{fig:ppm_plot}, primarily due to poorer observing conditions (see the higher seeing values in Table~\ref{tab:obs_setup}).
We also note increased contamination by residuals from the bright inner ring---compounded by the fainter planet signal---which biases the Gaussian fit center toward ring 3.
Given these limitations, we excluded the 2024 epoch from the orbital fit and used only the first ($H$-band 2023-10-19) and last ($K_s$-band 2025-04-26) epochs, which gives us the longest time baseline currently available for the system.
%%%%%edit: paragraph break

\begin{figure}
%\begin{figure}[htb!]
    \centering
    \includegraphics[width=0.48\textwidth]{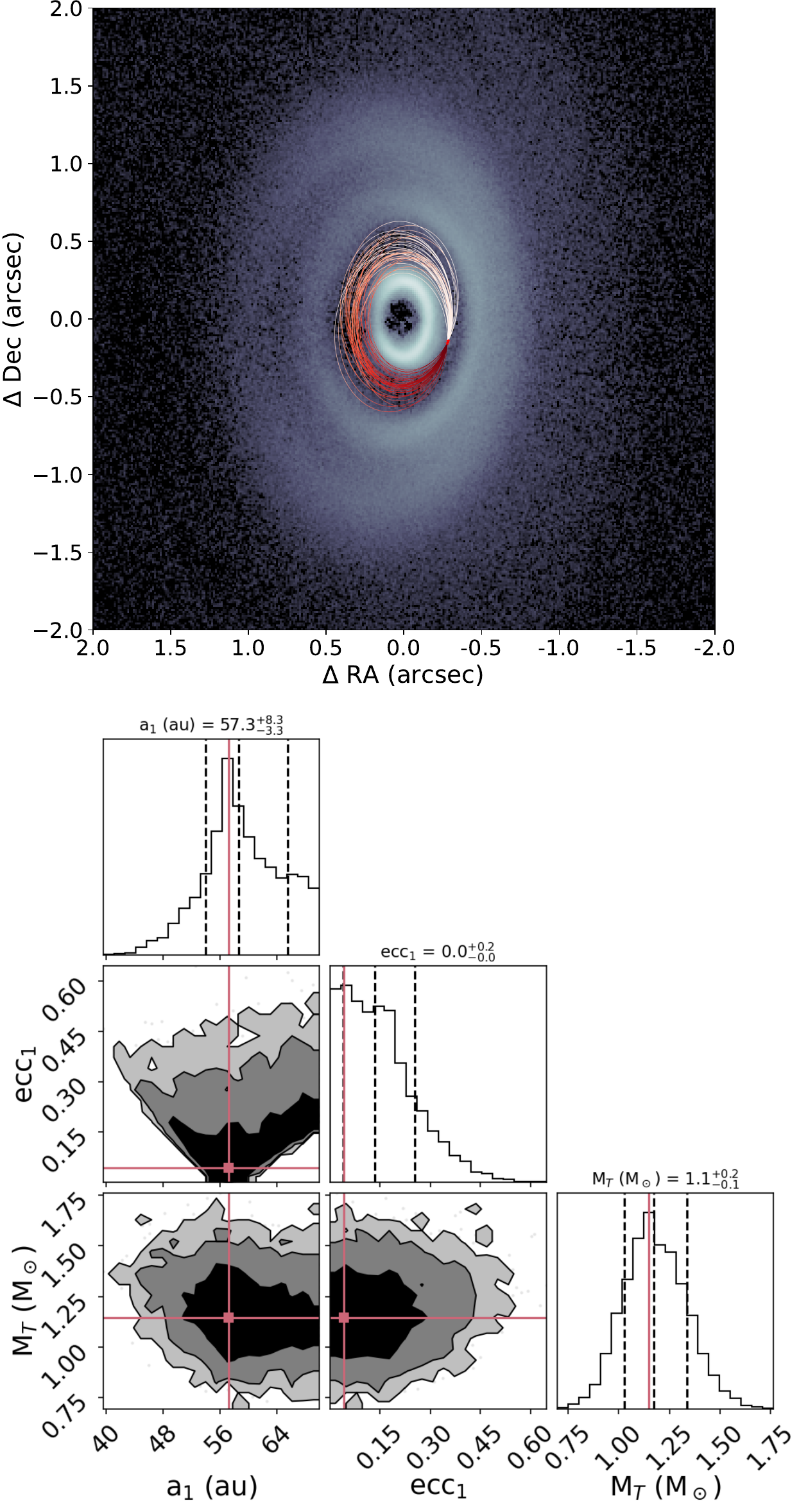}
    \caption{\textit{Top panel:} Predicted orbits of the planet (known astrometry denoted by red star) overlaid on the $H$-band polarized scattered light image.
    \textit{Bottom panel:} Extracted orbital elements and total system mass.}
    \label{fig:orbit_overlay}
    \end{figure}

The orbital fit was performed using the {\tt orbitize!} package \citep{Blunt_2020}, using the \ac{ofti} algorithm (\citealt{Blunt2017}) as it is particularly well-suited for systems where only a small fraction of the orbital motion has been observed, as is the case with \wisb.
As the planet is located in the disk gap, and the disk appears very symmetric and unperturbed in scattered light, we assume that the planet does not cross the disk.
For this initial orbital analysis, we further assume that the planet's orbit is co-planar and aligned with the disk.
These assumptions result in a fixed value for the orbital inclination of 135$^\circ$ (encoding also the clockwise orbital motion of the planet).
The angle of the ascending node was fixed at 0$^\circ$.
These values are consistent with our disk geometric fit presented in section~$\ref{sec:disk}$.
For the semi-major axis we used the default log-uniform prior with lower and upper bounds initially set to 40\,au and 100\,au, respectively.
The upper bound was purposefully chosen to be slightly larger than the outer edge of the disk gap to prevent a pile-up of solutions in the posterior distribution at the parameter boundary.
After the fits were concluded we then down-selected only solutions with a maximum semi-major axis of 70\,au, which are then fully contained within the disk gap.
Finally we chose the default Gaussian prior for the total system mass centered at 1\,M$_\odot$ and with a standard deviation of 0.1\,M$_\odot$.
A selection of orbital bundles from the fitting routine is shown in Figure~\ref{fig:orbit_overlay}.
According to the histogram in the bottom panel of Figure~\ref{fig:orbit_overlay}, the most probable semi-major axis is $\sim$ 57\,au, which falls in the inner region of the gap between ring 3 and the deepest part of gap 3.
Despite this clear peak, the distribution shows that this semi-major axis may extend upwards of 60\,au, which could place the planet at the deepest part of gap 3, where it may be the sole contributor to the dust depletion. 
With current constraints the distribution of eccentricities derived from the orbital fit is heavily skewed toward low values, with 93\% having $e<0.3$ and 77\% having $e<0.2$.

\begin{table*}
\caption{Astrometric and photometric measurements of \wisb.}
\label{tab:astrometry}
\centering
\def\arraystretch{1.2}
\setlength{\tabcolsep}{9pt}
\begin{tabular*}{\textwidth}{@{}llllll@{}}
\hline\hline
Observation date & Filter & Separation & Position angle & Magnitude contrast & Absolute magnitude \\
(yyyy-mm-dd) & & (mas) & (\textdegree) & (mag) & (mag) \\
\hline
2023-10-19 & $H$ & $318.9 \pm 2.2$ & $244.5 \pm 0.2$ & $9.8^{+0.4}_{-0.3}$ & $12.8^{+0.4}_{-0.3}$\\
2024-10-04 & $H$ & $313.7 \pm 3.5$ & $242.7 \pm 0.4$ & $10.6^{+0.6}_{-0.4}$ & $13.5^{+0.6}_{-0.4}$\\
2025-04-26 & $K_s$ & $320.1 \pm 2.1$ & $242.1 \pm 0.2$ & $9.0^{+0.04}_{-0.04}$ & $11.95^{+0.09}_{-0.09}$ \\
\hline
\end{tabular*}
\end{table*}

\subsection{Photometric analysis}
\label{sec:phot_planet}
The $K_s$-band flux was retrieved following the methods of \citet{Stolker2020b}, as described in Section~\ref{sec:astr_planet}.

To measure the planet flux in our short \wis $H$-band observations of 2023 and 2024 we used the \ac{rdi} reductions presented in Figure~\ref{fig:planet-images}.
For the 2023 epoch, we used the median-combined residuals from \ac{rdi}/\ac{pca} processing with 30 principal components.
Due to highly variable observing conditions during the 2024 epoch, we selected a subset of frames based on point source visibility and low RMS noise.
The final 2024 reduction uses median-combined residuals from frames 2, 4, 5 and 8 (of 8 total), processed with \ac{rdi}/\ac{pca} using 40 principal components.
As a calibrator we used the flux images taken of the central star in the same observation sequences, where the star was moved away from the coronagraph and a neutral density filter was inserted to prevent saturation.
The planet flux was measured inside an aperture with a radius of 4 pixels.
To account for \ac{rdi} over-subtraction we estimated the local background with the same aperture along the disk gap near the planet.
For the 2023 observation we find an $H$-band contrast between planet and central star of 9.8$^{+0.4}_{-0.3}$\,mag.
For the 2024 observation we measure 10.6$^{+0.6}_{-0.4}$\,mag.
While this is consistent within uncertainties with the magnitude contrast derived from the 2023 epoch, it is notably higher.
%%%% edit: paragraph break

In both epochs, a primary source of uncertainty is the background variation within the disk gap after the \ac{rdi} reduction.
However, in 2024 the main contributor to the uncertainty are the highly variable and overall poorer observing conditions (see Table~\ref{tab:obs_setup}), which led to non-detections in some frames and only marginal detections in others.
As a result, the embedded planet signal appears significantly fainter and is more likely to be partially removed in \ac{rdi} processing, as also discussed in Section~\ref{tab:astrometry} and visually evident in the residuals shown in Fig~\ref{fig:planet-images}.
Given these challenges associated with the 2024 data and its larger uncertainties, we adopt the magnitude derived from the 2023 epoch for further analysis.

\begin{figure}[t!]
    \centering
    \includegraphics[width=\hsize]{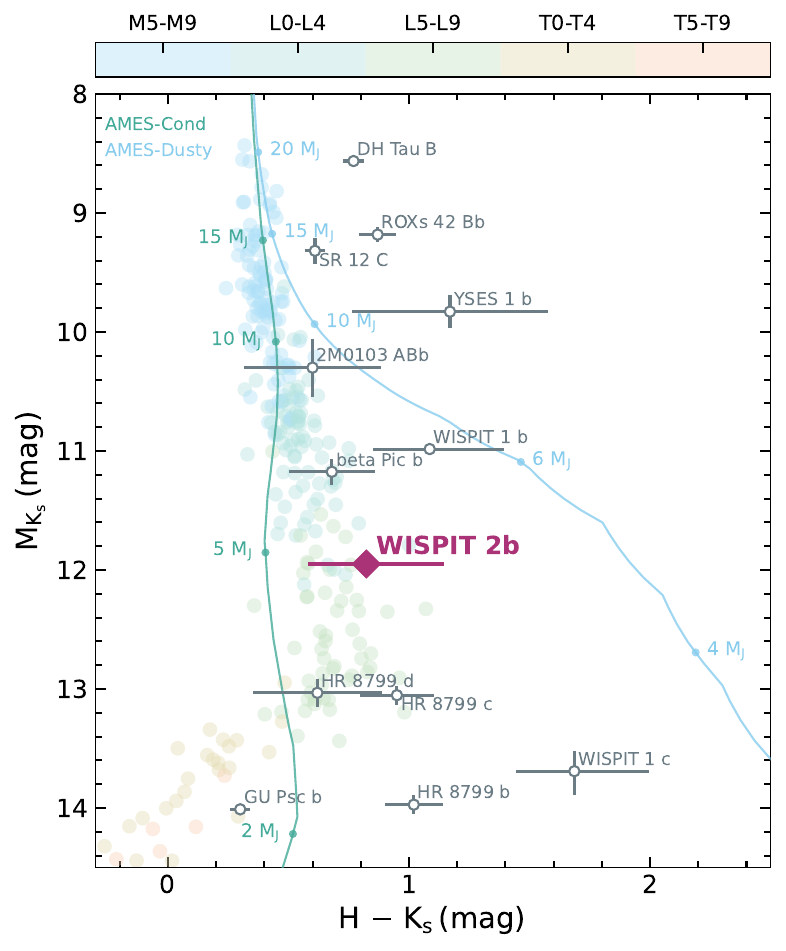}
    \caption{Color-magnitude diagram of \wisb, with field brown dwarfs of various spectral types and confirmed planetary companions. Teal and cyan tracks show 5.1 Myr AMES-COND and AMES-DUSTY isochrones respectively. \wisb is marked in purple.}
    \label{fig:cmd}
    \end{figure}

The color-magnitude of the companion is shown in Figure~\ref{fig:cmd}, along with $5.11$ Myr AMES-COND and AMES-DUSTY isochrones \citep{Allard2001,Chabrier2000} and other known planets with available $H$ and $K_s$ magnitudes.
To estimate the mass of the companion, we sampled from its asymmetric $H$-magnitude, $K_s$-magnitude, and age distributions.
%%%%% start edit
Here the age is the previously derived stellar age of $5.1^{+2.4}_{-1.3}$~Myr (see Section~\ref{sec:wispit2stars}).
%%%%% end edit
We used \texttt{species} \citep{Stolker2020b} to retrieve color-magnitude data from the AMES isochrones, and estimated the companion's mass by interpolating the sampled $K_s$-band magnitude onto the AMES-COND and AMES-DUSTY evolutionary model grids separately.
Whilst we note that both models give a consistent mass range for \wisb, we obtained a final mass estimate by interpolating the $H-K_s$ color between the two isochrones.
The derived masses from both models as well as the interpolated mass are presented in Table~\ref{tab:mass_models}; the final adopted mass is $4.9^{+0.9}_{-0.6}$~\mj.
Both the photometric measurements and the resulting mass are consistent with a planetary classification of the companion.

We caution that the reported uncertainty may be underestimated due to various factors that require additional data and further analysis, such as the age of the star---spectral analysis is necessary to more accurately characterize \wisa, which may, in turn, affect the mass estimate.
Additionally, there are factors that cannot be quantified with the current data. 
The most significant of these is the unknown extinction that could potentially be introduced by a \ac{cpd} or by the circumstellar disk itself, as was inferred in the case of PDS~70b (\citealt{Christiaens2019}). 
However, a comparison with PDS~70b suggests a lower degree of extinction in our case.
Using the $H$-band magnitude of $14.9 \pm 0.90$ from \citet{Mesa2019} and $K_s$-band magnitude of $11.37 \pm 0.30$ from \citet{Wahhaj2024}, PDS~70b has a significantly redder $H-K_s$ color of $ 3.53\pm0.95$, placing it far to the right of the AMES-DUSTY track in Figure~\ref{fig:cmd}, and even outside the plotted bounds.
In contrast, the color of \wisb lies between the AMES-COND and AMES-DUSTY isochrones, indicating significantly less reddening and pointing to a less substantial and/or more dust-depleted \ac{cpd}.
This is compounded by the lack of \ac{cpd} signal in polarized light from our $K_s$-band data, especially given the high SNR.
Both the color and lack of polarized emission suggest that the extinction is likely to be less severe than in the PDS~70b case.
Whilst we cannot reliably constrain this extinction, \wisb's low mass, which is well below the deuterium burning limit, makes it highly unlikely that additional extinction would alter the conclusion regarding its planetary nature.
Further observations and analysis are required to improve upon the companion's mass accuracy.

\begin{table}[h!]
\caption{Mass of \wisb from evolutionary models.}
\label{tab:mass_models}
\centering
\def\arraystretch{1.2}
\setlength{\tabcolsep}{55pt}
\begin{tabular}{@{}ll@{}}
\hline\hline
Model & Mass (\mj) \\
\hline
AMES-COND & $4.87^{+0.98}_{-0.59}$ \\
AMES-DUSTY & $4.78^{+0.94}_{-0.54}$ \\
\textbf{Interpolated} & $\mathbf{4.85^{+0.94}_{-0.58}}$ \\
\hline
\end{tabular}
\tablecomments{Mass estimates for \wisb\ using AMES-COND and AMES-DUSTY evolutionary tracks.
The final value in the \textit{Interpolated} row is derived by interpolating the $H-K_s$ color between both model grids.}
\end{table}

~
\newpage ~\newpage
\section{Planet-disk interaction}
\label{sec:planetdiskinteraction}

The opening of gaps in the planet-forming disk by embedded planets is predicted by theoretical models of planet-disk interaction and arises from the gravitational torques that the planet exerts on the surrounding disk gas and the resulting angular momentum exchange \citep[see e.g. ][]{Paardekooper2006, Bae2017}. 
The relation between the mass of the planet and the width of the gap that is opened by it has been the subject of detailed hydrodynamic studies.
To comparatively assess the estimated planet mass derived via photometric analysis, we implemented the \citet{Kanagawa_2016} and \citet{Zhang2018} models to investigate the relationship between gap width and the star-planet mass ratio for the case of the \wisa system.
The \citet{Kanagawa_2016} model focuses exclusively on the gas surface density, creating a more idealized framework.
In contrast, the \citet{Zhang2018} model incorporates both gas and dust as coupled but distinct fluids, simulating their interaction to directly model the dust emission gaps observed by ALMA.
Additionally, the two models adopt different definitions for the gap width parameter, $\Delta_{gap}$. 
\citet{Kanagawa_2016} define it as the absolute radial width $r_{out} - r_{in}$, while \citet{Zhang2018} normalize it to the gap location using $\frac{r_{out} - r_{in}}{r_{out}} $.
%%%% edit paragraph break 

\begin{figure}[htb!]
    \centering
    \includegraphics[width=\hsize]{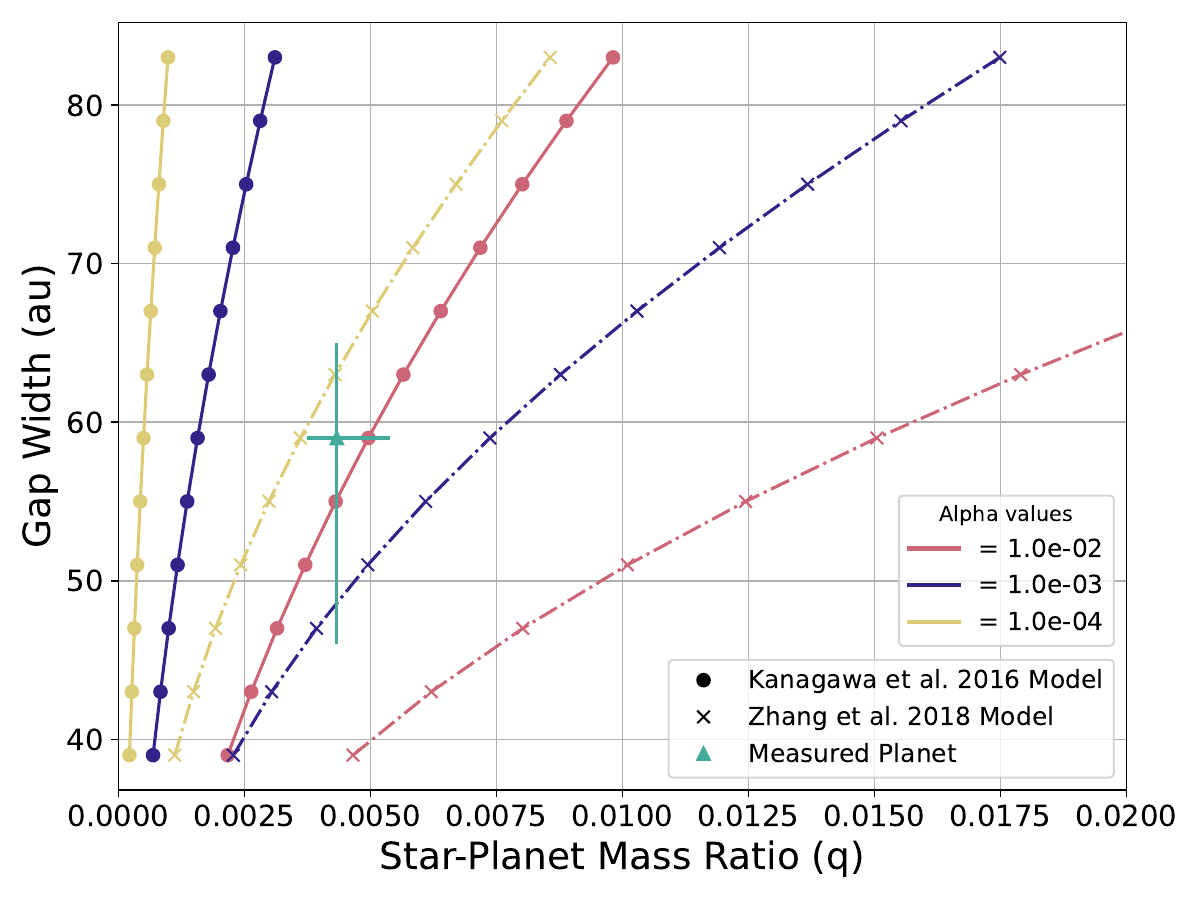}

    \caption{Relationship between the gap width and the mass ratio $q$.
    Here the \citet{Kanagawa_2016} model is represented by the dots and the \citet{Zhang2018} is represented by the crosses.
    A triangle marker has been inserted to show the star-planet mass ratio based on magnitude and age estimates.} 
    \label{fig:mass-width}
    \end{figure}

Figure~\ref{fig:mass-width} illustrates the increasing trend of the mass ratio $q$, with gap width for various disk viscosity parameters $\alpha$ \citep{Shakura1976}.
Using the gap width estimated from the Gaussian fitting of the dust rings (see Table ~\ref{tab:geometric_fitting}), $\sim$ 59\,au measured along the disk major axis in the $H$-band, we apply the respective models: 
\begin{equation}
    \label{eqn:zhangmodel1}
    \frac{M_p}{M_*} = 2.1 \times 10^{-3} \left(\frac{\Delta_{gap}}{R_p}\right)^2\left(\frac{h_p}{0.05R_p}\right)^\frac{3}{2}\left(\frac{\alpha}{10^{-3}}\right)^{\frac{1}{2}},
\end{equation}
and 
\begin{equation}
      \label{eqn:zhangmodel2}
      \frac{K'}{0.014} = \frac{q}{0.001}\left(\frac{\frac{h}{r}}{0.07}\right)^{-0.18}\left(\frac{\alpha}{10^{-3}}\right)^{-0.31},
\end{equation}

where the fitting parameter $K'$ is $K' = A\Delta^B$, and the best-fit gas surface density parameters of $A$ and $B$ are taken from Table~1 of \citet{Zhang2018}. 
%%%%% edit paragraph break

Considering the full range of viscosity $\alpha$ parameters from 10$^{-4}$ to 10$^{-2}$ we get mass ranges for the gap opening planet of 0.5\,\mj to 5.3\, \mj for the \citet{Kanagawa_2016} models and 4\,\mj to 16\,\mj for the \citet{Zhang2018} models.
Our photometrically estimated planet mass of $4.9^{+0.9}_{-0.6}$~\mj lies thus slightly above the upper end of the former and within the lower end of the latter models.
Thus in principle the scattered light gap is consistent with being opened by the detected planet, dependent on the disk viscosity.
The estimated planet mass, if assumed to sit in a gap 59\,au wide, aligns with a disk viscosity parameter $\alpha = 10^{-2}$ under the \citet{Kanagawa_2016} model, or $\alpha = 10^{-4}$ under the \citet{Zhang2018} model.
If we consider the lower bound of the measured gap width, approximately 46\,au, based on the \ac{fwhm} of the disk gap, the \citet{Zhang2018} model appears to more accurately capture the gap-opening efficiency of an embedded planet of this mass.
Conversely, if the upper bound of 65\,au (estimated from the $K_s$-band) more accurately represents the gap width, again the \citet{Zhang2018} model with $\alpha = 10^{-4}$ offers an excellent match.
Taken together, these results suggest that the disk viscosity might be well represented by at least $\alpha = 10^{-2}$ or higher.

%--------------------------------------------------------------------
%\vspace{-0.2cm}
\section{Comparison with other young planet-forming systems}
\label{sec:comparison}

In the following we will put the \wisa system in context of similar young systems that either host a (candidate) planet embedded in the disk or that show a disk with multiple ring structures.

\subsection{Systems with a detected proto-planet or candidate}
\label{sec:comparison_planets}

There are now several systems in the literature with resolved observations of the planet-forming environment and either direct or indirect detections of proto-planets or planet candidates.
Of these the most prominent is the PDS\,70 system \citep{Keppler2018, Haffert2019} in which two gas giants with masses of 0.5-10\,\mj and $<5$\,\mj are located in the gap of a transition disk around a K-type star \citep{Mueller2018, Stolker2020, Mesa2019, Wang2021}.
The size of the disk gap in the system (as measured from scattered light, near-infrared observations) is $\sim54$\,au.
The circumstellar disk consists of a single ring structure \citep{Keppler2018} with possibly some outer spiral arms detected toward the disk ansae \citep{Juillard2022} and has an outer extent of $\sim$100\,au in scattered light.
Compared with PDS\,70 the disk around \wisa is significantly more extended (roughly by a factor 3-4) and shows multiple concentric rings.
The vertical aspect ratio of the disk around PDS\,70 is 0.13 at 100\,au (\citealt{Keppler2018}, based on J-band data), while the disk around \wisa is flatter having an aspect ratio of 0.11.
The embedded planet \wisb is located at $\sim$55\,au, which is significantly further out than is the case for either of the PDS\,70 planets, which are located at 20.6\,au and 34.5\,au for b and c respectively \citep{Haffert2019}.
Given our photometric analysis, \wisb's mass appears to be close to the mass of PDS\,70\,b.
We do not yet have a clear picture of the circumplanetary environment of \wisb.
The observation of a significant H$\alpha$ signal by Close et al. (submitted; companion letter~2), seems to indicate the presence of a circumplanetary accretion disk, which is similar to the case of both PDS\,70 planets.  
%
%Given the seemingly similar stellar properties, this difference in disk aspect ratio might indicate that the \wisa system is somewhat older in evolutionary terms than the PDS\,70 system and is thus further along the dispersal of the disk gas.
%
%This would fit the picture given by the stellar age estimates where PDS\,70 appears to be XX\,Myr younger than \wisa.
%
%However we caution that we do not currently have a mass estimate for the disk in the \wisa system from mm-wavelength observations.

Besides the PDS\,70 system, there are currently two strong candidates for directly detected, embedded protoplanets in the AB\,Aur system \citep{Currie2022} as well as most recently in the HD\,169142 system \citep{Gratton2019, Hammond2023}.
Both of these orbit more massive Herbig stars compared to the \wisa system.
The main difference between both of these planet candidate detections and \wisb is that they both appear strongly embedded in local dust, with AB\,Aur\,b appearing as an extended source in the near-infrared and HD169142\,b showing a scattered stellar light spectrum \citep{Currie2022, Hammond2023}.
Neither of these have significant direct H$\alpha$ point source emission detected, but AB\,Aur\,b may have evidence of weak variable, somewhat extended, H$\alpha$~emission \citep{Bowler2025}.  
%The mass and nature of the object are still somewhat uncertain due to it being heavily embedded. The scattered light observations of AB\,Aur show a complex morphology, with the signal extending beyond the 6 arcsec field of view of SPHERE (\citealt{Boccaletti2020}). The system is likely to still interact with the star-forming environment (\citealt{Grady1999, Tang2012, Speedie2025}).
%Conversely the planet candidate in the HD\,169142 system was detected in a scattered light gap of the surrounding multi-ringed disk at an orbital separation of 37\,au. In the surrounding disk material outside of the gap the planet is observed to drive a spiral wake, as predicted by planet-disk interaction theory (\citealt{Hammond2023}). The estimated mass of the planet is based on the disk morphologies and models of circumplanetary disks and is of the order of 1-5\,M$_\mathrm{Jup}$. While there are clear indications of planet disk interatcion as well as orbital motion the planet spectrum is consistent with scattered stellar light suggesting heavy embedding by circumplanetary material. Consequently the planet signal is also detected in polarized light.\\
Of these two cases the disk morphology with a bright inner ring and fainter outer rings of HD\,169142 resembles most closely that of the \wisa system.
However, we do not detect significant polarized signal from the embedded planet position in \wisa which may suggest that the \ac{cpd} is more depleted of dust than is the case for HD\,169142.
It may also be possible that the gap in the \wisa system is generally more depleted of material.
An indication of this is that we see a signal in the inner disk ring that we speculate is the bottom side of the inner disk rim (see discussion in Appendix~\ref{app:inner disk bottom}).
This structure should only be visible if the gap is almost devoid of small dust particles \citep{George2025}.
The fact that the embedded planet in the \wisa system presents indeed as an unresolved point source with an $H-K_s$ color matching that of models of a young low-mass object (see Section~\ref{sec:planet}) reinforces the interpretation that we are receiving mostly thermal emission from the planet location and not scattered light emission, and that the \ac{cpd} is less pronounced in this case. 

%This could then also be an indication that indeed the \wisa system is further along in the evolution and eventual dispersal of the planet forming disk. However, mm-wavelength observations will be required to establish the disk mass and to see if it fits within this interpretation.

\subsection{Multi-ringed disks}

\begin{figure}
\center
\includegraphics[width=0.49\textwidth]{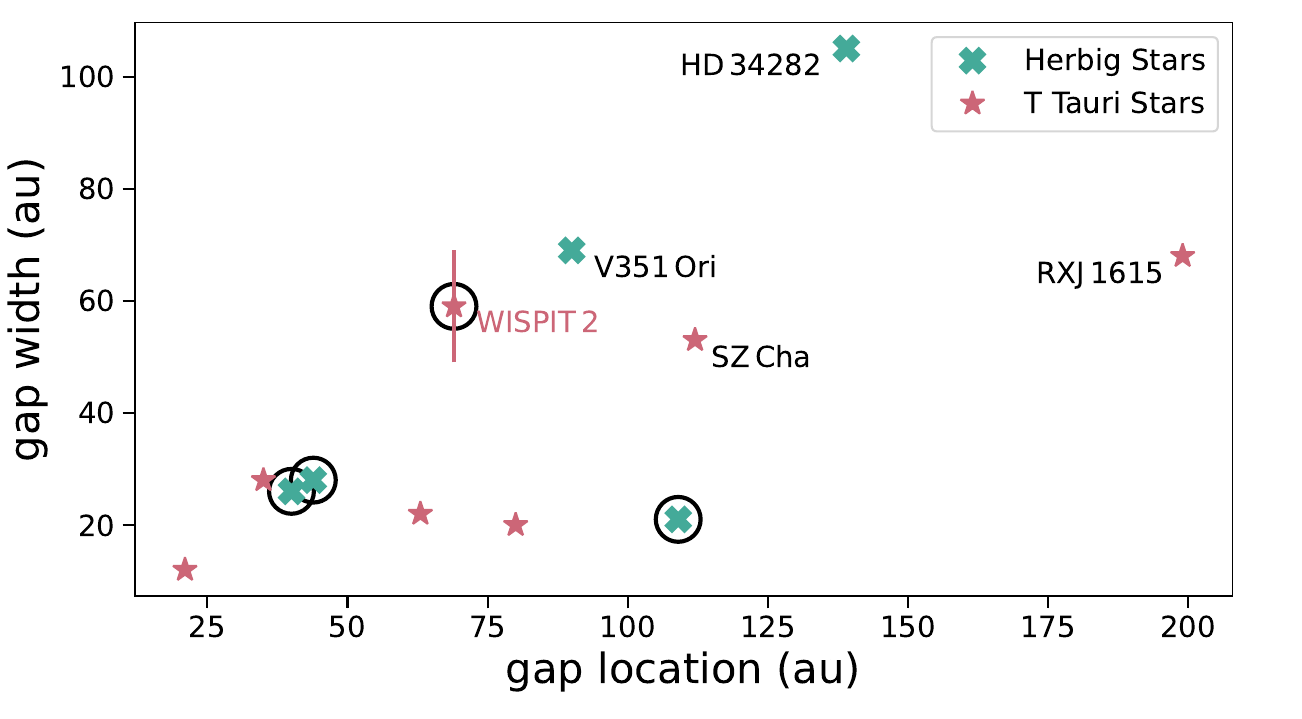} 
\caption{Full sample of all disk observations in near-infrared scattered light for which multiple rings were detected.
We show the location and width of the widest gap in each system as given in the literature.
When no specific gap widths were given, we used the difference between the peaks of adjoining rings as a measure for the gap width instead.
Positional uncertainties are typically low and on the order of $\sim$1\,au.
We distinguish between T\,Tauri and Herbig stars.
The systems with a similar or larger gap-width as \wisa are labeled.
The systems with a detected planet (\wisa) or planet candidates (HD\,169142, HD\,163296, HD\,97048) are indicated with a black circle around the marker.
} 
\label{fig: multi-ringed-sample}
\end{figure}

While rings in disks are among the more common sub-structures detected in scattered light \citep[see ][ for a recent review on disk demographics]{Benisty2023}, multi-ringed disks are somewhat rarer.
To the best of our knowledge there are currently 12 systems known in the literature that show at least two rings in scattered light (see Appendix~\ref{app:sec:multi-ringed-disks}).
To compare the disk in the \wisa system with the larger population we show in Figure~\ref{fig: multi-ringed-sample} the widest gap width and location in each system as taken from the literature.
The sample splits roughly evenly between low-mass T\,Tauri type stars and intermediate mass Herbig Stars.
Among the T\,Tauri star population \wisa is among the disks with the widest gap between rings, with only the RX\,1615 system showing a wider gap.
However, morphologically these two systems present quite differently, whereas \wisa has relatively broad individual rings, the rings in RXJ\,1615 are almost all very thin and radially not resolved \citep{deBoer2016}.
The two Herbig systems with wider gaps than \wisa are V351\,Ori and HD\,34282.
However, both of these objects show additional complex morphology possibly linked to large scale spiral arms, which makes the interpretation of their gap structure challenging \citep{Valegard2024, deBoer2021}.
Due to the significantly farther distance of these two systems compared to \wisa they are not well suited for the direct detection of planet thermal emission.

Besides \wisa there are three other multi-ringed disks that have indications of embedded planet-candidates: HD\,169142, HD\,163296 and HD\,97048. While HD\,169142 hosts a planet candidate also detected in the near-infrared (as discussed in section~\ref{sec:comparison_planets}), the planet candidates in HD\,163296 and HD\,97048 have been detected indirectly through local deviations of Keplerian motion of the disk gas at mm-wavelengths (\citealt{Teague2018, Pinte2019}). 
This highlights that multi-ringed disk structures might indeed be the signpost of ongoing planet formation.
However, all of these other systems are around more massive Herbig stars and show significantly smaller gap-widths.
Consequently the estimated planet masses are smaller than the inferred mass for \wisb and in none of these cases direct thermal emission has been detected from the planet photosphere.
Of particular note may be the comparison between the \wisa system and the HD\,97048 system.
By coincidence both of these systems are viewed under a similar inclination and position angle.
The overall morphology, with the large extent of scattered light signal as well as relatively broad individual rings, is very similar between them.
It is then interesting in the context of the occurrence rate of massive planets that of these two the lower-mass T Tauri star has an embedded massive wide orbit super-Jupiter, while the Herbig star may only have a slightly lower mass ($2-3$\,\mj) embedded planet \citep{Pinte2019}.   
%--------------------------------------------------------------------

\section{Conclusions}
\label{sec:conc}

In this study we present the discovery of a directly imaged wide orbit ($\sim$57\,au) gas giant embedded in a multi-ringed disk around the young solar analog \wisa.
The astrometry of the planet relative to the central star across three observational epochs shows that the companion is inconsistent with a distant stationary background source. %BAM GONE
While additional high precision astrometric measurements (e.g. with VLT/GRAVITY) are needed to constrain the dynamical mass of the planet-star system, the present data are compatible with a co-planar Keplerian orbit for \wisb inside the disk gap.

Photometric analysis places the companion between AMES-COND and AMES-DUSTY isochrones in color-magnitude space.
Using the derived system age of $5.1^{+2.4}_{-1.3}$~Myr, we interpolated the absolute magnitudes of $M_{K_s}=11.95^{+0.09}_{-0.09}$\,mag and $M_{H}=12.8^{+0.4}_{-0.3}$\,mag to the AMES isochrone grids, and derived a companion mass of $4.9^{+0.9}_{-0.6}$~\mj---consistent with a planetary-mass object.

A detection of H$\alpha$ emission (Close et al. submitted; companion letter~2) further confirms this planet and provides evidence of accretion, indicating the presence of a \ac{cpd}.
However, the lack of polarized signal of \wisb may indicate that the \ac{cpd} is more depleted of dust.
The relatively blue $H-K_s$ color, apparent lack of dust in the surrounding disk gap, and the absence of polarized \ac{cpd} emission leads us to conclude that \ac{cpd} extinction is unlikely to significantly affect the derived photometry and the resulting mass estimate.

While the presence of embedded planets has long since been speculated to be a driver for sub-structure in ringed disks, the unambiguous detection of \wisb in the newly resolved disk surrounding \wisa provides us with the best laboratory to study planet-disk interaction in detail.
Our preliminary analysis of the width of the disk gap in which the embedded planet resides, not only shows a general agreement with hydrodynamic models, but it also points us toward the possibility of using the embedded planet to measure the disk viscosity for the first time.
This is a key parameter in the evolution of the planet-forming environment. 

As the planet resides in the cleared gap and its mass is consistent with the modeled planet mass required to open such a gap, we argue that it likely formed in-situ through core accretion and that there is no rapid migration on dynamical timescales.
%
%Although the planet could be undergoing type II migration, the well-structured disk and lack of strong radial streamers or spirals suggest that there is no ongoing migration acting on short timescales.
%
Future follow-up observations of \wisb with ALMA and JWST will enable studies of its atmosphere and the impact of the embedded planet on the disk's gas kinematics and surface density structure.
This will allow us to calibrate ALMA observations of other embedded planet candidates, to unlock the full potential of this complementary technique.
These future observations will also allow to directly measure isotopologue ratios for CO in the disk to compare and contrast with spectroscopic isotopologue ratios measured directly from the planet \citep[e.g. ][]{Zhang21b,Zhang24,Hoch2025}.

The discovery of \wisb embedded in the gap of a seemingly unperturbed disk, demonstrates for the first time that wide separation gas giants, discovered by direct imaging around older systems, can indeed form in-situ.
Thus \wisb marks a promising starting point to study wide separation planets in time.

%TC:ignore
\acknowledgments

MB has received funding from the European Research Council (ERC) under the European Union's Horizon 2020 research and innovation programme (PROTOPLANETS, grant agreement No. 101002188).
R.T. was supported by JSPS KAKENHI Grant Number JP25K07351 and JP25K01049.
The research was carried out at the Jet Propulsion Laboratory, California Institute of Technology, under a contract with the National Aeronautics and Space Administration (80NM0018D0004).
The WISPIT team would like to thank Donna E. Keeley for her contributions to the survey program, and the anonymous reviewer for their helpful feedback to this letter.
\vspace{2mm}
\facilities{VLT(SPHERE)}

\software{Python (\url{https://www.python.org/}),
SciPy \citep{2020SciPy-NMeth},
NumPy \citep{oliphant2006guide},
Photutils \citep{photutils},
Matplotlib \citep{astropy_1},
astropy \citep{astropy_2}
}


%\newpage
\clearpage
\FloatBarrier
\appendix

\section{Analysis of stellar parameters and environs of \wisa}
\label{app:stellar_params}

The following section contains a detailed analysis of the stellar parameters and environs of \wisa.
Section \ref{app:reddening} discusses the reddening and extinction towards \wisa.
As it is not a member of a large young association, section \ref{app:membership} discusses membership to young stellar groups.
The age of this group is analyzed in \ref{app:age} and compared to the age fit to \wisa based on its \ac{sed}.

\subsection{Interstellar reddening towards \wisa}
\label{app:reddening}
\begin{figure*}%[t!] 
\centering
\includegraphics[width=0.6\textwidth]{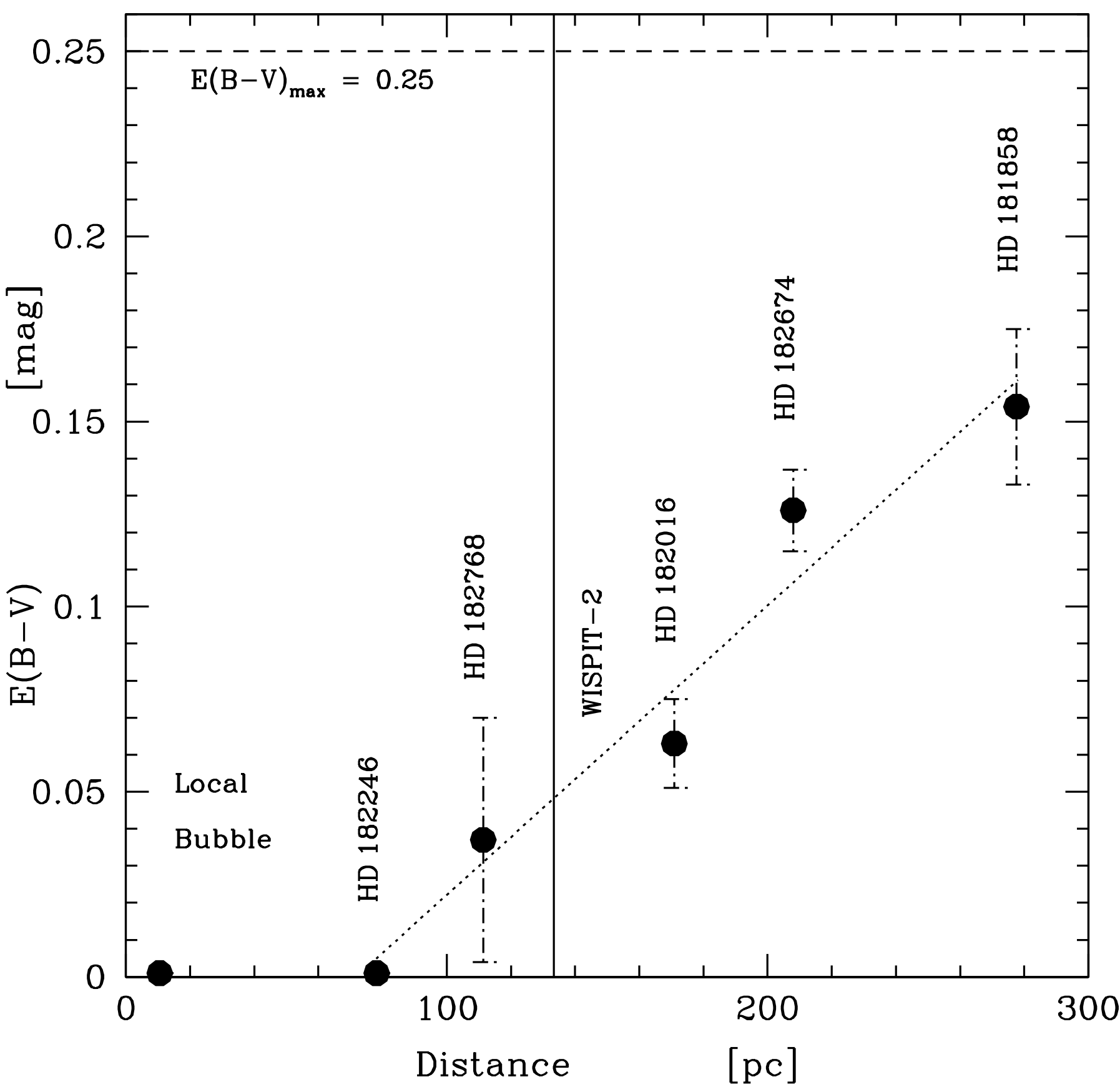} \\
\caption{
Distance versus reddening $E(B-V)$ for stars within 1$^{\circ}$, with distances based on Gaia DR3 parallaxes and reddening from \citet{Paunzen2024}.} 
\label{fig:reddening}
\end{figure*}
The reported values for reddening and extinction by Gaia DR3 of $A_0 = 1.0855^{+0.0764}_{-0.0381}$, 
 $A_G = 0.8601^{+0.0634}_{-0.0315}$ and $E(B_P - R_P) = 0.4706^{+0.0351}_{-0.0172}$, are unusually high for the region of \wisa.
 In Figure~\ref{fig:reddening}, we plot the $E(B-V)$ reddening values from the recent analysis of \citet{Paunzen2024} for several stars within 1$^{\circ}$ of \wisa at a range of distances. 
The maximum interstellar reddening due to the full Galactic column of dust is estimated to be $E(B-V)_{max} = 0.25$ mag from the \citet{Schlafly2010} update to the \citet{Schlegel1998} IRAS dust maps, accessed via IRSA\footnote{\url{https://irsa.ipac.caltech.edu/applications/DUST/}}, and indeed a distant B giant at $d\simeq$ 761\,pc, HD 182411, has reddening $E(B-V) = 0.263$ in the \citet{Paunzen2024} catalog -- the highest value within 0$^{\circ}$.8 of \wisa. 
In Figure~\ref{fig:reddening}, the reddening values for the stars within 300\,pc within 1$^{\circ}$ of \wisa in the \citet{Paunzen2024} catalog can be summarized by the following trends: (1) negligible reddening within $\sim$80\,pc, due to the Local Bubble, and (2) a roughly linear trend between 80\,pc\,$<d<$\,280\,pc. 
At the distance of \wisa ($d$ $\simeq$ 133\,pc), interpolation of the trend would predict interstellar reddening of $E(B-V) \simeq 0.043$\,mag. 
This is consistent with the reddening estimate $E(B-V)$ quoted in the TESS Input Catalog \citep[$E(B-V) =0.0439\pm0.0279$\,mag;][]{Stassun2019}, which are based upon the \citet{Green2018} Pan-STARRS 3D dust maps.
This corresponds to $A_V=0.1361\pm0.0865$\,mag, which we adopt for our analysis of the \ac{sed}.

\subsection{Memberships to young stellar associations}
\label{app:membership}

Multiple authors have analyzed Gaia astrometric and photometric data and added \wisa to membership lists of recently identified young stellar groups, compiled in Table~\ref{tab:Theia53}.
\citet{Kounkel2019} included \wisa in the membership list for their new group Theia~53, containing 44 members, and for which they estimate mean age of log(age/yr) = 7.40 (25\,Myr), mean parallax $\varpi$ = 7.394\,mas ($d = 135$\,pc).
\citet{Kerr2021} used HDBSCAN clustering algorithm on Gaia DR2 data and identified 27 ``top level clusters'' within 333\,pc, including a sample of 30 associated stars (including WISPIT-2) which they designate ``TLC 7'' or ``Aquila East''\footnote{Use of ``Aquila East'' for the young stellar association TLC 7 should be deprecated as multiple studies refer to ``Aquila East'' as an active star-formation region in the Aquila Rift complex \citep[e.g.,][]{Park2012,Fiorellino2021}.}. 
\citet{Kerr2021} quote a centroid position for the group at ($\alpha$, $\delta$) = ($297^{\circ}.8$, $-9^{\circ}.0$; ICRS), ($\ell$, $b$) = ($31^{\circ}.6 $, $-17^{\circ}.3$), covering  $12^{\circ}.2$ $\times$ $3^{\circ}.8$ degrees of sky, at distance $d = 136.6\,\pm\,6.5$\,pc, and mean proper motion  (\pmra, \pmdec) = ($8.5, -25.9$) \masyr, which all compare well to \wisa. 
\citet{Kerr2021} also quote a mean age of $20.2\,\pm\,1.5$ Myr.

\citet{Qin2023} identifies a similar young group (OCSN 8) in the area, although doesn't alias it with Theia 53. 
They list a membership of 60 member stars in Gaia DR3 for OCSN 8 , with the brightest members being the bright B-type pair 57\,Aql A and B.
They estimate the mean parallax for OCSN 8 of $\varpi$ = $7.25\pm0.36$\,mas (implying mean distance $d$ $\simeq$ $138\pm7$ pc) and mean age log(age/yr) = 7.95, or 89\,Myr.
\citet{Hunt2024} detected this group again (entry 5493), cross-referencing it with both Theia~53 and OCSN~8, and increasing the membership list to 123 stars, among which is \wisa.
The quote a mean group parallax of $\varpi$ = 7.696 ($\pm$0.581 mas st.dev; $\pm$0.055 s.e.m.), implying a mean distance of $d$ = 130\,pc. 
They estimate Theia 53's age to be log(age/yr) = $7.55^{+0.26}_{-0.28}$ or 35.7$^{+28.5}_{-17.0}$\,Myr.

It is noteworthy that the T~Tauri star BZ~Sgr (PDS 101, IRAS 19558-1405, Gaia DR3 6878598726815263488) was included in the membership lists for both TLC-7 \citep{Kerr2021} and OCSN~8 \citep{Qin2023}. 
BZ~Sgr shows obvious youth indicators like strong H$\alpha$ emission (EW(H$\alpha$) = -52\,$\mathrm{\AA}$), strong \ion{Li}{1} absorption (EW(Li$\lambda$6707) = 0.33$\mathrm{\AA}$).
It is also situated next to the MBM~159 molecular cloud \citep{Magnani1985}, with a distance estimate of $d$ = 144\,pc \citep{Schlafly2014}.
Since previous characterizations of this group have a range of age estimates between $\sim$20 and $\sim$90 Myr, with the recent estimate by \citet{Hunt2024} including Gaia DR3 data for both Theia~53 and OCSN~8 defining the `state of the art'. 
An updated age analysis of this group is included in Section~\ref{app:age}, but we note that further analysis is required to improve upon the group's age accuracy.

\begin{table*}[t!]
\centering
\def\arraystretch{1.2}
\setlength{\tabcolsep}{5pt}
\caption{Mean Properties for Theia 53 Group and its Aliases From Different Studies}\label{tab:Theia53}
%\resizebox{\columnwidth}{!}{%
\begin{tabular}{lccccccccc}
\hline\hline
Ref. & ID & $\alpha$ & $\delta$ & \pmra & \pmdec & $\varpi$ & $N_{mem}$ & log(age/yr) & Lucida\\
... & ... & deg & deg & \masyr & \masyr & mas & ... & ... & ...\\
\hline\hline
1 & Theia 53 & $299.41$ & $-7.14$ & ... & ... & $7.39$ & 44 & 7.40 & 57\,Aql\,A\\
2 & TLC-7 & $297.80$ & $-9.00$ & $8.5$ & $-25.9$ & $7.32\pm0.35$ & 30 & $7.305\pm0.032$ & 57\,Aql\,A\\
3 & OCSN 8 & $298.74$ & $-8.18$ & $9.73\pm3.77$ & $-25.44\pm3.95$ & $7.25\pm0.36$ & 60 & 7.95 & 57\,Aql\,A\\
4 & Theia 53/OCSN 8 & $298.75$ & $-8.15$ & $8.78\pm3.93$ & $-25.00\pm4.14$ & $7.70\pm0.58$ & 123 & $7.553^{+0.255}_{-0.281}$ & 57\,Aql\,A\\
\hline
\hline
\end{tabular}%
\tablecomments{The $\pm$ values in proper motions and parallax are standard deviations, not standard errors. References:
(1) \citet{Kounkel2019},
(2) \citet{Kerr2021},
(3) \citet{Qin2023},
(4) \citet{Hunt2024}. 
} 
\end{table*}

\subsection{Age classification of \wisa}
\label{app:age}
\begin{table*}%[h]
\centering
\def\arraystretch{1.2}
\setlength{\tabcolsep}{10pt}
\caption{Lithium equivalent widths and effective temperatures of co-moving stars.}
\label{tab:ewli}
\begin{tabular}{lccr}
\hline\hline
ID & $T_{\rm eff}$ [K] & EW(Li) [m\AA] & Ref. EW(Li)\\
\hline\hline
UCAC4 416-129178 & 4135 & $490 \pm 24.5$ & \citet{Zerjal2021} \\
TYC 5736-0649-1 & 5590 & $250 \pm 12.5$ & \citet{Torres2006} \\
BZ Sgr (= UCAC2 27000662) & 5280 & $400 \pm 20.0$ & \citet{Torres2006} \\
\hline\hline
\end{tabular}
\tablecomments{An error of 5\% was assumed on the lithium equivalent width measurement.
}
\end{table*}

\begin{figure*}%[h]
    \centering
%    \begin{subfigure}[t]{0.49\textwidth}
%        \centering
%        \includegraphics[width=0.99\textwidth]{outputtheia_iso.pdf}
%        \caption{Best-fit isochrone from \ac{eagles}.}
%    \end{subfigure}%
%    ~ 
%    \begin{subfigure}[t]{0.49\textwidth}
%        \centering
%        \includegraphics[width=0.99\textwidth]{outputtheia_prob.pdf}
%        \caption{Posterior probability distribution from \ac{eagles}.}
%    \end{subfigure}
%    ~
%        \begin{subfigure}[t]{0.49\textwidth}
%        \centering
%        \includegraphics[width=0.99\textwidth]{output_theia_iso_v2.pdf}
%        \caption{Best-fit isochrone from \ac{eagles} v2.0.}
%    \end{subfigure}%
%    ~ 
%    \begin{subfigure}[t]{0.49\textwidth}
%        \centering
%        \includegraphics[width=0.99\textwidth]{output_theia_prob_v2.pdf}
%        \caption{Posterior probability distribution from \ac{eagles} v2.0.}
%    \end{subfigure}
\includegraphics[width=\textwidth]{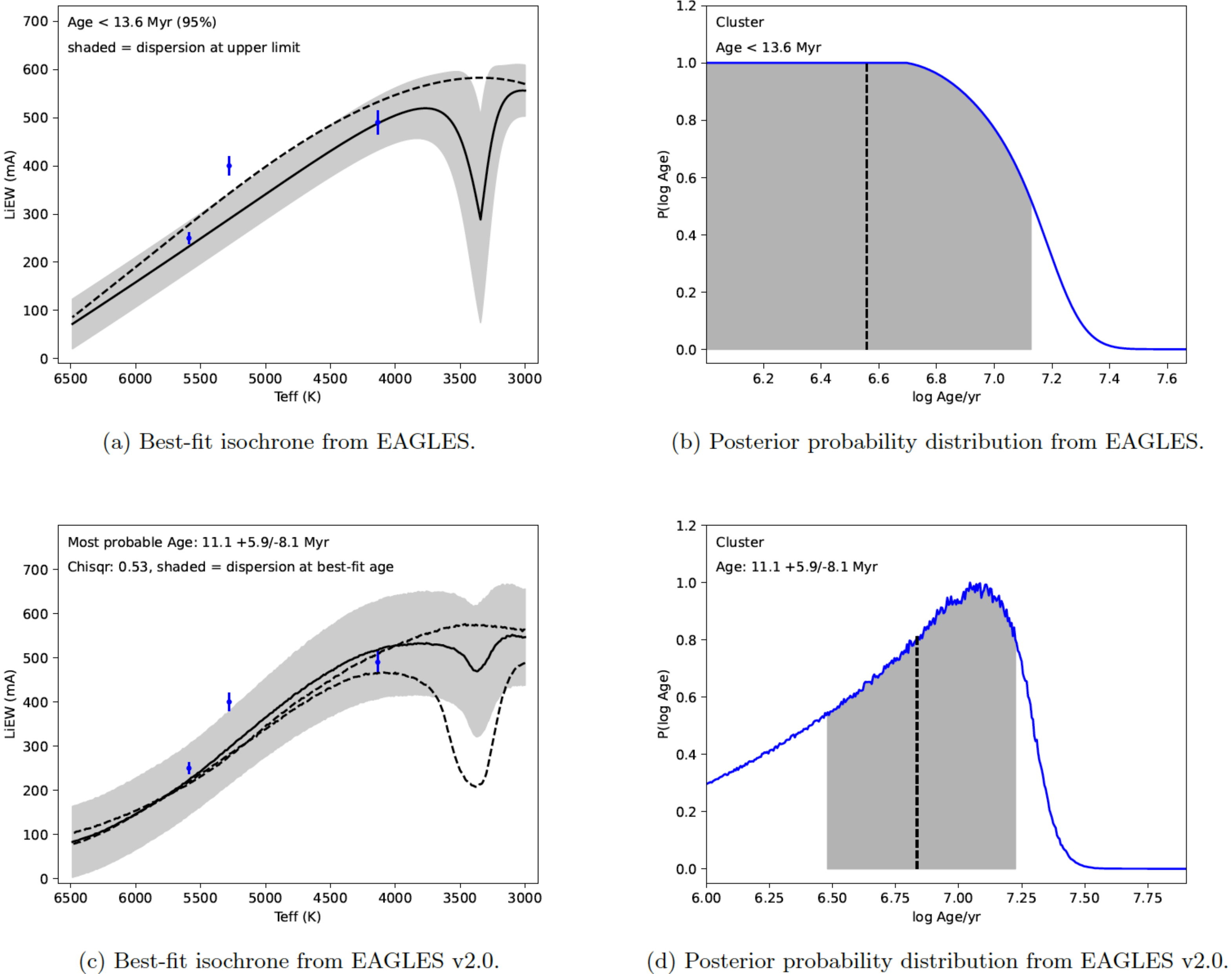} \\
    \caption{Results from fitting the age of the cluster using the EW(Li) and $T_\mathrm{eff}$ values listed in Table \ref{tab:ewli} with \ac{eagles} (upper panels) and \ac{eagles}~v2.0 (lower panels).}
    \label{fig:ewlifit}
\end{figure*}

There are three co-moving co-distant stars that are on any of the membership lists for either Theia~53, OCSN~8 or TLC-7 with reported equivalent width measurement of lithium absorption lines. 
These targets are listed in Table~\ref{tab:ewli}. 
To estimate the age of the group, we used \ac{eagles}, a software implementation of an empirical model that predicts the lithium equivalent width of a star as a function of age and effective temperature, as described in \citet{Jeffries2023}.
The results provide an upper limit of 13.6 Myr for the age of the group, as shown in Figure~\ref{fig:ewlifit}.
A newer version of this software, \ac{eagles} v2.0 \citep{Weaver2024}, implements an artificial neural network (ANN) model for the relationship between EW(Li), $T_\mathrm{eff}$ and age.
This model is free from the constraints of an arbitrary analytical model, and provides better accuracy in reproducing the relationship between EW(Li) and its dispersion with age.
It provides a cluster age of $11.1^{+5.9}_{-8.1}$~Myr, consistent with the upper limit provided by \ac{eagles}, with the probability distribution in Fig.~\ref{fig:ewlifit} showing that it is indeed likely to be younger rather than older.

\begin{figure*}%[h!]
\centering
\includegraphics[width=\textwidth]{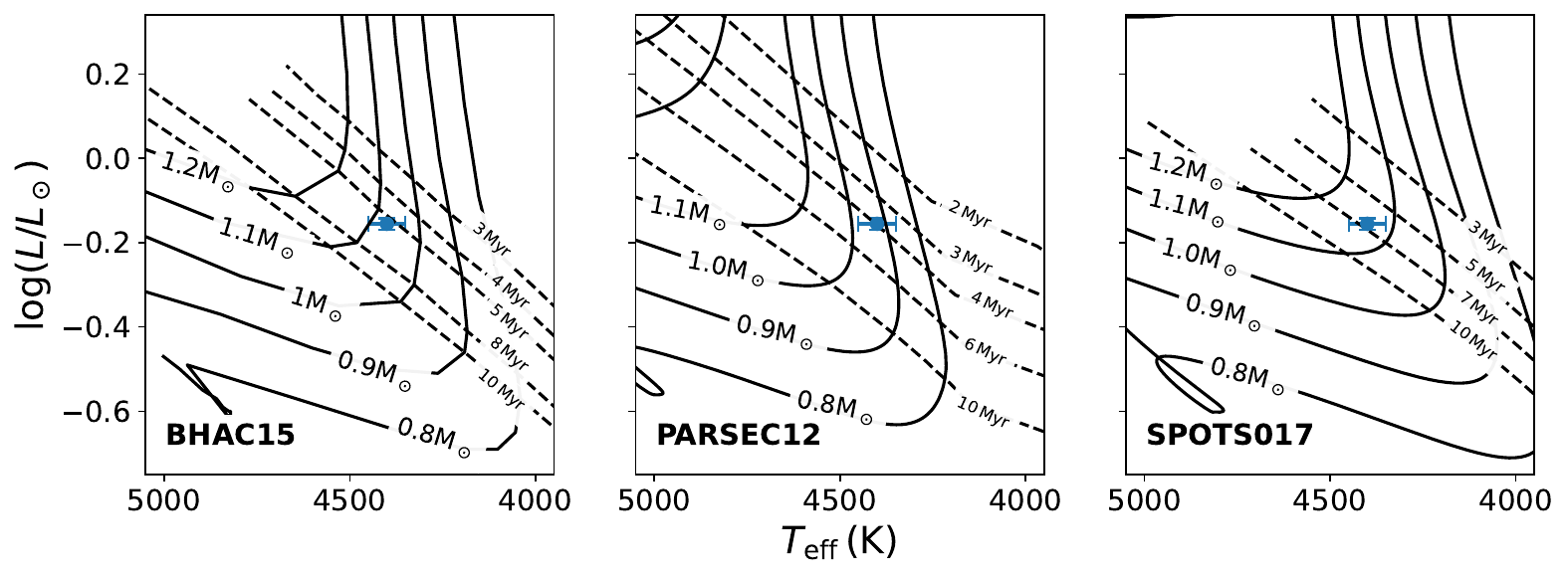} \\
\caption{Stellar evolution isochrones for various masses (solid lines) and ages (dashed lines). 
The models are, from left to right, BHAC15, PARSEC 1.2, and SPOTS ($f=0.17$).
\wisa is indicated with a blue marker.} 
\label{fig:stellar_iso}
\end{figure*}

While this provides an upper limit (\ac{eagles}) or rough constraints (\ac{eagles} v2.0) on the age of the young group to which \wisa likely belongs, it does not directly constrain the age of the star itself.
%%%%%%%%%%%%%% start edit
To address this, we used the bolometric luminosity and effective temperature resulting from the \ac{vosa} \ac{sed} fit and used \ac{vosa} to interpolate them to the BHAC15 stellar isochrone grid \citep{Baraffe2015}.
The resulting age and mass are $5.11^{+1.77}_{-0.61}$~Myr and $1.08^{+0.02}_{-0.07}\,\mathrm{M_\odot}$, respectively.
To investigate the systemic errors introduced by the evolutionary models, we have repeated this process for two other models: PARSEC~1.2 \citep{Bressan2012} and SPOTS with a spot coverage fraction of $f=0.17$ \citep{Somers2020}. 
Given the available SPOTS models, the choice of $f=0.17$ is reasonable for T-Tauri stars, which generally have spot coverage fractions between approximately 0.1 and 0.25 \citep[e.g.][]{Bouvier1989,Lanza2016,Long2011}.
As these models span a reasonable range around our preferred model (BHAC15), see Table~\ref{tab:star_isochrones} and Figure~\ref{fig:stellar_iso}, we adopt the derived age of $5.1$~Myr as our fiducial value.
We adjust the errors to account for systemic uncertainties due to the choice of evolutionary model by combining the model dependent variation in age with the BHAC15 fit errors in quadrature.
This yields a final adopted age of $5.1^{+2.4}_{-1.3}$~Myr.
Applying the same procedure to the mass results in a final adopted stellar mass of $1.08^{+0.06}_{-0.17}\,\mathrm{M_\odot}$.

\begin{table*}
\caption{Mass and age of \wisa from stellar isochrones.}
\label{tab:star_isochrones}
\centering
\def\arraystretch{1.2}
\setlength{\tabcolsep}{12pt}
\begin{tabular}{lcc}
\hline\hline
Model & Age & Mass \\
 & Myr & \mj \\
\hline
BHAC15 & $5.11^{+1.77}_{-0.61}$ & $1.08^{+0.02}_{-0.07}$ \\
PARSEC 1.2 & $4.00^{+0.98}_{-0.91}$ & $0.93^{+0.06}_{-0.07}$ \\
SPOTS (f=0.17) &$6.72^{+1.22}_{-1.10}$ & $1.14^{+0.01}_{-0.04}$  \\
\hline
\end{tabular}
\end{table*}

%%%%%%%%%%%%%%% end edit
We note that the resulting age of $5.1^{+2.4}_{-1.3}$ Myr is consistent with the lower end of our derived age constraints for the Theia~53/OCSN~8/TLC-7 group.

\FloatBarrier
\section{Stokes Q and U images}
\label{app:stokes}

We show the Stokes $Q$ and $U$ images along with the derived $Q_\phi$ and $U_\phi$ images for both the $H$ and $K_s$-band polarimetric observation epoch.
Both observation epochs are flux calibrated, using the respective 2MASS $H$ and $K_s$-band magnitudes of the central star as reference.
We show both observation epochs at the same absolute scale, demonstrating that the disk is significantly fainter in $K_s$-band compared to $H$-band polarized scattered light.

\begin{figure*}[h!]
   \centering
   \includegraphics[width=\textwidth]{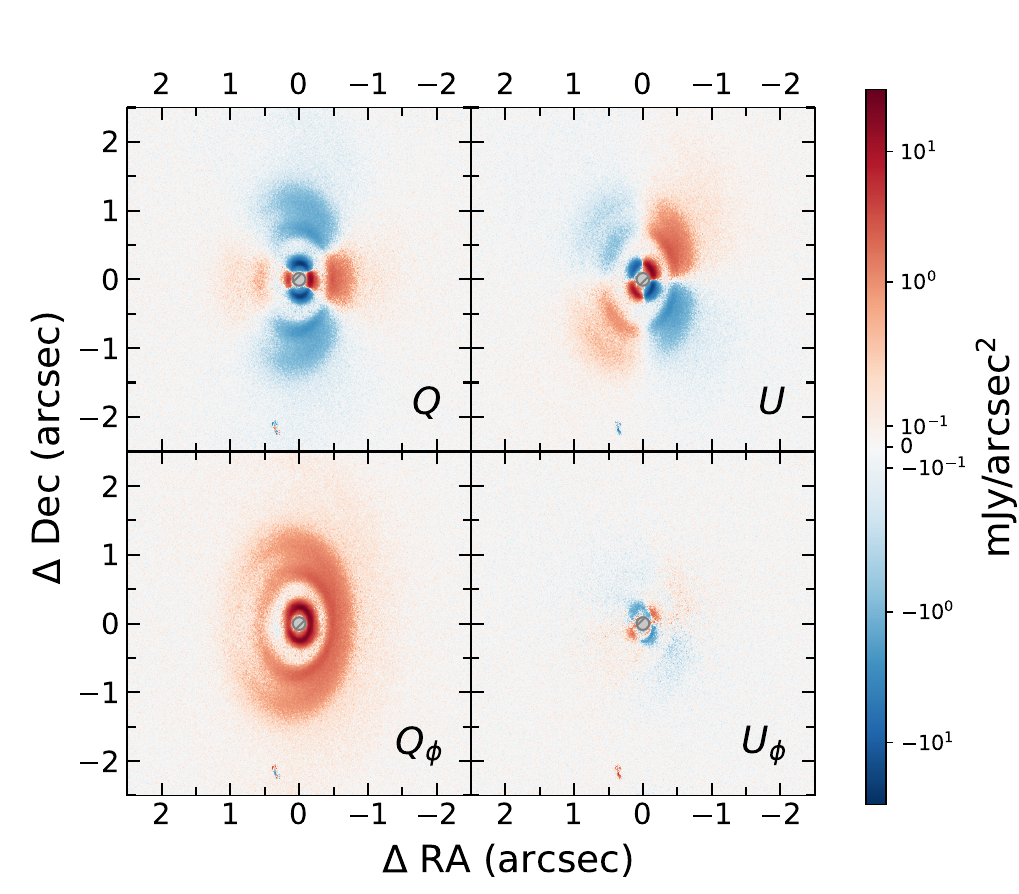}
      \caption{$H$-band Stokes $Q$ and $U$ and derived $Q_\phi$ and $U_\phi$ images of the \wisa disk.
      We note that polarized light observations are typically not sensitive to planet thermal emission, which is predominantly unpolarized.
      Consequently the planet \wisb is not visible in these images.}
         \label{app:fig:pol_H}
\end{figure*}

\begin{figure*}[h!]
   \centering
   \includegraphics[width=\textwidth]{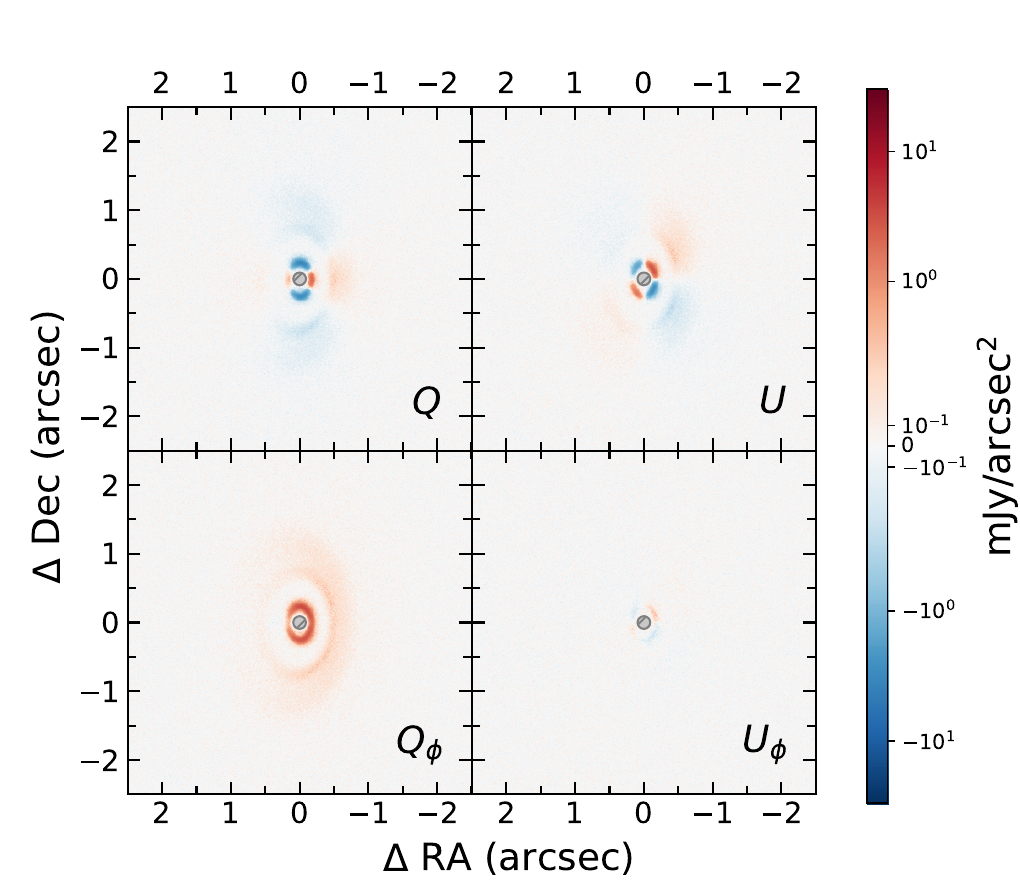}
      \caption{$K_s$-band Stokes $Q$ and $U$ and derived $Q_\phi$ and $U_\phi$ images of the \wisa disk.
      We note that polarized light observations are typically not sensitive to planet thermal emission, which is predominantly unpolarized.
      Consequently the planet \wisb is not visible in these images.}
         \label{app:fig:pol_K}
\end{figure*}

\FloatBarrier
\section{Discussion of the inner disk bottom side}
\label{app:inner disk bottom}

\begin{figure*}[t!]
   \centering
   \includegraphics[width=\textwidth]{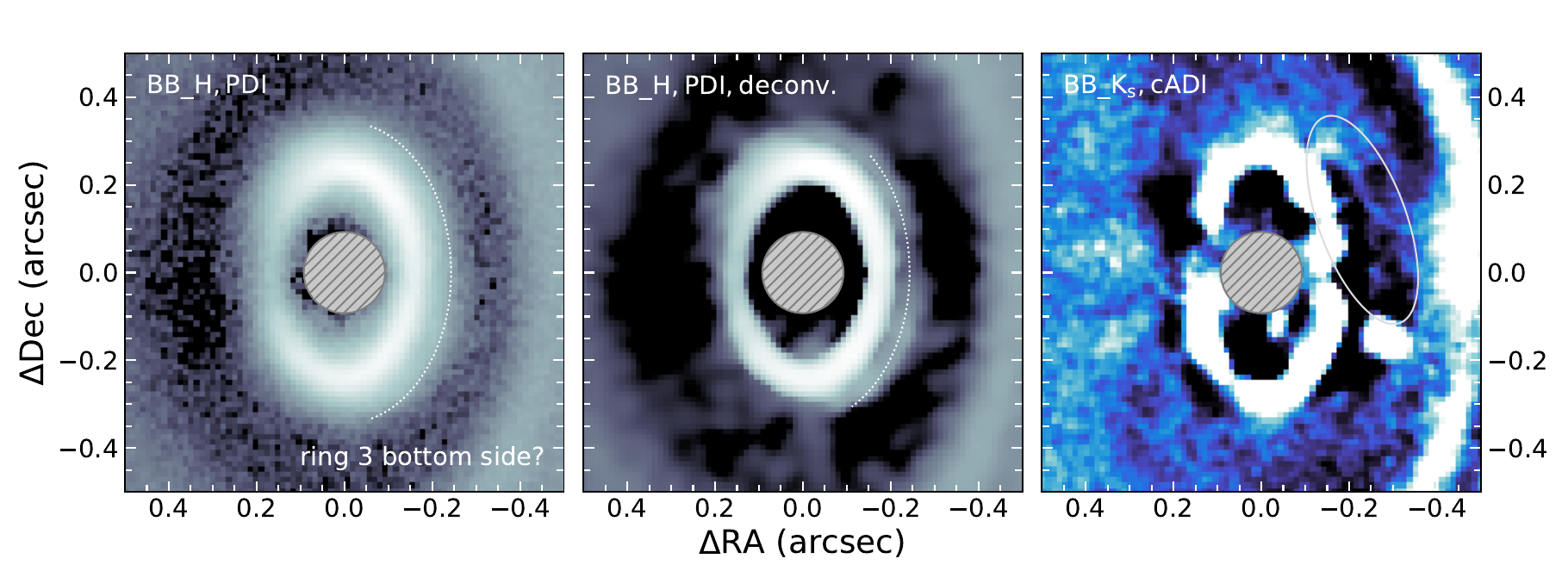}
      \caption{Zoom-in on the innermost disk ring (ring 3).
      We show polarized light $Q_\phi$ images in the gray color scheme in the left and middle panel and the total intensity cADI $K_s$-band image in the blue color scheme in the right panel.
      The middle panel shows the $Q_\phi$ image after image deconvolution was applied.
      Left and middle panel are displayed on a log scale due to the large dynamic range of the inner disk region while the right panel is on a linear scale.
      The gray hashed circle in the image center marks the area covered by the coronagraphic mask.
      We indicate with dotted lines (left and middle panel) and with a solid encircling ellipse (right panel) the signature of the inner disk bottom side.}
         \label{app:fig:disk bottom}
\end{figure*}
Scattered light images trace the upper vertical layers of the disk.
This leads to an illuminated surface that is visible from our line of sight.
If the disk is seen under moderate or large inclination then it is possible to spot the dark lane, which demarks the disk mid-plane and which appears dark since no star light reaches it to produce a scattered light signal.
Following from the dark lane there are some disks for which we can see the illuminated forward scattering rim of the disk bottom side.
For a more detailed description of the appearance of scattered light disks we refer to the recent review by \citet{Benisty2023}.
The forward scattering rim of the disk bottom side has been detected for several disks.
Archetypal examples would be the disk around the IM\,Lup system \citep{Avenhaus2018} or more recently the disk in the PDS\,111 system \citep{Derkink2024}.
However, the disk bottom side is not detected for all inclined disks.
This can be the case if there is an extended disk with low dust surface density obscuring our view along the line of sight.
This is likely a common configuration as we routinely find that the gas disks detected in tracers such as $^{12}$CO with ALMA are significantly more extended than the disks seen in scattered light.
If there is even a moderate amount of small dust particles still entrained in the disk gas, then it can easily make this diffuse outer region optically thick and thus obscure the view on the rim of the disk bottom side.
This problem was recently investigated by George, Dominik \& Ginski (accepted).
They found that in order to see the disk bottom side the outer disk needs to be sharply truncated, possibly by a planetary or stellar companion or fly-by, so that little or no material remains to obscure the line of sight on the disk bottom side.

While the detection of the forward scattering rim of the disk bottom side is reasonably common on the outer edge of inclined disks, this has to our knowledge not been the case for the outer edge of an inner disk component seen through a gap inside the disk.
To enable such a detection would mean that the gap itself needs to be near devoid of small dust particles, which would otherwise make the gap optically thick in the near infrared.
For the innermost disk ring (ring 3) in the \wisa system we report the tentative detection of the forward scattering rim of the disk bottom side.
We highlight the signal in question in Figure~\ref{app:fig:disk bottom}.
We see evidence for the disk bottom side, both in the $H$-band polarized light image as well as in the $K_s$-band total intensity image.
For the $H$-band polarized light image we can tell from Figure~\ref{app:fig:disk bottom} (left panel) that the bright ridge of the ring has an extended ``halo'' in a crescent shape centered along the disk minor axis towards the west.
Such a crescent shape would indeed be the expected morphology of the forward scattering rim of the disk bottom side.
To highlight this feature we performed an image deconvolution using the AIDA \citep[Adaptive Image Deconvolution Algorithm; ][]{Hom2007} Python package.
We used a flux calibration image (in which no disk signal was detected due to the short integration time) as realization of the instrumental point spread function as input for the deconvolution.
A similar analysis was recently performed to highlight the narrow gaps in the PDS\,111 system by \citet{Derkink2024}.
The resulting deconvolved image is shown in Figure~\ref{app:fig:disk bottom} (middle panel).
The crescent shaped signal is pronounced and appears distinct for the main ring ridge.

We also tentatively detect a signature of the disk bottom side in the $K_s$-band classical \ac{adi} image shown in Figure~\ref{app:fig:disk bottom} (right panel).
The \ac{adi} procedure effectively acts as a high pass filter, highlighting sharp disk structures while self-subtracting more diffuse signal \citep[see e.g. the discussion in ][]{Milli2012, Stapper2022}.
In the cADI image we tentatively see a faint crescent-shaped ridge again to the West of the main ring (highlighted by the elliptical aperture).
This signal is in line with what might be expected from the ridge of the visible disk bottom side.

While our interpretation of this signal is strengthened by its detection in two independent data sets and reductions, we do note that we do not see a clear detection of the separating dark lane that one might expect between the forward scattering rims of the disk top and bottom sides.
This is likely due to too low spatial resolution of our observations.
Based on our geometric fitting of the inner ring (ring 3), we find that it has a height of the scattering surface of 1.7\,au.
This means the dark lane separating the disk top and bottom sides should have a thickness of 3.4\,au.
Given the distance and inclination of the system this translates to a thickness of 18\,mas, i.e. less than half of the resolution element we have in the $H$-band.
Thus our non-detection of the disk mid-plane signature is consistent with our interpretation.
To solidify this interpretation follow-up observations in the optical with SPHERE/ZIMPOL in the $R$ or $V$-band might be able to resolve the disk mid-plane.

\section{Gaussian fitting routine for astrometric measurement}
\label{app:gauss_fit_astrometry}
To obtain astrometric measurements we fitted the {\tt Astropy} \citep{astropy_1,astropy_2,astropy_3} 2D Gaussian model to the approximate position of the companion.
We used the \texttt{TRFLSQFitter}, a Trust Region Reflective algorithm with bound constraints and least-squares statistics. 
The center of the Gaussian fit was constrained to lie within 3 pixels (i.e., a 6$\times$6 pixel region) of the manually estimated ($x$, $y$) position.
Additionally, to ensure that the fitted \ac{fwhm} is consistent with that of SPHERE/IRDIS observations, we constrained the standard deviation of the Gaussian be within 2.0 pixels of the median \ac{psf} standard deviation in $H$-band.
This median \ac{psf} was constructed from \ac{yses} observations---all contributing observations are listed in Table~\ref{tab:med_psf_H} in Appendix~\ref{app:median_psfs}.
These constraints were implemented to ensure that only the physical planet signal was fitted.
However, to avoid biasing the results with the imposed bounds, we removed a fit from the results if a bound of a constraint was touched, even if that fit formally converged.

\section{Geometric Fitting Results Table}
\label{app:geometric_fitting_table}

\begin{figure*}%[h]
    \centering
%    \begin{subfigure}[t]{0.49\textwidth}
%        \centering
%        \includegraphics[width=0.99\textwidth]{Forward-Scattering.pdf}
%        \caption{Position angle convention.}
%    \end{subfigure}%
%    ~ 
%    \begin{subfigure}[t]{0.49\textwidth}
%        \centering
%        \includegraphics[width=0.99\textwidth]{deprojected_figure.pdf}
%        \caption{De-projected disk.}
%    \end{subfigure}
\includegraphics[width=\textwidth]{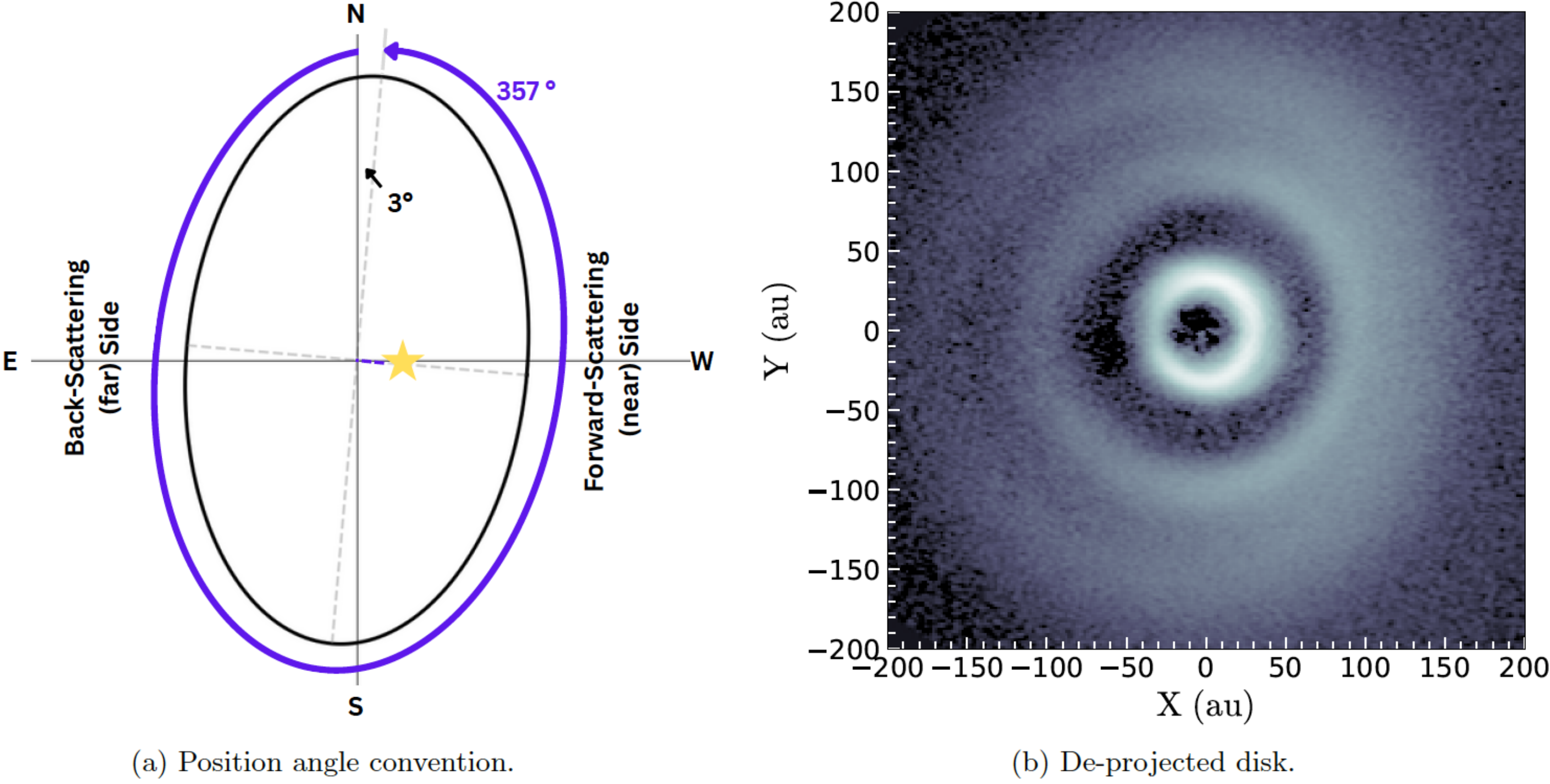} \\
\caption{Panel (a): Definition of the PA, measured counterclockwise from North to East. In this convention, the PA is measured from North to the major axis and contains the forward scattering side, corresponding to a PA of 357\textdegree\ in this diagram. Panel (b): Image of the disk de-projected using the height and inclination measurements extracted from individual rings following this convention, with angular size on sky translated to au.} 
\label{app:fig:disk_deprojection}
\end{figure*}

%\begin{figure*}[b!]
%\centering
%\includegraphics[width=0.55\textwidth]{deprojected_figure.pdf} \\
%\caption{Disk de-projected using the height and inclination measurements extracted from individual rings, with angular size on sky translated to au.} 
%\label{app:fig:disk_deprojection}
%\end{figure*}
In Table~\ref{tab:geometric_fitting}, we present the results of our ellipse fitting approach (discussed in Section~\ref{sec:disk}) to the individual sub-structures.
As we fit each structure individually we obtain a small range of disk inclinations and position angles.
Here the position angle follows the convention that it is measured from the North in direction of East (counter-clockwise) towards the disk major axis on the disk near side, see Fig.~\ref{app:fig:disk_deprojection}.
This could indicate in principle that there are small relative misalignments between different disk structures.
We note that these differences in inclination and position angle appear to be somewhat consistent between the $H$ and $K_s$-band images, e.g. among the rings 1--3, ring 1 has in both cases the largest measured inclination and ring 2 the smallest.
However, this result is of very low significance at this stage and detailed modeling of the disk (beyond the scope of this study) might be needed to verify such a small misalignment or warp in the disk.
In figure~\ref{app:fig:disk_deprojection} we show the disk de-projected using our height measurements as well as the inclination and major axis position angle.
We find that the de-projection shows reasonable circular rings as expected for a non-eccentric disk.

\begin{table*}
\caption{Results of the geometric fitting of \wisa}
\label{tab:geometric_fitting}
\centering
\def\arraystretch{1.2}
\setlength{\tabcolsep}{9pt}
\begin{tabular*}{\textwidth}{@{}llllllll@{}}
\hline\hline
& & Separation (au) & Height (au) & Aspect Ratio $(h/r)$ & Inclination ($^\circ$) & Position Angle ($^\circ$) \\
\hline
$H$-band & Ring 0 & $316.361 \pm 4.499$ & $76.083 \pm 6.315$ & $0.240 \pm 0.030$ & $43.618 \pm 1.543$ & $357.811 \pm 1.556$ \\
& Ring 1 & $163.554 \pm 2.928$ & $24.048 \pm 1.148$ & $0.147 \pm 0.007$ & $45.397 \pm 1.074$ & $356.511 \pm 1.311$ \\
& Ring 2 & $96.729 \pm 0.634$ & $10.444 \pm 0.623$ & $0.108 \pm 0.006$ & $41.797 \pm 0.564$ & $357.097 \pm 1.548$ \\
& Gap 3 & $68.989 \pm 0.600$ & $4.098 \pm 0.665$ & $0.059 \pm 0.010$ & $44.185 \pm 0.757$ & $0.472 \pm 0.917$ \\
& Ring 3 & $38.441 \pm 0.078$ & $1.700 \pm 0.154$ & $0.044 \pm 0.004$ & $44.953 \pm 0.393$ & $1.540 \pm 0.279$ \\

$K_s$-band & Ring 1 & $156.298 \pm 1.763$ & $17.934 \pm 2.731$ & $0.115 \pm 0.018$ & $47.517 \pm 1.266$ & $356.816 \pm 1.496$ \\
& Ring 2 & $102.745 \pm 1.363$ & $6.621 \pm 1.576$ & $0.064 \pm 0.015$ & $42.642 \pm 1.272$ & $359.343 \pm 4.458$ \\
& Gap 3 & $68.005 \pm 1.449$ & $2.585 \pm 1.351$ & $0.038 \pm 0.020$ & $49.322 \pm 1.662$ & $0.828 \pm 2.871$ \\
& Ring 3 & $37.224 \pm 0.125$ & $1.093 \pm 0.485$ & $0.029 \pm 0.013$ & $43.972 \pm 0.876$ & $0.243 \pm 0.482$ \\

\hline
\end{tabular*}
\tablecomments{Results for fitting the rings present in our data.
Further details on the different fitting methods are found in Section~\ref{sec:disk}.
Ring 0 was not detectable in $K_s$-band. }
\end{table*}

\section{Multi-ringed disk population}
\label{app:sec:multi-ringed-disks}

To put the \wisa system in context of previous observations of planet-forming disks, we have assembled from the literature all disks that show a similar morphology, i.e. a multiple ring structure in near-infrared scattered light.
We note that we for now exclude systems with multiple rings detected with sub-mm observations, unless they also have multiple rings detected in scattered light.
This is mainly due to the fact that we do not currently have sub-mm observations of the \wisa system and thus cannot directly compare to other such observations.
In Table~\ref{tab: app: multi-ringed-disks} we give the full literature sample of such disks, the reference publication as well as the values for the width an location of the widest gap in the system used for Figure~\ref{fig: multi-ringed-sample}.

\begin{table*}
\caption{Literature sample of multi-ringed scattered light disks}
\label{tab: app: multi-ringed-disks}
\centering
\def\arraystretch{1.2}
\setlength{\tabcolsep}{9pt}
\begin{tabular*}{\textwidth}{@{}llcc@{}}
\hline\hline
system name& reference & gap width (au) & gap location (au) \\
\hline
%HD\,141569 & \citet{Weinberger1999, Perrot2016} & 60 & 250 \\
HD\,163296 & \citet{Grady2000, Monnier2017} & 26 & 40 \\
HD\,169142 & \citet{Quanz2013} & 28 & 44 \\
HD\,34282 & \citet{deBoer2021} & 105 & 139 \\
HD\,34700 & \citet{Monnier2019, Columba2024} & - & - \\
HD\,97048 & \citet{Ginski2016} & 21 & 109 \\
PDS\,111 & \citet{Derkink2024} & 22 & 63 \\
RXJ\,1615.3-3255 & \citet{deBoer2016} & 68 & 199 \\
SZ\,Cha & \citet{Ginski2024} & 53 & 112 \\
TW\,Hya & \citet{vanBoekel2017} & 20 & 80 \\
V\,351\,Ori & \citet{Valegard2024} & 69 & 90 \\
V\,4046\,Sgr & \citet{Avenhaus2018} & 12 & 21 \\
WRAY\,15-788 & \citet{Bohn2019} & 28 & 35 \\
\hline
\end{tabular*}
\tablecomments{As reference we always give the first observation in which the multi-ringed sub-structure was detected and or measured in detail.
In cases where subsequent substructure was detected over the course of time, we give the detection of the first ring-like structure and the first paper in which then a second ring was discovered.
Gap width and location refer to the widest detected gap in the system.
If no specific center gap location was fitted in the reference literature, then we refer to the geometric center between the two adjacent rings as gap location.
We only include Class II disks in this sample.
For the HD\,34700 system we do not give a gap width and location as the disk structure is strongly asymmetric.}
\end{table*}

\section{Constructing the composite image}
\label{app:composit_fig}
\begin{figure*}[!htb]
\center
\includegraphics[width=\textwidth]{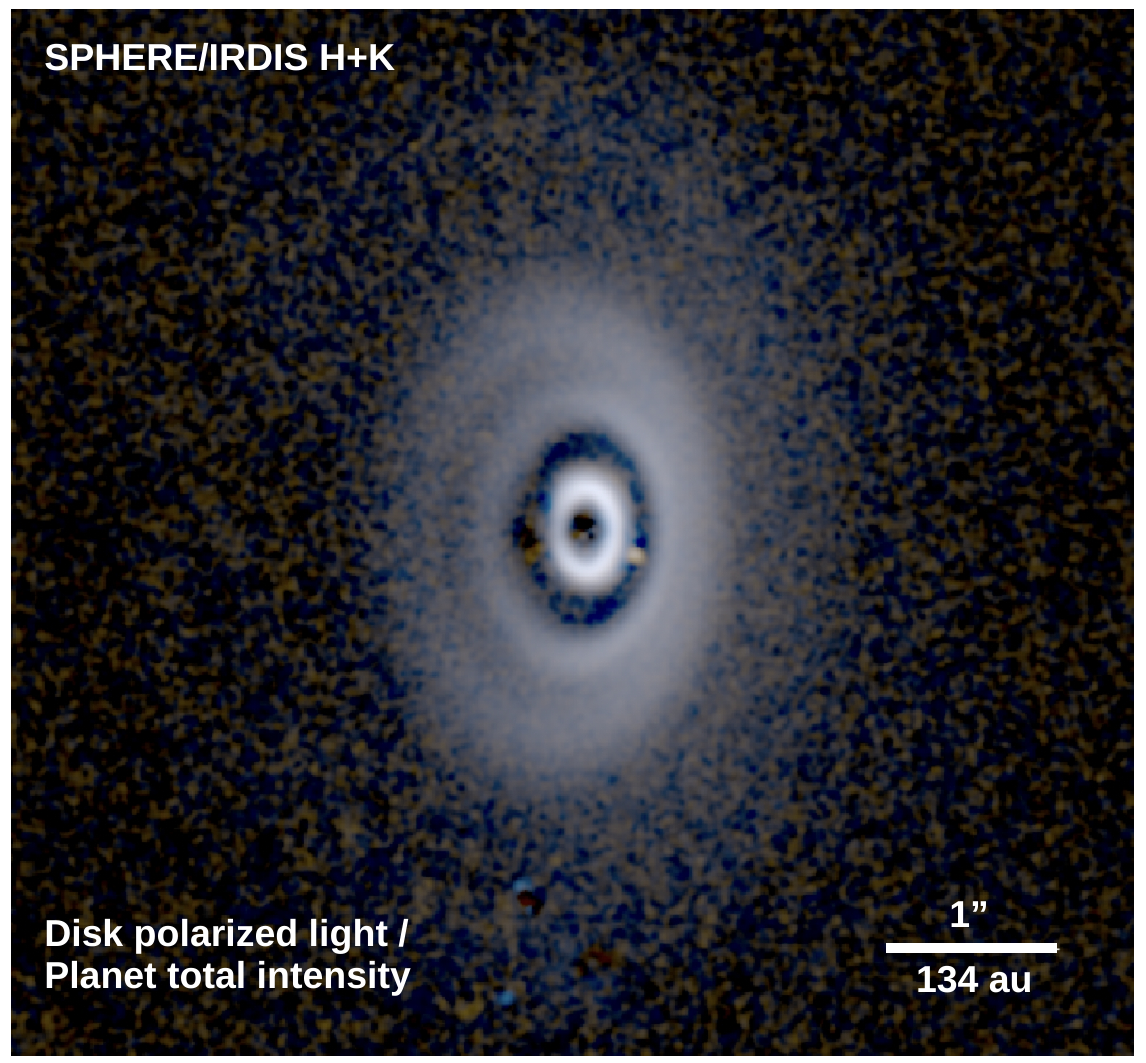} 
\caption{SPHERE/IRDIS multi-band image of the \wisa system.
The $H$-band $Q_\phi$ image was added as blue channel and the median combination of $H$-band and $K_s$-band $Q_\phi$ images was added as the green channel.
The red channel is a combination of $K_s$-band $Q_\phi$ image and $K_s$-band cADI image in which we masked all but the gap containing the thermal emission from \wisb.
} 
\label{fig:sphere_rgb_fullsize}
\end{figure*}

We show the page-wide version of Figure~\ref{fig:sphere_rgb} in Figure~\ref{fig:sphere_rgb_fullsize}.
This composite figure is constructed using the polarized-light $H$ and $K_s$-band images for the disk signal and masked versions of the 2023 \ac{rdi} $H$ and 2025 cADI $K_s$-band total intensity images for the planet.
The polarized light images are not sensitive to the thermal emission of the planet (since thermal emission is unpolarized) and thus the planet is not visible in these images. Conversely these are the images that best highlight the disk structure without any morphological artifacts introduced by e.g. \ac{adi}.
The $H$ and $K_s$ images were taken with the same single exposure settings (64\,s).
We use the $H$-band image for the blue channel of the image and the $K_s$-band image for the red channel of the image.
Since we do not have a third wavelength for the disk at this point we use the average of the $H$ and $K_s$- band data for the green channel.
The image is displayed on a logarithmic color scale to highlight the full disk features and extend.
For the planet signal we used a similar approach, i.e. we used the $H$-band data for the blue image channel, the average of $H$ and $K_s$-band for the green channel and the $K_s$-band data for the red channel.
to suppress the somewhat distorted disk signal in these images we cut out a small square area around the planet in the $H$-band image and an annulus centered on the disk gap for the $K_s$-band image.
Since the underlying $K_s$-band data for the polarized image of the disk and total intensity image of the disk are identical, the relative position of the planet to the disk is accurate.
Since the $H$-band total intensity image of the planet was taken about two years prior to the $K_s$-band data, the planet showed significant orbital motion (as we discuss in section~\ref{sec:astr_planet}).
To avoid a smeared appearance of the planet for the purpose of this illustrative figure, we thus centered the 2023 $H$-band observation of the planet on the same position as the 2025 $K_s$-band observation.
The figure therefore depicts the accurate position of the planet relative to the disk and the central star as of 2025-04-26.

\newpage
\section{Reference library for {\em H}-band RDI reduction}
\label{app:reflib}

In Table~\ref{tab:reference_library} we provide all observations that were used for the \ac{rdi} library for PCA-based reference differential imaging processing of the \wisa $H$-band data taken in 2023 and 2024. 
This library is a subset of all \ac{yses} observations, filtered to remove low-quality observations, misaligned frames, frames with sources in the field of view and close stellar binaries. 
The resulting library consists of 340 frames of which, for each observation, the 275 frames with the highest correlation were selected based on \ac{mse}.

\def\arraystretch{1.2}
\setlength{\tabcolsep}{55pt}
\begin{longtable}{@{}llr@{}}
\label{tab:reference_library}
\\
\hline\hline
Target name & Archive name & Observation date \\
\hline
\endfirsthead

\hline
Target name & Archive name & Observation date \\
\hline
\endhead

\hline
\endfoot

\hline
\endlastfoot

ASAS J114452-6438.9	&	2MASS J11445217-6438548	&	2018-05-14	\\
ASAS J114452-6438.9	&	2MASS J11445217-6438548	&	2023-04-20	\\
1RXS J114519.6-574925	&	2MASS J11452016-5749094	&	2018-05-14	\\
1RXS J114542.7-573928	&	2MASS J11454278-5739285	&	2018-06-05	\\
1RXS J114542.7-573928	&	2MASS J11454278-5739285	&	2019-01-13	\\
1RXS J114542.7-573928	&	2MASS J11454278-5739285	&	2023-04-20	\\
1RXS J120652.1-504448	&	2MASS J12065276-5044463	&	2017-04-02	\\
1RXS J121010.3-485538	&	2MASS J12101065-4855476	&	2017-04-18	\\
CD-57 4328	&	2MASS J12113142-5816533	&	2018-12-22	\\
CD-57 4328	&	2MASS J12113142-5816533	&	2019-02-18	\\
PM J12160-5614	&	2MASS J12160114-5614068	&	2018-12-27	\\
RX J1216.6-7007A	&	2MASS J12164023-7007361	&	2018-12-23	\\
RX J1216.6-7007A	&	2MASS J12164023-7007361	&	2019-02-15	\\
RX J1216.6-7007A	&	2MASS J12164023-7007361	&	2023-12-21	\\
CPD-64 1859	&	2MASS J12192161-6454101	&	2023-06-17	\\
RX J1220.0-5018A	&	2MASS J12195938-5018404	&	2018-12-30	\\
RX J1220.0-5018A	&	2MASS J12195938-5018404	&	2023-06-17	\\
CD-47 7559	&	2MASS J12220430-4841248	&	2017-04-18	\\
ASAS J122648-5214.8	&	2MASS J12264842-5215070	&	2018-12-30	\\
ASAS J122648-5214.8	&	2MASS J12264842-5215070	&	2023-05-28	\\
RX J1230.5-5222	&	2MASS J12302957-5222269	&	2018-12-30	\\
RX J1230.5-5222	&	2MASS J12302957-5222269	&	2022-03-30	\\
CPD-56 5307	&	2MASS J12333381-5714066	&	2019-01-01	\\
CPD-56 5307	&	2MASS J12333381-5714066	&	2019-01-14	\\
CPD-56 5307	&	2MASS J12333381-5714066	&	2023-05-28	\\
CPD-50 5313	&	2MASS J12361767-5042421	&	2018-12-30	\\
2MASS J12374883-5209463	&	2MASS J12374883-5209463	&	2018-12-30	\\
2MASS J12374883-5209463	&	2MASS J12374883-5209463	&	2023-07-14	\\
1RXS J123834.9-591645	&	2MASS J12383556-5916438	&	2019-01-03	\\
1RXS J123834.9-591645	&	2MASS J12383556-5916438	&	2019-01-12	\\
1RXS J123834.9-591645	&	2MASS J12383556-5916438	&	2023-07-14	\\
CD-56 4581	&	2MASS J12393796-5731406	&	2017-06-17	\\
CD-51 6900	&	2MASS J12404664-5211046	&	2018-04-30	\\
CD-51 6900	&	2MASS J12404664-5211046	&	2023-05-30	\\
ASAS J124547-5411.0	&	2MASS J12454884-5410583	&	2018-04-30	\\
HD 111227	&	2MASS J12480778-4439167	&	2017-04-18	\\
1RXS J124830.1-594449	&	2MASS J12483152-5944493	&	2023-08-07	\\
V1257 Cen	&	2MASS J12505143-5156353	&	2019-01-12	\\
CPD-52 6110	&	2MASS J13015069-5304581	&	2019-01-08	\\
ASAS J130550-5304.2	&	2MASS J13055087-5304181	&	2018-06-11	\\
ASAS J130550-5304.2	&	2MASS J13055087-5304181	&	2018-07-05	\\
ASAS J130550-5304.2	&	2MASS J13055087-5304181	&	2022-04-02	\\
CD-51 7268	&	2MASS J13064012-5159386	&	2018-04-30	\\
CD-51 7268	&	2MASS J13064012-5159386	&	2023-06-15	\\
2MASS J13065439-4541313	&	2MASS J13065439-4541313	&	2018-04-08	\\
2MASS J13065439-4541313	&	2MASS J13065439-4541313	&	2023-07-08	\\
UCAC2 12444765	&	2MASS J13095880-4527388	&	2018-05-01	\\
ASAS J131033-4816.9	&	2MASS J13103245-4817036	&	2018-05-01	\\
2MASS J13121764-5508258	&	2MASS J13121764-5508258	&	2017-08-31	\\
2MASS J13121764-5508258	&	2MASS J13121764-5508258	&	2018-05-15	\\
UNSW-V 514	&	2MASS J13174687-4456534	&	2018-05-28	\\
2MASS J13334410-6359345	&	2MASS J13334410-6359345	&	2017-07-05	\\
2MASS J13334410-6359345	&	2MASS J13334410-6359345	&	2023-06-16	\\
CD-41 7947	&	2MASS J13343188-4209305	&	2017-04-02	\\
CD-41 7947	&	2MASS J13343188-4209305	&	2023-08-07	\\
TYC 8265-229-1	&	2MASS J13354082-4818124	&	2017-04-02	\\
TYC 7800-858-1	&	2MASS J13380596-4344564	&	2017-04-02	\\
CD-56 4581	&	CD-56 4581	&	2024-06-13	\\
HD 304428	&	HD 304428	&	2024-06-10	\\
RX J1216.6-7007A	&	TYC 9231-1185-1	&	2024-06-10	\\
UCAC4 186-087394	&	UCAC4 186-087394	&	2024-06-13	\\

\end{longtable}

%\newpage
\section{Median PSFs}
\label{app:median_psfs}
Table \ref{tab:med_psf_H} contains all observations used to create the normalized median flux PSF in $H$-band.
Here `Target name' denotes the commonly used designation for the source, and `Archive name' corresponds to the name it is registered under in the ESO Science Archive Facility.

\def\arraystretch{1.2}
\setlength{\tabcolsep}{55pt}
\begin{longtable}{@{}llr@{}}
\caption{Flux observations used for constructing the median $H$-band PSF.} \label{tab:med_psf_H} \\
\hline\hline
Target name & Archive name & Observation date \\
\hline
\endfirsthead

\hline
Target name & Archive name & Observation date \\
\hline
\endhead

\hline
\endfoot

\hline
\endlastfoot
1RXS J114519.6-574925 & 2MASS J11452016-5749094 & 2018-05-14 \\
1RXS J114542.7-573928 & 2MASS J11454278-5739285 & 2019-01-13 \\
1RXS J114542.7-573928 & 2MASS J11454278-5739285 & 2023-04-20 \\
1RXS J121010.3-485538 & 2MASS J12101065-4855476 & 2017-04-18 \\
1RXS J123834.9-591645 & 2MASS J12383556-5916438 & 2019-01-03 \\
1RXS J123834.9-591645 & 2MASS J12383556-5916438 & 2019-01-12 \\
1RXS J123834.9-591645 & 2MASS J12383556-5916438 & 2023-07-14 \\
1RXS J124830.1-594449 & 2MASS J12483152-5944493 & 2023-08-07 \\
2MASS J12374883-5209463 & 2MASS J12374883-5209463 & 2018-12-30 \\
2MASS J12374883-5209463 & 2MASS J12374883-5209463 & 2023-07-14 \\
2MASS J13065439-4541313 & 2MASS J13065439-4541313 & 2018-04-08 \\
2MASS J13065439-4541313 & 2MASS J13065439-4541313 & 2023-07-08 \\
2MASS J13121764-5508258 & 2MASS J13121764-5508258 & 2018-05-15 \\
2MASS J13334410-6359345 & 2MASS J13334410-6359345 & 2023-06-16 \\
ASAS J114452-6438.9 & 2MASS J11445217-6438548 & 2018-05-14 \\
ASAS J114452-6438.9 & 2MASS J11445217-6438548 & 2023-04-20 \\
ASAS J122648-5214.8 & 2MASS J12264842-5215070 & 2018-12-30 \\
ASAS J122648-5214.8 & 2MASS J12264842-5215070 & 2023-05-28 \\
ASAS J124547-5411.0 & 2MASS J12454884-5410583 & 2018-04-30 \\
ASAS J130550-5304.2 & 2MASS J13055087-5304181 & 2022-04-02 \\
ASAS J131033-4816.9 & 2MASS J13103245-4817036 & 2018-05-01 \\
CD-41 7947 & 2MASS J13343188-4209305 & 2023-08-07 \\
CD-47 7559 & 2MASS J12220430-4841248 & 2017-04-18 \\
CD-51 6900 & 2MASS J12404664-5211046 & 2018-04-30 \\
CD-51 6900 & 2MASS J12404664-5211046 & 2023-05-30 \\
CD-51 7268 & 2MASS J13064012-5159386 & 2018-04-30 \\
CD-51 7268 & 2MASS J13064012-5159386 & 2023-06-15 \\
CD-57 4328 & 2MASS J12113142-5816533 & 2018-12-22 \\
CD-57 4328 & 2MASS J12113142-5816533 & 2019-02-18 \\
CPD-50 5313 & 2MASS J12361767-5042421 & 2018-12-30 \\
CPD-52 6110 & 2MASS J13015069-5304581 & 2019-01-08 \\
CPD-56 5307 & 2MASS J12333381-5714066 & 2019-01-01 \\
CPD-56 5307 & 2MASS J12333381-5714066 & 2019-01-14 \\
CPD-56 5307 & 2MASS J12333381-5714066 & 2023-05-28 \\
CPD-64 1859 & 2MASS J12192161-6454101 & 2023-06-17 \\
PM J12160-5614 & 2MASS J12160114-5614068 & 2018-12-27 \\
RX J1216.6-7007A & 2MASS J12164023-7007361 & 2018-12-23 \\
RX J1216.6-7007A & 2MASS J12164023-7007361 & 2019-02-15 \\
RX J1216.6-7007A & 2MASS J12164023-7007361 & 2023-12-21 \\
RX J1216.6-7007A & TYC 9231-1185-1 & 2024-06-10 \\
RX J1220.0-5018A & 2MASS J12195938-5018404 & 2018-12-30 \\
RX J1220.0-5018A & 2MASS J12195938-5018404 & 2023-06-17 \\
RX J1230.5-5222 & 2MASS J12302957-5222269 & 2018-12-30 \\
RX J1230.5-5222 & 2MASS J12302957-5222269 & 2022-03-30 \\
UCAC2 12444765 & 2MASS J13095880-4527388 & 2018-05-01 \\
UCAC4 186-087394 & UCAC4 186-087394 & 2024-06-13 \\
UNSW-V 514 & 2MASS J13174687-4456534 & 2018-05-28 \\
V1257 Cen & 2MASS J12505143-5156353 & 2019-01-12 \\
\end{longtable}

\bibliography{MyBibFMe}{}
\bibliographystyle{aasjournal}

%TC:endignore
\end{document}

%% file: authorlist.tex
\author[0009-0002-6729-646X]{Richelle F. van Capelleveen}
\affiliation{Leiden Observatory, Leiden University, Postbus 9513, 2300 RA Leiden, The Netherlands}

\author[0000-0002-4438-1971]{Christian Ginski}
\affiliation{School of Natural Sciences, Center for Astronomy, University of Galway, Galway, H91 CF50, Ireland}

\author[0000-0002-7064-8270]{Matthew A. Kenworthy}
\affiliation{Leiden Observatory, Leiden University, Postbus 9513, 2300 RA Leiden, The Netherlands}

\author[0009-0002-6096-9617]{Jake Byrne}
\affiliation{School of Natural Sciences, Center for Astronomy, University of Galway, Galway, H91 CF50, Ireland}

\author[0009-0001-0368-1062]{Chloe Lawlor}
\affiliation{School of Natural Sciences, Center for Astronomy, University of Galway, Galway, H91 CF50, Ireland}

\author[0009-0002-4783-6529]{Dan McLachlan}
\affiliation{School of Natural Sciences, Center for Astronomy, University of Galway, Galway, H91 CF50, Ireland}

\author[0000-0003-2008-1488]{Eric E. Mamajek}
\affiliation{Jet Propulsion Laboratory, California Institute of Technology, 4800 Oak Grove Drive, M/S 321-162, Pasadena, CA, 91109, USA}

\author[0000-0002-5823-3072]{Tomas Stolker}
\affiliation{Leiden Observatory, Leiden University, Postbus 9513, 2300 RA Leiden, The Netherlands}

\author[0000-0002-7695-7605]{Myriam Benisty}
\affiliation{Max-Planck-Institut für Astronomie, Königstuhl 17, 69117 Heidelberg, Germany}

\author[0000-0003-1401-9952]{Alexander J. Bohn}
\affiliation{Leiden Observatory, Leiden University, Postbus 9513, 2300 RA Leiden, The Netherlands}

\author[0000-0002-2167-8246]{Laird M. Close}
\affiliation{Steward Observatory, University of Arizona, 933 N. Cherry Avenue, Tucson, AZ 85719, USA}

\author[0000-0002-3393-2459]{Carsten Dominik}
\affiliation{Anton Pannekoek Institute for Astronomy, University of Amsterdam, Science Park 904, 1098 XH Amsterdam, The Netherlands}

\author[0000-0001-5130-9153]{Sebastiaan Haffert}
\affiliation{Leiden Observatory, Leiden University, Postbus 9513, 2300 RA Leiden, The Netherlands}
\affiliation{Steward Observatory, University of Arizona, 933 N. Cherry Avenue, Tucson, AZ 85719, USA}

\author[0000-0002-7261-8083]{Rico Landman}
\affiliation{Leiden Observatory, Leiden University, Postbus 9513, 2300 RA Leiden, The Netherlands}

\author[0000-0003-3583-6652]{Jie Ma}
\affiliation{Universit\'{e} Grenoble Alpes, CNRS, Institut de Plan\'{e}tologie et d’Astrophysique (IPAG), F-38000}

\author[0000-0003-1624-3667]{Ignas Snellen}
\affiliation{Leiden Observatory, Leiden University, Postbus 9513, 2300 RA Leiden, The Netherlands}

\author[0000-0003-1451-6836]{Ryo Tazaki}
\affiliation{Department of Earth Science and Astronomy, The University of Tokyo, Tokyo 153-8902, Japan}

\author[0000-0003-2458-9756]{Nienke van der Marel}
\affiliation{Leiden Observatory, Leiden University, Postbus 9513, 2300 RA Leiden, The Netherlands}

\author[0009-0008-8810-6577]{Lukas Welzel}
\affiliation{Leiden Observatory, Leiden University, Postbus 9513, 2300 RA Leiden, The Netherlands}

\author[0000-0003-0097-4414]{Yapeng Zhang}
\affiliation{Department of Astronomy, California Institute of Technology, Pasadena, CA 91125, USA}

%% file: main.bbl
\begin{thebibliography}{}
\expandafter\ifx\csname natexlab\endcsname\relax\def\natexlab#1{#1}\fi
\providecommand{\url}[1]{\href{#1}{#1}}
\providecommand{\dodoi}[1]{doi:~\href{http://doi.org/#1}{\nolinkurl{#1}}}
\providecommand{\doeprint}[1]{\href{http://ascl.net/#1}{\nolinkurl{http://ascl.net/#1}}}
\providecommand{\doarXiv}[1]{\href{https://arxiv.org/abs/#1}{\nolinkurl{https://arxiv.org/abs/#1}}}

\bibitem[{Allard {et~al.}(2001)Allard, Hauschildt, Alexander, Tamanai, \& Schweitzer}]{Allard2001}
Allard, F., Hauschildt, P.~H., Alexander, D.~R., Tamanai, A., \& Schweitzer, A. 2001, The Astrophysical Journal, 556, 357, \dodoi{10.1086/321547}

\bibitem[{{Allard} {et~al.}(2013){Allard}, {Homeier}, {Freytag}, {Schaffenberger}, \& {Rajpurohit}}]{Allard2013}
{Allard}, F., {Homeier}, D., {Freytag}, B., {Schaffenberger}, W., \& {Rajpurohit}, A.~S. 2013, Memorie della Societa Astronomica Italiana Supplementi, 24, 128, \dodoi{10.48550/arXiv.1302.6559}

\bibitem[{{Amara} \& {Quanz}(2012)}]{Amara2012}
{Amara}, A., \& {Quanz}, S.~P. 2012, \mnras, 427, 948, \dodoi{10.1111/j.1365-2966.2012.21918.x}

\bibitem[{{Astropy Collaboration} {et~al.}(2013){Astropy Collaboration}, {Robitaille}, {Tollerud}, {Greenfield}, {Droettboom}, {Bray}, {Aldcroft}, {Davis}, {Ginsburg}, {Price-Whelan}, {Kerzendorf}, {Conley}, {Crighton}, {Barbary}, {Muna}, {Ferguson}, {Grollier}, {Parikh}, {Nair}, {Unther}, {Deil}, {Woillez}, {Conseil}, {Kramer}, {Turner}, {Singer}, {Fox}, {Weaver}, {Zabalza}, {Edwards}, {Azalee Bostroem}, {Burke}, {Casey}, {Crawford}, {Dencheva}, {Ely}, {Jenness}, {Labrie}, {Lim}, {Pierfederici}, {Pontzen}, {Ptak}, {Refsdal}, {Servillat}, \& {Streicher}}]{astropy_1}
{Astropy Collaboration}, {Robitaille}, T.~P., {Tollerud}, E.~J., {et~al.} 2013, \aap, 558, A33, \dodoi{10.1051/0004-6361/201322068}

\bibitem[{{Astropy Collaboration} {et~al.}(2018){Astropy Collaboration}, {Price-Whelan}, {Sip{\H{o}}cz}, {G{\"u}nther}, {Lim}, {Crawford}, {Conseil}, {Shupe}, {Craig}, {Dencheva}, {Ginsburg}, {Vand erPlas}, {Bradley}, {P{\'e}rez-Su{\'a}rez}, {de Val-Borro}, {Aldcroft}, {Cruz}, {Robitaille}, {Tollerud}, {Ardelean}, {Babej}, {Bach}, {Bachetti}, {Bakanov}, {Bamford}, {Barentsen}, {Barmby}, {Baumbach}, {Berry}, {Biscani}, {Boquien}, {Bostroem}, {Bouma}, {Brammer}, {Bray}, {Breytenbach}, {Buddelmeijer}, {Burke}, {Calderone}, {Cano Rodr{\'\i}guez}, {Cara}, {Cardoso}, {Cheedella}, {Copin}, {Corrales}, {Crichton}, {D'Avella}, {Deil}, {Depagne}, {Dietrich}, {Donath}, {Droettboom}, {Earl}, {Erben}, {Fabbro}, {Ferreira}, {Finethy}, {Fox}, {Garrison}, {Gibbons}, {Goldstein}, {Gommers}, {Greco}, {Greenfield}, {Groener}, {Grollier}, {Hagen}, {Hirst}, {Homeier}, {Horton}, {Hosseinzadeh}, {Hu}, {Hunkeler}, {Ivezi{\'c}}, {Jain}, {Jenness}, {Kanarek}, {Kendrew}, {Kern}, {Kerzendorf}, {Khvalko}, {King}, {Kirkby}, {Kulkarni},
  {Kumar}, {Lee}, {Lenz}, {Littlefair}, {Ma}, {Macleod}, {Mastropietro}, {McCully}, {Montagnac}, {Morris}, {Mueller}, {Mumford}, {Muna}, {Murphy}, {Nelson}, {Nguyen}, {Ninan}, {N{\"o}the}, {Ogaz}, {Oh}, {Parejko}, {Parley}, {Pascual}, {Patil}, {Patil}, {Plunkett}, {Prochaska}, {Rastogi}, {Reddy Janga}, {Sabater}, {Sakurikar}, {Seifert}, {Sherbert}, {Sherwood-Taylor}, {Shih}, {Sick}, {Silbiger}, {Singanamalla}, {Singer}, {Sladen}, {Sooley}, {Sornarajah}, {Streicher}, {Teuben}, {Thomas}, {Tremblay}, {Turner}, {Terr{\'o}n}, {van Kerkwijk}, {de la Vega}, {Watkins}, {Weaver}, {Whitmore}, {Woillez}, {Zabalza}, \& {Astropy Contributors}}]{astropy_2}
{Astropy Collaboration}, {Price-Whelan}, A.~M., {Sip{\H{o}}cz}, B.~M., {et~al.} 2018, \aj, 156, 123, \dodoi{10.3847/1538-3881/aabc4f}

\bibitem[{{Astropy Collaboration} {et~al.}(2022){Astropy Collaboration}, {Price-Whelan}, {Lim}, {Earl}, {Starkman}, {Bradley}, {Shupe}, {Patil}, {Corrales}, {Brasseur}, {N{"o}the}, {Donath}, {Tollerud}, {Morris}, {Ginsburg}, {Vaher}, {Weaver}, {Tocknell}, {Jamieson}, {van Kerkwijk}, {Robitaille}, {Merry}, {Bachetti}, {G{"u}nther}, {Aldcroft}, {Alvarado-Montes}, {Archibald}, {B{'o}di}, {Bapat}, {Barentsen}, {Baz{'a}n}, {Biswas}, {Boquien}, {Burke}, {Cara}, {Cara}, {Conroy}, {Conseil}, {Craig}, {Cross}, {Cruz}, {D'Eugenio}, {Dencheva}, {Devillepoix}, {Dietrich}, {Eigenbrot}, {Erben}, {Ferreira}, {Foreman-Mackey}, {Fox}, {Freij}, {Garg}, {Geda}, {Glattly}, {Gondhalekar}, {Gordon}, {Grant}, {Greenfield}, {Groener}, {Guest}, {Gurovich}, {Handberg}, {Hart}, {Hatfield-Dodds}, {Homeier}, {Hosseinzadeh}, {Jenness}, {Jones}, {Joseph}, {Kalmbach}, {Karamehmetoglu}, {Ka{l}uszy{'n}ski}, {Kelley}, {Kern}, {Kerzendorf}, {Koch}, {Kulumani}, {Lee}, {Ly}, {Ma}, {MacBride}, {Maljaars}, {Muna}, {Murphy}, {Norman}, {O'Steen},
  {Oman}, {Pacifici}, {Pascual}, {Pascual-Granado}, {Patil}, {Perren}, {Pickering}, {Rastogi}, {Roulston}, {Ryan}, {Rykoff}, {Sabater}, {Sakurikar}, {Salgado}, {Sanghi}, {Saunders}, {Savchenko}, {Schwardt}, {Seifert-Eckert}, {Shih}, {Jain}, {Shukla}, {Sick}, {Simpson}, {Singanamalla}, {Singer}, {Singhal}, {Sinha}, {Sip{H{o}}cz}, {Spitler}, {Stansby}, {Streicher}, {{{S}}umak}, {Swinbank}, {Taranu}, {Tewary}, {Tremblay}, {Val-Borro}, {Van Kooten}, {Vasovi{'c}}, {Verma}, {de Miranda Cardoso}, {Williams}, {Wilson}, {Winkel}, {Wood-Vasey}, {Xue}, {Yoachim}, {Zhang}, {Zonca}, \& {Astropy Project Contributors}}]{astropy_3}
{Astropy Collaboration}, {Price-Whelan}, A.~M., {Lim}, P.~L., {et~al.} 2022, \apj, 935, 167, \dodoi{10.3847/1538-4357/ac7c74}

\bibitem[{{Avenhaus} {et~al.}(2018){Avenhaus}, {Quanz}, {Garufi}, {Perez}, {Casassus}, {Pinte}, {Bertrang}, {Caceres}, {Benisty}, \& {Dominik}}]{Avenhaus2018}
{Avenhaus}, H., {Quanz}, S.~P., {Garufi}, A., {et~al.} 2018, \apj, 863, 44, \dodoi{10.3847/1538-4357/aab846}

\bibitem[{{Bae} {et~al.}(2023){Bae}, {Isella}, {Zhu}, {Martin}, {Okuzumi}, \& {Suriano}}]{Bae2023}
{Bae}, J., {Isella}, A., {Zhu}, Z., {et~al.} 2023, in Astronomical Society of the Pacific Conference Series, Vol. 534, Protostars and Planets VII, ed. S.~{Inutsuka}, Y.~{Aikawa}, T.~{Muto}, K.~{Tomida}, \& M.~{Tamura}, 423, \dodoi{10.48550/arXiv.2210.13314}

\bibitem[{Bae {et~al.}(2017)Bae, Zhu, \& Hartmann}]{Bae2017}
Bae, J., Zhu, Z., \& Hartmann, L. 2017, The Astrophysical Journal, 850, 201, \dodoi{10.3847/1538-4357/aa9705}

\bibitem[{{Bailer-Jones} {et~al.}(2021){Bailer-Jones}, {Rybizki}, {Fouesneau}, {Demleitner}, \& {Andrae}}]{BailerJones2021}
{Bailer-Jones}, C.~A.~L., {Rybizki}, J., {Fouesneau}, M., {Demleitner}, M., \& {Andrae}, R. 2021, \aj, 161, 147, \dodoi{10.3847/1538-3881/abd806}

\bibitem[{{Baraffe} {et~al.}(2015){Baraffe}, {Homeier}, {Allard}, \& {Chabrier}}]{Baraffe2015}
{Baraffe}, I., {Homeier}, D., {Allard}, F., \& {Chabrier}, G. 2015, \aap, 577, A42, \dodoi{10.1051/0004-6361/201425481}

\bibitem[{{Bayo} {et~al.}(2008){Bayo}, {Rodrigo}, {Barrado Y Navascu{\'e}s}, {Solano}, {Guti{\'e}rrez}, {Morales-Calder{\'o}n}, \& {Allard}}]{Bayo2008}
{Bayo}, A., {Rodrigo}, C., {Barrado Y Navascu{\'e}s}, D., {et~al.} 2008, \aap, 492, 277, \dodoi{10.1051/0004-6361:200810395}

\bibitem[{{Benisty} {et~al.}(2021){Benisty}, {Bae}, {Facchini}, {Keppler}, {Teague}, {Isella}, {Kurtovic}, {P{\'e}rez}, {Sierra}, {Andrews}, {Carpenter}, {Czekala}, {Dominik}, {Henning}, {Menard}, {Pinilla}, \& {Zurlo}}]{Benisty2021}
{Benisty}, M., {Bae}, J., {Facchini}, S., {et~al.} 2021, \apjl, 916, L2, \dodoi{10.3847/2041-8213/ac0f83}

\bibitem[{{Benisty} {et~al.}(2023){Benisty}, {Dominik}, {Follette}, {Garufi}, {Ginski}, {Hashimoto}, {Keppler}, {Kley}, \& {Monnier}}]{Benisty2023}
{Benisty}, M., {Dominik}, C., {Follette}, K., {et~al.} 2023, in Astronomical Society of the Pacific Conference Series, Vol. 534, Protostars and Planets VII, ed. S.~{Inutsuka}, Y.~{Aikawa}, T.~{Muto}, K.~{Tomida}, \& M.~{Tamura}, 605, \dodoi{10.48550/arXiv.2203.09991}

\bibitem[{{Beuzit} {et~al.}(2019){Beuzit}, {Vigan}, {Mouillet}, {Dohlen}, {Gratton}, {Boccaletti}, {Sauvage}, {Schmid}, {Langlois}, {Petit}, {Baruffolo}, {Feldt}, {Milli}, {Wahhaj}, {Abe}, {Anselmi}, {Antichi}, {Barette}, {Baudrand}, {Baudoz}, {Bazzon}, {Bernardi}, {Blanchard}, {Brast}, {Bruno}, {Buey}, {Carbillet}, {Carle}, {Cascone}, {Chapron}, {Charton}, {Chauvin}, {Claudi}, {Costille}, {De Caprio}, {de Boer}, {Delboulb{\'e}}, {Desidera}, {Dominik}, {Downing}, {Dupuis}, {Fabron}, {Fantinel}, {Farisato}, {Feautrier}, {Fedrigo}, {Fusco}, {Gigan}, {Ginski}, {Girard}, {Giro}, {Gisler}, {Gluck}, {Gry}, {Henning}, {Hubin}, {Hugot}, {Incorvaia}, {Jaquet}, {Kasper}, {Lagadec}, {Lagrange}, {Le Coroller}, {Le Mignant}, {Le Ruyet}, {Lessio}, {Lizon}, {Llored}, {Lundin}, {Madec}, {Magnard}, {Marteaud}, {Martinez}, {Maurel}, {M{\'e}nard}, {Mesa}, {M{\"o}ller-Nilsson}, {Moulin}, {Moutou}, {Orign{\'e}}, {Parisot}, {Pavlov}, {Perret}, {Pragt}, {Puget}, {Rabou}, {Ramos}, {Reess}, {Rigal}, {Rochat}, {Roelfsema}, {Rousset},
  {Roux}, {Saisse}, {Salasnich}, {Santambrogio}, {Scuderi}, {Segransan}, {Sevin}, {Siebenmorgen}, {Soenke}, {Stadler}, {Suarez}, {Tiph{\`e}ne}, {Turatto}, {Udry}, {Vakili}, {Waters}, {Weber}, {Wildi}, {Zins}, \& {Zurlo}}]{Beuzit2019}
{Beuzit}, J.~L., {Vigan}, A., {Mouillet}, D., {et~al.} 2019, \aap, 631, A155, \dodoi{10.1051/0004-6361/201935251}

\bibitem[{{Bianchi} {et~al.}(2011){Bianchi}, {Herald}, {Efremova}, {Girardi}, {Zabot}, {Marigo}, {Conti}, \& {Shiao}}]{Bianchi2011}
{Bianchi}, L., {Herald}, J., {Efremova}, B., {et~al.} 2011, \apss, 335, 161, \dodoi{10.1007/s10509-010-0581-x}

\bibitem[{{Blunt} {et~al.}(2017){Blunt}, {Nielsen}, {De Rosa}, {Konopacky}, {Ryan}, {Wang}, {Pueyo}, {Rameau}, {Marois}, {Marchis}, {Macintosh}, {Graham}, {Duch{\^e}ne}, \& {Schneider}}]{Blunt2017}
{Blunt}, S., {Nielsen}, E.~L., {De Rosa}, R.~J., {et~al.} 2017, \aj, 153, 229, \dodoi{10.3847/1538-3881/aa6930}

\bibitem[{Blunt {et~al.}(2020)Blunt, Wang, Angelo, Ngo, Cody, De~Rosa, Graham, Hirsch, Nagpal, Nielsen, Pearce, Rice, \& Tejada}]{Blunt_2020}
Blunt, S., Wang, J.~J., Angelo, I., {et~al.} 2020, The Astronomical Journal, 159, 89, \dodoi{10.3847/1538-3881/ab6663}

\bibitem[{{Bohn} {et~al.}(2019){Bohn}, {Kenworthy}, {Ginski}, {Benisty}, {de Boer}, {Keller}, {Mamajek}, {Meshkat}, {Muro-Arena}, {Pecaut}, {Snik}, {Wolff}, \& {Reggiani}}]{Bohn2019}
{Bohn}, A.~J., {Kenworthy}, M.~A., {Ginski}, C., {et~al.} 2019, \aap, 624, A87, \dodoi{10.1051/0004-6361/201834523}

\bibitem[{{Bohn} {et~al.}(2021){Bohn}, {Ginski}, {Kenworthy}, {Mamajek}, {Pecaut}, {Mugrauer}, {Vogt}, {Adam}, {Meshkat}, {Reggiani}, \& {Snik}}]{Bohn2021}
{Bohn}, A.~J., {Ginski}, C., {Kenworthy}, M.~A., {et~al.} 2021, \aap, 648, A73, \dodoi{10.1051/0004-6361/202140508}

\bibitem[{{Boss}(1997)}]{Boss1997}
{Boss}, A.~P. 1997, Science, 276, 1836, \dodoi{10.1126/science.276.5320.1836}

\bibitem[{{Bouvier} \& {Bertout}(1989)}]{Bouvier1989}
{Bouvier}, J., \& {Bertout}, C. 1989, \aap, 211, 99

\bibitem[{{Bowler} {et~al.}(2025){Bowler}, {Zhou}, {Biddle}, {Jiang}, {Bae}, {Close}, {Follette}, {Franson}, {Kraus}, {Sanghi}, {Tran}, {Ward-Duong}, {Wu}, \& {Zhu}}]{Bowler2025}
{Bowler}, B.~P., {Zhou}, Y., {Biddle}, L.~I., {et~al.} 2025, \aj, 169, 258, \dodoi{10.3847/1538-3881/adb6a1}

\bibitem[{{Bradley} {et~al.}(2016){Bradley}, {Sipocz}, {Robitaille}, {Tollerud}, {Deil}, {Vin{\'\i}cius}, {Barbary}, {G{\"u}nther}, {Bostroem}, {Droettboom}, {Bray}, {Bratholm}, {Pickering}, {Craig}, {Pascual}, {Greco}, {Donath}, {Kerzendorf}, {Littlefair}, {Barentsen}, {D'Eugenio}, \& {Weaver}}]{photutils}
{Bradley}, L., {Sipocz}, B., {Robitaille}, T., {et~al.} 2016, {Photutils: Photometry tools}.
\newblock \doeprint{1609.011}

\bibitem[{{Bressan} {et~al.}(2012){Bressan}, {Marigo}, {Girardi}, {Salasnich}, {Dal Cero}, {Rubele}, \& {Nanni}}]{Bressan2012}
{Bressan}, A., {Marigo}, P., {Girardi}, L., {et~al.} 2012, \mnras, 427, 127, \dodoi{10.1111/j.1365-2966.2012.21948.x}

\bibitem[{{Chabrier} {et~al.}(2000){Chabrier}, {Baraffe}, {Allard}, \& {Hauschildt}}]{Chabrier2000}
{Chabrier}, G., {Baraffe}, I., {Allard}, F., \& {Hauschildt}, P. 2000, \apj, 542, 464, \dodoi{10.1086/309513}

\bibitem[{{Chambers} {et~al.}(2016){Chambers}, {Magnier}, {Metcalfe}, {Flewelling}, {Huber}, {Waters}, {Denneau}, {Draper}, {Farrow}, {Finkbeiner}, {Holmberg}, {Koppenhoefer}, {Price}, {Saglia}, {Schlafly}, {Smartt}, {Sweeney}, {Wainscoat}, {Burgett}, {Grav}, {Heasley}, {Hodapp}, {Jedicke}, {Kaiser}, {Kudritzki}, {Luppino}, {Lupton}, {Monet}, {Morgan}, {Onaka}, {Stubbs}, {Tonry}, {Banados}, {Bell}, {Bender}, {Bernard}, {Botticella}, {Casertano}, {Chastel}, {Chen}, {Chen}, {Cole}, {Deacon}, {Frenk}, {Fitzsimmons}, {Gezari}, {Goessl}, {Goggia}, {Goldman}, {Grebel}, {Hambly}, {Hasinger}, {Heavens}, {Heckman}, {Henderson}, {Henning}, {Holman}, {Hopp}, {Ip}, {Isani}, {Keyes}, {Koekemoer}, {Kotak}, {Long}, {Lucey}, {Liu}, {Martin}, {McLean}, {Morganson}, {Murphy}, {Nieto-Santisteban}, {Norberg}, {Peacock}, {Pier}, {Postman}, {Primak}, {Rae}, {Rest}, {Riess}, {Riffeser}, {Rix}, {Roser}, {Schilbach}, {Schultz}, {Scolnic}, {Szalay}, {Seitz}, {Shiao}, {Small}, {Smith}, {Soderblom}, {Taylor}, {Thakar}, {Thiel},
  {Thilker}, {Urata}, {Valenti}, {Walter}, {Watters}, {Werner}, {White}, {Wood-Vasey}, \& {Wyse}}]{Chambers2016}
{Chambers}, K.~C., {Magnier}, E.~A., {Metcalfe}, N., {et~al.} 2016, ArXiv e-prints.
\newblock \doarXiv{1612.05560}

\bibitem[{{Christiaens} {et~al.}(2019){Christiaens}, {Cantalloube}, {Casassus}, {Price}, {Absil}, {Pinte}, {Girard}, \& {Montesinos}}]{Christiaens2019}
{Christiaens}, V., {Cantalloube}, F., {Casassus}, S., {et~al.} 2019, \apjl, 877, L33, \dodoi{10.3847/2041-8213/ab212b}

\bibitem[{{Close} {et~al.}(2025){Close}, {Males}, {Li}, {Haffert}, {Long}, {Hedglen}, {Weinberger}, {Follette}, {Apai}, {Doyon}, {Foster}, {Gasho}, {Van Gorkom}, {Guyon}, {Kautz}, {Kueny}, {Lumbres}, {McLeod}, {McEwen}, {Pavao}, {Pearce}, {Perez}, {Schatz}, {Szul{\'a}gyi}, {Wagner}, \& {Wu}}]{Close2025a}
{Close}, L.~M., {Males}, J.~R., {Li}, J., {et~al.} 2025, \aj, 169, 35, \dodoi{10.3847/1538-3881/ad8648}

\bibitem[{{Columba} {et~al.}(2024){Columba}, {Rigliaco}, {Gratton}, {Mesa}, {D'Orazi}, {Ginski}, {Engler}, {Williams}, {Bae}, {Benisty}, {Birnstiel}, {Delorme}, {Dominik}, {Facchini}, {Menard}, {Pinilla}, {Rab}, {Ribas}, {Squicciarini}, {van Holstein}, \& {Zurlo}}]{Columba2024}
{Columba}, G., {Rigliaco}, E., {Gratton}, R., {et~al.} 2024, \aap, 681, A19, \dodoi{10.1051/0004-6361/202347109}

\bibitem[{Currie {et~al.}(2019)Currie, Marois, Cieza, Mulders, Lawson, Caceres, Rodriguez-Ruiz, Wisniewski, Guyon, Brandt, Kasdin, Groff, Lozi, Chilcote, Hodapp, Jovanovic, Martinache, Skaf, Lyra, Tamura, Asensio-Torres, Dong, Grady, Gerard, Fukagawa, Hand, Hayashi, Henning, Kudo, Kuzuhara, Kwon, McElwain, \& Uyama}]{Currie2019}
Currie, T., Marois, C., Cieza, L., {et~al.} 2019, The Astrophysical Journal Letters, 877, L3, \dodoi{10.3847/2041-8213/ab1b42}

\bibitem[{{Currie} {et~al.}(2022){Currie}, {Lawson}, {Schneider}, {Lyra}, {Wisniewski}, {Grady}, {Guyon}, {Tamura}, {Kotani}, {Kawahara}, {Brandt}, {Uyama}, {Muto}, {Dong}, {Kudo}, {Hashimoto}, {Fukagawa}, {Wagner}, {Lozi}, {Chilcote}, {Tobin}, {Groff}, {Ward-Duong}, {Januszewski}, {Norris}, {Tuthill}, {van der Marel}, {Sitko}, {Deo}, {Vievard}, {Jovanovic}, {Martinache}, \& {Skaf}}]{Currie2022}
{Currie}, T., {Lawson}, K., {Schneider}, G., {et~al.} 2022, Nature Astronomy, 6, 751, \dodoi{10.1038/s41550-022-01634-x}

\bibitem[{{Cutri} {et~al.}(2003){Cutri}, {Skrutskie}, {van Dyk}, {Beichman}, {Carpenter}, {Chester}, {Cambresy}, {Evans}, {Fowler}, {Gizis}, {Howard}, {Huchra}, {Jarrett}, {Kopan}, {Kirkpatrick}, {Light}, {Marsh}, {McCallon}, {Schneider}, {Stiening}, {Sykes}, {Weinberg}, {Wheaton}, {Wheelock}, \& {Zacarias}}]{Cutri2003}
{Cutri}, R.~M., {Skrutskie}, M.~F., {van Dyk}, S., {et~al.} 2003, {2MASS All Sky Catalog of point sources.} (Infrared Science Archive)

\bibitem[{{Cutri} {et~al.}(2012){Cutri}, {Wright}, {Conrow}, {Bauer}, {Benford}, {Brandenburg}, {Dailey}, {Eisenhardt}, {Evans}, {Fajardo-Acosta}, {Fowler}, {Gelino}, {Grillmair}, {Harbut}, {Hoffman}, {Jarrett}, {Kirkpatrick}, {Leisawitz}, {Liu}, {Mainzer}, {Marsh}, {Masci}, {McCallon}, {Padgett}, {Ressler}, {Royer}, {Skrutskie}, {Stanford}, {Wyatt}, {Tholen}, {Tsai}, {Wachter}, {Wheelock}, {Yan}, {Alles}, {Beck}, {Grav}, {Masiero}, {McCollum}, {McGehee}, {Papin}, \& {Wittman}}]{Cutri2012}
{Cutri}, R.~M., {Wright}, E.~L., {Conrow}, T., {et~al.} 2012, {Explanatory Supplement to the WISE All-Sky Data Release Products}, Explanatory Supplement to the WISE All-Sky Data Release Products

\bibitem[{{Dawson} \& {Fabrycky}(2010)}]{Dawson2010}
{Dawson}, R.~I., \& {Fabrycky}, D.~C. 2010, \apj, 722, 937, \dodoi{10.1088/0004-637X/722/1/937}

\bibitem[{{de Boer} {et~al.}(2016){de Boer}, {Salter}, {Benisty}, {Vigan}, {Boccaletti}, {Pinilla}, {Ginski}, {Juhasz}, {Maire}, {Messina}, {Desidera}, {Cheetham}, {Girard}, {Wahhaj}, {Langlois}, {Bonnefoy}, {Beuzit}, {Buenzli}, {Chauvin}, {Dominik}, {Feldt}, {Gratton}, {Hagelberg}, {Isella}, {Janson}, {Keller}, {Lagrange}, {Lannier}, {Menard}, {Mesa}, {Mouillet}, {Mugrauer}, {Peretti}, {Perrot}, {Sissa}, {Snik}, {Vogt}, {Zurlo}, \& {SPHERE Consortium}}]{deBoer2016}
{de Boer}, J., {Salter}, G., {Benisty}, M., {et~al.} 2016, \aap, 595, A114, \dodoi{10.1051/0004-6361/201629267}

\bibitem[{{de Boer} {et~al.}(2020){de Boer}, {Langlois}, {van Holstein}, {Girard}, {Mouillet}, {Vigan}, {Dohlen}, {Snik}, {Keller}, {Ginski}, {Stam}, {Milli}, {Wahhaj}, {Kasper}, {Schmid}, {Rabou}, {Gluck}, {Hugot}, {Perret}, {Martinez}, {Weber}, {Pragt}, {Sauvage}, {Boccaletti}, {Le Coroller}, {Dominik}, {Henning}, {Lagadec}, {M{\'e}nard}, {Turatto}, {Udry}, {Chauvin}, {Feldt}, \& {Beuzit}}]{deBoer2020}
{de Boer}, J., {Langlois}, M., {van Holstein}, R.~G., {et~al.} 2020, \aap, 633, A63, \dodoi{10.1051/0004-6361/201834989}

\bibitem[{{de Boer} {et~al.}(2021){de Boer}, {Ginski}, {Chauvin}, {M{\'e}nard}, {Benisty}, {Dominik}, {Maaskant}, {Girard}, {van der Plas}, {Garufi}, {Perrot}, {Stolker}, {Avenhaus}, {Bohn}, {Delboulb{\'e}}, {Jaquet}, {Buey}, {M{\"o}ller-Nilsson}, {Pragt}, \& {Fusco}}]{deBoer2021}
{de Boer}, J., {Ginski}, C., {Chauvin}, G., {et~al.} 2021, \aap, 649, A25, \dodoi{10.1051/0004-6361/201936787}

\bibitem[{{Derkink} {et~al.}(2024){Derkink}, {Ginski}, {Pinilla}, {Kurtovic}, {Kaper}, {de Koter}, {Valeg{\r{a}}rd}, {Mamajek}, {Backs}, {Benisty}, {Birnstiel}, {Columba}, {Dominik}, {Garufi}, {Hogerheijde}, {van Holstein}, {Huang}, {M{\'e}nard}, {Rab}, {Ram{\'\i}rez-Tannus}, {Ribas}, {Williams}, \& {Zurlo}}]{Derkink2024}
{Derkink}, A., {Ginski}, C., {Pinilla}, P., {et~al.} 2024, \aap, 688, A149, \dodoi{10.1051/0004-6361/202348555}

\bibitem[{{Dohlen} {et~al.}(2008){Dohlen}, {Langlois}, {Saisse}, {Hill}, {Origne}, {Jacquet}, {Fabron}, {Blanc}, {Llored}, {Carle}, {Moutou}, {Vigan}, {Boccaletti}, {Carbillet}, {Mouillet}, \& {Beuzit}}]{Dohlen2008}
{Dohlen}, K., {Langlois}, M., {Saisse}, M., {et~al.} 2008, in Society of Photo-Optical Instrumentation Engineers (SPIE) Conference Series, Vol. 7014, Ground-based and Airborne Instrumentation for Astronomy II, ed. I.~S. {McLean} \& M.~M. {Casali}, 70143L, \dodoi{10.1117/12.789786}

\bibitem[{Dupuy {et~al.}(2018)Dupuy, Liu, Allers, Biller, Kratter, Mann, Shkolnik, Kraus, \& Best}]{Dupuy2018}
Dupuy, T.~J., Liu, M.~C., Allers, K.~N., {et~al.} 2018, The Astronomical Journal, 156, 57, \dodoi{10.3847/1538-3881/aacbc2}

\bibitem[{{Epchtein} {et~al.}(1999){Epchtein}, {Deul}, {Derriere}, {Borsenberger}, {Egret}, {Simon}, {Alard}, {Bal{\'a}zs}, {de Batz}, {Cioni}, {Copet}, {Dennefeld}, {Forveille}, {Fouqu{\'e}}, {Garz{\'o}n}, {Habing}, {Holl}, {Hron}, {Kimeswenger}, {Lacombe}, {Le Bertre}, {Loup}, {Mamon}, {Omont}, {Paturel}, {Persi}, {Robin}, {Rouan}, {Tiph{\`e}ne}, {Vauglin}, \& {Wagner}}]{Epchtein1999}
{Epchtein}, N., {Deul}, E., {Derriere}, S., {et~al.} 1999, \aap, 349, 236

\bibitem[{{Fiorellino} {et~al.}(2021){Fiorellino}, {Elia}, {Andr{\'e}}, {Men'shchikov}, {Pezzuto}, {Schisano}, {K{\"o}nyves}, {Arzoumanian}, {Benedettini}, {Ward-Thompson}, {Bracco}, {Di Francesco}, {Bontemps}, {Kirk}, {Motte}, \& {Molinari}}]{Fiorellino2021}
{Fiorellino}, E., {Elia}, D., {Andr{\'e}}, P., {et~al.} 2021, \mnras, 500, 4257, \dodoi{10.1093/mnras/staa3420}

\bibitem[{{Fitzpatrick}(1999)}]{Fitzpatrick1999}
{Fitzpatrick}, E.~L. 1999, \pasp, 111, 63, \dodoi{10.1086/316293}

\bibitem[{{Follette} {et~al.}(2017){Follette}, {Rameau}, {Dong}, {Pueyo}, {Close}, {Duch{\^e}ne}, {Fung}, {Leonard}, {Macintosh}, {Males}, {Marois}, {Millar-Blanchaer}, {Morzinski}, {Mullen}, {Perrin}, {Spiro}, {Wang}, {Ammons}, {Bailey}, {Barman}, {Bulger}, {Chilcote}, {Cotten}, {De Rosa}, {Doyon}, {Fitzgerald}, {Goodsell}, {Graham}, {Greenbaum}, {Hibon}, {Hung}, {Ingraham}, {Kalas}, {Konopacky}, {Larkin}, {Maire}, {Marchis}, {Metchev}, {Nielsen}, {Oppenheimer}, {Palmer}, {Patience}, {Poyneer}, {Rajan}, {Rantakyr{\"o}}, {Savransky}, {Schneider}, {Sivaramakrishnan}, {Song}, {Soummer}, {Thomas}, {Vega}, {Wallace}, {Ward-Duong}, {Wiktorowicz}, \& {Wolff}}]{Folette2017}
{Follette}, K.~B., {Rameau}, J., {Dong}, R., {et~al.} 2017, \aj, 153, 264, \dodoi{10.3847/1538-3881/aa6d85}

\bibitem[{{Gaia Collaboration} {et~al.}(2023){Gaia Collaboration}, {Vallenari}, {Brown}, {Prusti}, {de Bruijne}, {Arenou}, {Babusiaux}, {Biermann}, {Creevey}, {Ducourant}, {Evans}, {Eyer}, {Guerra}, {Hutton}, {Jordi}, {Klioner}, {Lammers}, {Lindegren}, {Luri}, {Mignard}, {Panem}, {Pourbaix}, {Randich}, {Sartoretti}, {Soubiran}, {Tanga}, {Walton}, {Bailer-Jones}, {Bastian}, {Drimmel}, {Jansen}, {Katz}, {Lattanzi}, {van Leeuwen}, {Bakker}, {Cacciari}, {Casta{\~n}eda}, {De Angeli}, {Fabricius}, {Fouesneau}, {Fr{\'e}mat}, {Galluccio}, {Guerrier}, {Heiter}, {Masana}, {Messineo}, {Mowlavi}, {Nicolas}, {Nienartowicz}, {Pailler}, {Panuzzo}, {Riclet}, {Roux}, {Seabroke}, {Sordo}, {Th{\'e}venin}, {Gracia-Abril}, {Portell}, {Teyssier}, {Altmann}, {Andrae}, {Audard}, {Bellas-Velidis}, {Benson}, {Berthier}, {Blomme}, {Burgess}, {Busonero}, {Busso}, {C{\'a}novas}, {Carry}, {Cellino}, {Cheek}, {Clementini}, {Damerdji}, {Davidson}, {de Teodoro}, {Nu{\~n}ez Campos}, {Delchambre}, {Dell'Oro}, {Esquej},
  {Fern{\'a}ndez-Hern{\'a}ndez}, {Fraile}, {Garabato}, {Garc{\'\i}a-Lario}, {Gosset}, {Haigron}, {Halbwachs}, {Hambly}, {Harrison}, {Hern{\'a}ndez}, {Hestroffer}, {Hodgkin}, {Holl}, {Jan{\ss}en}, {Jevardat de Fombelle}, {Jordan}, {Krone-Martins}, {Lanzafame}, {L{\"o}ffler}, {Marchal}, {Marrese}, {Moitinho}, {Muinonen}, {Osborne}, {Pancino}, {Pauwels}, {Recio-Blanco}, {Reyl{\'e}}, {Riello}, {Rimoldini}, {Roegiers}, {Rybizki}, {Sarro}, {Siopis}, {Smith}, {Sozzetti}, {Utrilla}, {van Leeuwen}, {Abbas}, {{\'A}brah{\'a}m}, {Abreu Aramburu}, {Aerts}, {Aguado}, {Ajaj}, {Aldea-Montero}, {Altavilla}, {{\'A}lvarez}, {Alves}, {Anders}, {Anderson}, {Anglada Varela}, {Antoja}, {Baines}, {Baker}, {Balaguer-N{\'u}{\~n}ez}, {Balbinot}, {Balog}, {Barache}, {Barbato}, {Barros}, {Barstow}, {Bartolom{\'e}}, {Bassilana}, {Bauchet}, {Becciani}, {Bellazzini}, {Berihuete}, {Bernet}, {Bertone}, {Bianchi}, {Binnenfeld}, {Blanco-Cuaresma}, {Blazere}, {Boch}, {Bombrun}, {Bossini}, {Bouquillon}, {Bragaglia}, {Bramante}, {Breedt},
  {Bressan}, {Brouillet}, {Brugaletta}, {Bucciarelli}, {Burlacu}, {Butkevich}, {Buzzi}, {Caffau}, {Cancelliere}, {Cantat-Gaudin}, {Carballo}, {Carlucci}, {Carnerero}, {Carrasco}, {Casamiquela}, {Castellani}, {Castro-Ginard}, {Chaoul}, {Charlot}, {Chemin}, {Chiaramida}, {Chiavassa}, {Chornay}, {Comoretto}, {Contursi}, {Cooper}, {Cornez}, {Cowell}, {Crifo}, {Cropper}, {Crosta}, {Crowley}, {Dafonte}, {Dapergolas}, {David}, {David}, {de Laverny}, {De Luise}, \& {De March}}]{Gaia_DR3}
{Gaia Collaboration}, {Vallenari}, A., {Brown}, A.~G.~A., {et~al.} 2023, \aap, 674, A1, \dodoi{10.1051/0004-6361/202243940}

\bibitem[{{Gapp} {et~al.}(2025){Gapp}, {Evans-Soma}, {Barstow}, {Lothringer}, {Sing}, {Ruseva}, {Ahrer}, {Goyal}, {Christie}, {Kreidberg}, \& {Mayne}}]{Gapp2025}
{Gapp}, C., {Evans-Soma}, T.~M., {Barstow}, J.~K., {et~al.} 2025, \aj, 169, 341, \dodoi{10.3847/1538-3881/ad9c6e}

\bibitem[{{George} {et~al.}(2025){George}, {Dominik}, \& {Ginski}}]{George2025}
{George}, J., {Dominik}, C., \& {Ginski}, C. 2025, arXiv e-prints, arXiv:2506.03624, \dodoi{10.48550/arXiv.2506.03624}

\bibitem[{{Ginski} {et~al.}(2016){Ginski}, {Stolker}, {Pinilla}, {Dominik}, {Boccaletti}, {de Boer}, {Benisty}, {Biller}, {Feldt}, {Garufi}, {Keller}, {Kenworthy}, {Maire}, {M{\'e}nard}, {Mesa}, {Milli}, {Min}, {Pinte}, {Quanz}, {van Boekel}, {Bonnefoy}, {Chauvin}, {Desidera}, {Gratton}, {Girard}, {Keppler}, {Kopytova}, {Lagrange}, {Langlois}, {Rouan}, \& {Vigan}}]{Ginski2016}
{Ginski}, C., {Stolker}, T., {Pinilla}, P., {et~al.} 2016, \aap, 595, A112, \dodoi{10.1051/0004-6361/201629265}

\bibitem[{{Ginski} {et~al.}(2021){Ginski}, {Facchini}, {Huang}, {Benisty}, {Vaendel}, {Stapper}, {Dominik}, {Bae}, {M{\'e}nard}, {Muro-Arena}, {Hogerheijde}, {McClure}, {van Holstein}, {Birnstiel}, {Boehler}, {Bohn}, {Flock}, {Mamajek}, {Manara}, {Pinilla}, {Pinte}, \& {Ribas}}]{Ginski2021}
{Ginski}, C., {Facchini}, S., {Huang}, J., {et~al.} 2021, \apjl, 908, L25, \dodoi{10.3847/2041-8213/abdf57}

\bibitem[{{Ginski, C.} {et~al.}(2024){Ginski, C.}, {Garufi, A.}, {Benisty, M.}, {Tazaki, R.}, {Dominik, C.}, {Ribas, Á.}, {Engler, N.}, {Birnstiel, T.}, {Chauvin, G.}, {Columba, G.}, {Facchini, S.}, {Goncharov, A.}, {Hagelberg, J.}, {Henning, T.}, {Hogerheijde, M.}, {van Holstein, R. G.}, {Huang, J.}, {Muto, T.}, {Pinilla, P.}, {Kanagawa, K.}, {Kim, S.}, {Kurtovic, N.}, {Langlois, M.}, {Manara, C.}, {Milli, J.}, {Momose, M.}, {Orihara, R.}, {Pawellek, N.}, {Pinte, C.}, {Rab, C.}, {Schmidt, T. O. B.}, {Snik, F.}, {Wahhaj, Z.}, {Williams, J.}, \& {Zurlo, A.}}]{Ginski2024}
{Ginski, C.}, {Garufi, A.}, {Benisty, M.}, {et~al.} 2024, A\&A, 685, A52, \dodoi{10.1051/0004-6361/202244005}

\bibitem[{{Goffo} {et~al.}(2023){Goffo}, {Gandolfi}, {Egger}, {Mustill}, {Albrecht}, {Hirano}, {Kochukhov}, {Astudillo-Defru}, {Barragan}, {Serrano}, {Hatzes}, {Alibert}, {Guenther}, {Dai}, {Lam}, {Csizmadia}, {Smith}, {Fossati}, {Luque}, {Rodler}, {Winther}, {R{\o}rsted}, {Alarcon}, {Bonfils}, {Cochran}, {Deeg}, {Jenkins}, {Korth}, {Livingston}, {Meech}, {Murgas}, {Orell-Miquel}, {Osborne}, {Palle}, {Persson}, {Redfield}, {Ricker}, {Seager}, {Vanderspek}, {Van Eylen}, \& {Winn}}]{Goffo2023}
{Goffo}, E., {Gandolfi}, D., {Egger}, J.~A., {et~al.} 2023, \apjl, 955, L3, \dodoi{10.3847/2041-8213/ace0c7}

\bibitem[{{Grady} {et~al.}(2000){Grady}, {Devine}, {Woodgate}, {Kimble}, {Bruhweiler}, {Boggess}, {Linsky}, {Plait}, {Clampin}, \& {Kalas}}]{Grady2000}
{Grady}, C.~A., {Devine}, D., {Woodgate}, B., {et~al.} 2000, \apj, 544, 895, \dodoi{10.1086/317222}

\bibitem[{{Gratton} {et~al.}(2019){Gratton}, {Ligi}, {Sissa}, {Desidera}, {Mesa}, {Bonnefoy}, {Chauvin}, {Cheetham}, {Feldt}, {Lagrange}, {Langlois}, {Meyer}, {Vigan}, {Boccaletti}, {Janson}, {Lazzoni}, {Zurlo}, {De Boer}, {Henning}, {D'Orazi}, {Gluck}, {Madec}, {Jaquet}, {Baudoz}, {Fantinel}, {Pavlov}, \& {Wildi}}]{Gratton2019}
{Gratton}, R., {Ligi}, R., {Sissa}, E., {et~al.} 2019, \aap, 623, A140, \dodoi{10.1051/0004-6361/201834760}

\bibitem[{{Green} {et~al.}(2018){Green}, {Schlafly}, {Finkbeiner}, {Rix}, {Martin}, {Burgett}, {Draper}, {Flewelling}, {Hodapp}, {Kaiser}, {Kudritzki}, {Magnier}, {Metcalfe}, {Tonry}, {Wainscoat}, \& {Waters}}]{Green2018}
{Green}, G.~M., {Schlafly}, E.~F., {Finkbeiner}, D., {et~al.} 2018, \mnras, 478, 651, \dodoi{10.1093/mnras/sty1008}

\bibitem[{{Haffert} {et~al.}(2019){Haffert}, {Bohn}, {de Boer}, {Snellen}, {Brinchmann}, {Girard}, {Keller}, \& {Bacon}}]{Haffert2019}
{Haffert}, S.~Y., {Bohn}, A.~J., {de Boer}, J., {et~al.} 2019, Nature Astronomy, 3, 749, \dodoi{10.1038/s41550-019-0780-5}

\bibitem[{{Hammond} {et~al.}(2023){Hammond}, {Christiaens}, {Price}, {Toci}, {Pinte}, {Juillard}, \& {Garg}}]{Hammond2023}
{Hammond}, I., {Christiaens}, V., {Price}, D.~J., {et~al.} 2023, \mnras, 522, L51, \dodoi{10.1093/mnrasl/slad027}

\bibitem[{Hoch {et~al.}(2025)Hoch, Rowland, Petrus, Nasedkin, Ingebretsen, Kammerer, Perrin, D'Orazi, Balmer, Barman, Bonnefoy, Chauvin, Chen, De~Rosa, Girard, Gonzales, Kenworthy, Konopacky, Macintosh, Moran, Morley, Palma-Bifani, Pueyo, Ren, Rickman, Ruffio, Theissen, Ward-Duong, \& Zhang}]{Hoch2025}
Hoch, K. K.~W., Rowland, M., Petrus, S., {et~al.} 2025, Nature, \dodoi{10.1038/s41586-025-09174-w}

\bibitem[{{H{\o}g} {et~al.}(2000){H{\o}g}, {Fabricius}, {Makarov}, {Urban}, {Corbin}, {Wycoff}, {Bastian}, {Schwekendiek}, \& {Wicenec}}]{Hog2000}
{H{\o}g}, E., {Fabricius}, C., {Makarov}, V.~V., {et~al.} 2000, \aap, 363, 385

\bibitem[{{Hom} {et~al.}(2007){Hom}, {Marchis}, {Lee}, {Haase}, {Agard}, \& {Sedat}}]{Hom2007}
{Hom}, E. F.~Y., {Marchis}, F., {Lee}, T.~K., {et~al.} 2007, Journal of the Optical Society of America A, 24, 1580, \dodoi{10.1364/JOSAA.24.001580}

\bibitem[{{Hunt} \& {Reffert}(2024)}]{Hunt2024}
{Hunt}, E.~L., \& {Reffert}, S. 2024, \aap, 686, A42, \dodoi{10.1051/0004-6361/202348662}

\bibitem[{{Indebetouw} {et~al.}(2005){Indebetouw}, {Mathis}, {Babler}, {Meade}, {Watson}, {Whitney}, {Wolff}, {Wolfire}, {Cohen}, {Bania}, {Benjamin}, {Clemens}, {Dickey}, {Jackson}, {Kobulnicky}, {Marston}, {Mercer}, {Stauffer}, {Stolovy}, \& {Churchwell}}]{Indebetouw2005}
{Indebetouw}, R., {Mathis}, J.~S., {Babler}, B.~L., {et~al.} 2005, \apj, 619, 931, \dodoi{10.1086/426679}

\bibitem[{{Janson} {et~al.}(2021){Janson}, {Gratton}, {Rodet}, {Vigan}, {Bonnefoy}, {Delorme}, {Mamajek}, {Reffert}, {Stock}, {Marleau}, {Langlois}, {Chauvin}, {Desidera}, {Ringqvist}, {Mayer}, {Viswanath}, {Squicciarini}, {Meyer}, {Samland}, {Petrus}, {Helled}, {Kenworthy}, {Quanz}, {Biller}, {Henning}, {Mesa}, {Engler}, \& {Carson}}]{Janson2021}
{Janson}, M., {Gratton}, R., {Rodet}, L., {et~al.} 2021, \nat, 600, 231, \dodoi{10.1038/s41586-021-04124-8}

\bibitem[{{Jeffries} {et~al.}(2023){Jeffries}, {Jackson}, {Wright}, {Weaver}, {Gilmore}, {Randich}, {Bragaglia}, {Korn}, {Smiljanic}, {Biazzo}, {Casey}, {Frasca}, {Gonneau}, {Guiglion}, {Morbidelli}, {Prisinzano}, {Sacco}, {Tautvai{\v{s}}ien{\.{e}}}, {Worley}, \& {Zaggia}}]{Jeffries2023}
{Jeffries}, R.~D., {Jackson}, R.~J., {Wright}, N.~J., {et~al.} 2023, \mnras, 523, 802, \dodoi{10.1093/mnras/stad1293}

\bibitem[{{Juillard} {et~al.}(2022){Juillard}, {Christiaens}, \& {Absil}}]{Juillard2022}
{Juillard}, S., {Christiaens}, V., \& {Absil}, O. 2022, \aap, 668, A125, \dodoi{10.1051/0004-6361/202244402}

\bibitem[{Kanagawa {et~al.}(2016)Kanagawa, Muto, Tanaka, Tanigawa, Takeuchi, Tsukagoshi, \& Momose}]{Kanagawa_2016}
Kanagawa, K.~D., Muto, T., Tanaka, H., {et~al.} 2016, Publications of the Astronomical Society of Japan, 68, \dodoi{10.1093/pasj/psw037}

\bibitem[{{Keppler} {et~al.}(2018){Keppler}, {Benisty}, {M{\"u}ller}, {Henning}, {van Boekel}, {Cantalloube}, {Ginski}, {van Holstein}, {Maire}, {Pohl}, {Samland }, {Avenhaus}, {Baudino}, {Boccaletti}, {de Boer}, {Bonnefoy}, {Chauvin}, {Desidera}, {Langlois}, {Lazzoni}, {Marleau}, {Mordasini}, {Pawellek}, {Stolker}, {Vigan}, {Zurlo}, {Birnstiel}, {Brandner}, {Feldt}, {Flock}, {Girard}, {Gratton}, {Hagelberg}, {Isella}, {Janson}, {Juhasz}, {Kemmer}, {Kral}, {Lagrange}, {Launhardt}, {Matter}, {M{\'e}nard}, {Milli}, {Molli{\`e}re}, {Olofsson}, {P{\'e}rez}, {Pinilla}, {Pinte}, {Quanz}, {Schmidt}, {Udry}, {Wahhaj}, {Williams}, {Buenzli}, {Cudel}, {Dominik}, {Galicher}, {Kasper}, {Lannier}, {Mesa}, {Mouillet}, {Peretti}, {Perrot}, {Salter}, {Sissa}, {Wildi}, {Abe}, {Antichi}, {Augereau}, {Baruffolo}, {Baudoz}, {Bazzon}, {Beuzit}, {Blanchard}, {Brems}, {Buey}, {De Caprio}, {Carbillet}, {Carle}, {Cascone}, {Cheetham}, {Claudi}, {Costille}, {Delboulb{\'e}}, {Dohlen}, {Fantinel}, {Feautrier}, {Fusco}, {Giro}, {Gluck},
  {Gry}, {Hubin}, {Hugot}, {Jaquet}, {Le Mignant}, {Llored}, {Madec}, {Magnard}, {Martinez}, {Maurel}, {Meyer}, {M{\"o}ller-Nilsson}, {Moulin}, {Mugnier}, {Orign{\'e}}, {Pavlov}, {Perret}, {Petit}, {Pragt}, {Puget}, {Rabou}, {Ramos}, {Rigal}, {Rochat}, {Roelfsema}, {Rousset}, {Roux}, {Salasnich}, {Sauvage}, {Sevin}, {Soenke}, {Stadler}, {Suarez}, {Turatto}, \& {Weber}}]{Keppler2018}
{Keppler}, M., {Benisty}, M., {M{\"u}ller}, A., {et~al.} 2018, \aap, 617, A44, \dodoi{10.1051/0004-6361/201832957}

\bibitem[{{Kerr} {et~al.}(2021){Kerr}, {Rizzuto}, {Kraus}, \& {Offner}}]{Kerr2021}
{Kerr}, R. M.~P., {Rizzuto}, A.~C., {Kraus}, A.~L., \& {Offner}, S. S.~R. 2021, \apj, 917, 23, \dodoi{10.3847/1538-4357/ac0251}

\bibitem[{{Kostov} {et~al.}(2020){Kostov}, {Orosz}, {Feinstein}, {Welsh}, {Cukier}, {Haghighipour}, {Quarles}, {Martin}, {Montet}, {Torres}, {Triaud}, {Barclay}, {Boyd}, {Briceno}, {Cameron}, {Correia}, {Gilbert}, {Gill}, {Gillon}, {Haqq-Misra}, {Hellier}, {Dressing}, {Fabrycky}, {Furesz}, {Jenkins}, {Kane}, {Kopparapu}, {Hod{\v{z}}i{\'c}}, {Latham}, {Law}, {Levine}, {Li}, {Lintott}, {Lissauer}, {Mann}, {Mazeh}, {Mardling}, {Maxted}, {Eisner}, {Pepe}, {Pepper}, {Pollacco}, {Quinn}, {Quintana}, {Rowe}, {Ricker}, {Rose}, {Seager}, {Santerne}, {S{\'e}gransan}, {Short}, {Smith}, {Standing}, {Tokovinin}, {Trifonov}, {Turner}, {Twicken}, {Udry}, {Vanderspek}, {Winn}, {Wolf}, {Ziegler}, {Ansorge}, {Barnet}, {Bergeron}, {Huten}, {Pappa}, \& {van der Straeten}}]{Kostov2020}
{Kostov}, V.~B., {Orosz}, J.~A., {Feinstein}, A.~D., {et~al.} 2020, \aj, 159, 253, \dodoi{10.3847/1538-3881/ab8a48}

\bibitem[{{Kounkel} \& {Covey}(2019)}]{Kounkel2019}
{Kounkel}, M., \& {Covey}, K. 2019, \aj, 158, 122, \dodoi{10.3847/1538-3881/ab339a}

\bibitem[{{Kroupa}(1995)}]{Kroupa1995}
{Kroupa}, P. 1995, \mnras, 277, 1491, \dodoi{10.1093/mnras/277.4.1491}

\bibitem[{{Kuhn} {et~al.}(2001){Kuhn}, {Potter}, \& {Parise}}]{Kuhn2001}
{Kuhn}, J.~R., {Potter}, D., \& {Parise}, B. 2001, \apjl, 553, L189, \dodoi{10.1086/320686}

\bibitem[{{Lafreni{\`e}re} {et~al.}(2009){Lafreni{\`e}re}, {Marois}, {Doyon}, \& {Barman}}]{Lafreniere2009}
{Lafreni{\`e}re}, D., {Marois}, C., {Doyon}, R., \& {Barman}, T. 2009, \apjl, 694, L148, \dodoi{10.1088/0004-637X/694/2/L148}

\bibitem[{{Lagrange} {et~al.}(2009){Lagrange}, {Gratadour}, {Chauvin}, {Fusco}, {Ehrenreich}, {Mouillet}, {Rousset}, {Rouan}, {Allard}, {Gendron}, {Charton}, {Mugnier}, {Rabou}, {Montri}, \& {Lacombe}}]{Lagrange2009}
{Lagrange}, A.~M., {Gratadour}, D., {Chauvin}, G., {et~al.} 2009, \aap, 493, L21, \dodoi{10.1051/0004-6361:200811325}

\bibitem[{{Lanza, A. F.} {et~al.}(2016){Lanza, A. F.}, {Flaccomio, E.}, {Messina, S.}, {Micela, G.}, {Pagano, I.}, \& {Leto, G.}}]{Lanza2016}
{Lanza, A. F.}, {Flaccomio, E.}, {Messina, S.}, {et~al.} 2016, \AA, 592, A140, \dodoi{10.1051/0004-6361/201628382}

\bibitem[{Long {et~al.}(2011)Long, Romanova, Kulkarni, \& Donati}]{Long2011}
Long, M., Romanova, M.~M., Kulkarni, A.~K., \& Donati, J.-F. 2011, Monthly Notices of the Royal Astronomical Society, 413, 1061, \dodoi{10.1111/j.1365-2966.2010.18193.x}

\bibitem[{{MacDonald} \& {Madhusudhan}(2017)}]{MacDonald2017}
{MacDonald}, R.~J., \& {Madhusudhan}, N. 2017, \mnras, 469, 1979, \dodoi{10.1093/mnras/stx804}

\bibitem[{MacKay(2003)}]{MacKay03}
MacKay, D. J.~C. 2003, Information Theory, Inference, and Learning Algorithms (Copyright Cambridge University Press)

\bibitem[{{Magnani} {et~al.}(1985){Magnani}, {Blitz}, \& {Mundy}}]{Magnani1985}
{Magnani}, L., {Blitz}, L., \& {Mundy}, L. 1985, \apj, 295, 402, \dodoi{10.1086/163385}

\bibitem[{{Maire} {et~al.}(2016){Maire}, {Langlois}, {Dohlen}, {Lagrange}, {Gratton}, {Chauvin}, {Desidera}, {Girard}, {Milli}, {Vigan}, {Zins}, {Delorme}, {Beuzit}, {Claudi}, {Feldt}, {Mouillet}, {Puget}, {Turatto}, \& {Wildi}}]{Maire2016}
{Maire}, A.-L., {Langlois}, M., {Dohlen}, K., {et~al.} 2016, in Society of Photo-Optical Instrumentation Engineers (SPIE) Conference Series, Vol. 9908, Ground-based and Airborne Instrumentation for Astronomy VI, ed. C.~J. {Evans}, L.~{Simard}, \& H.~{Takami}, 990834, \dodoi{10.1117/12.2233013}

\bibitem[{{Maire} {et~al.}(2021){Maire}, {Langlois}, {Delorme}, {Chauvin}, {Gratton}, {Vigan}, {Girard}, {Wahhaj}, {Pott}, {Burtscher}, {Boccaletti}, {Carlotti}, {Henning}, {Kenworthy}, {Kervella}, {Rickman}, \& {Schmidt}}]{Maire2021}
{Maire}, A.-L., {Langlois}, M., {Delorme}, P., {et~al.} 2021, Journal of Astronomical Telescopes, Instruments, and Systems, 7, 035004, \dodoi{10.1117/1.JATIS.7.3.035004}

\bibitem[{{Marois} {et~al.}(2006){Marois}, {Lafreni{\`e}re}, {Doyon}, {Macintosh}, \& {Nadeau}}]{Marois2006}
{Marois}, C., {Lafreni{\`e}re}, D., {Doyon}, R., {Macintosh}, B., \& {Nadeau}, D. 2006, \apj, 641, 556, \dodoi{10.1086/500401}

\bibitem[{{Mesa} {et~al.}(2019){Mesa}, {Keppler}, {Cantalloube}, {Rodet}, {Charnay}, {Gratton}, {Langlois}, {Boccaletti}, {Bonnefoy}, {Vigan}, {Flasseur}, {Bae}, {Benisty}, {Chauvin}, {de Boer}, {Desidera}, {Henning}, {Lagrange}, {Meyer}, {Milli}, {M{\"u}ller}, {Pairet}, {Zurlo}, {Antoniucci}, {Baudino}, {Brown Sevilla}, {Cascone}, {Cheetham}, {Claudi}, {Delorme}, {D'Orazi}, {Feldt}, {Hagelberg}, {Janson}, {Kral}, {Lagadec}, {Lazzoni}, {Ligi}, {Maire}, {Martinez}, {Menard}, {Meunier}, {Perrot}, {Petrus}, {Pinte}, {Rickman}, {Rochat}, {Rouan}, {Samland}, {Sauvage}, {Schmidt}, {Udry}, {Weber}, \& {Wildi}}]{Mesa2019}
{Mesa}, D., {Keppler}, M., {Cantalloube}, F., {et~al.} 2019, \aap, 632, A25, \dodoi{10.1051/0004-6361/201936764}

\bibitem[{{Milli} {et~al.}(2012){Milli}, {Mouillet}, {Lagrange}, {Boccaletti}, {Mawet}, {Chauvin}, \& {Bonnefoy}}]{Milli2012}
{Milli}, J., {Mouillet}, D., {Lagrange}, A.~M., {et~al.} 2012, \aap, 545, A111, \dodoi{10.1051/0004-6361/201219687}

\bibitem[{{Monnier} {et~al.}(2017){Monnier}, {Harries}, {Aarnio}, {Adams}, {Andrews}, {Calvet}, {Espaillat}, {Hartmann}, {Hinkley}, {Kraus}, {McClure}, {Oppenheimer}, {Perrin}, \& {Wilner}}]{Monnier2017}
{Monnier}, J.~D., {Harries}, T.~J., {Aarnio}, A., {et~al.} 2017, \apj, 838, 20, \dodoi{10.3847/1538-4357/aa6248}

\bibitem[{{Monnier} {et~al.}(2019){Monnier}, {Harries}, {Bae}, {Setterholm}, {Laws}, {Aarnio}, {Adams}, {Andrews}, {Calvet}, {Espaillat}, {Hartmann}, {Kraus}, {McClure}, {Miller}, {Oppenheimer}, {Wilner}, \& {Zhu}}]{Monnier2019}
{Monnier}, J.~D., {Harries}, T.~J., {Bae}, J., {et~al.} 2019, \apj, 872, 122, \dodoi{10.3847/1538-4357/aafe87}

\bibitem[{{M{\"u}ller} {et~al.}(2018){M{\"u}ller}, {Keppler}, {Henning}, {Samland}, {Chauvin}, {Beust}, {Maire}, {Molaverdikhani}, {van Boekel}, {Benisty}, {Boccaletti}, {Bonnefoy}, {Cantalloube}, {Charnay}, {Baudino}, {Gennaro}, {Long}, {Cheetham}, {Desidera}, {Feldt}, {Fusco}, {Girard}, {Gratton}, {Hagelberg}, {Janson}, {Lagrange}, {Langlois}, {Lazzoni}, {Ligi}, {M{\'e}nard}, {Mesa}, {Meyer}, {Molli{\`e}re}, {Mordasini}, {Moulin}, {Pavlov}, {Pawellek}, {Quanz}, {Ramos}, {Rouan}, {Sissa}, {Stadler}, {Vigan}, {Wahhaj}, {Weber}, \& {Zurlo}}]{Mueller2018}
{M{\"u}ller}, A., {Keppler}, M., {Henning}, T., {et~al.} 2018, \aap, 617, L2, \dodoi{10.1051/0004-6361/201833584}

\bibitem[{Oliphant(2006)}]{oliphant2006guide}
Oliphant, T.~E. 2006, A guide to NumPy, Vol.~1 (Trelgol Publishing USA)

\bibitem[{Paardekooper \& Mellema(2006)}]{Paardekooper2006}
Paardekooper, S.-J., \& Mellema, G. 2006, A\&A, 453, 1129–1140, \dodoi{10.1051/0004-6361:20054449}

\bibitem[{{Park} {et~al.}(2012){Park}, {Min}, {Seon}, {Han}, {Lee}, \& {Edelstein}}]{Park2012}
{Park}, S.~J., {Min}, K.~W., {Seon}, K.~I., {et~al.} 2012, \apj, 754, 10, \dodoi{10.1088/0004-637X/754/1/10}

\bibitem[{{Paunzen} {et~al.}(2024){Paunzen}, {Netopil}, {Pri{\v{s}}egen}, \& {Faltov{\'a}}}]{Paunzen2024}
{Paunzen}, E., {Netopil}, M., {Pri{\v{s}}egen}, M., \& {Faltov{\'a}}, N. 2024, \aap, 689, A270, \dodoi{10.1051/0004-6361/202347768}

\bibitem[{{Pinte} {et~al.}(2019){Pinte}, {van der Plas}, {M{\'e}nard}, {Price}, {Christiaens}, {Hill}, {Mentiplay}, {Ginski}, {Choquet}, {Boehler}, {Duch{\^e}ne}, {Perez}, \& {Casassus}}]{Pinte2019}
{Pinte}, C., {van der Plas}, G., {M{\'e}nard}, F., {et~al.} 2019, Nature Astronomy, 419, \dodoi{10.1038/s41550-019-0852-6}

\bibitem[{{Pollack} {et~al.}(1996){Pollack}, {Hubickyj}, {Bodenheimer}, {Lissauer}, {Podolak}, \& {Greenzweig}}]{Pollack1996}
{Pollack}, J.~B., {Hubickyj}, O., {Bodenheimer}, P., {et~al.} 1996, \icarus, 124, 62, \dodoi{10.1006/icar.1996.0190}

\bibitem[{{Prato} \& {Simon}(2023)}]{Prato2023}
{Prato}, L., \& {Simon}, M. 2023, Research Notes of the American Astronomical Society, 7, 150, \dodoi{10.3847/2515-5172/ace88e}

\bibitem[{{Qin} {et~al.}(2023){Qin}, {Zhong}, {Tang}, \& {Chen}}]{Qin2023}
{Qin}, S., {Zhong}, J., {Tang}, T., \& {Chen}, L. 2023, \apjs, 265, 12, \dodoi{10.3847/1538-4365/acadd6}

\bibitem[{{Quanz} {et~al.}(2013){Quanz}, {Avenhaus}, {Buenzli}, {Garufi}, {Schmid}, \& {Wolf}}]{Quanz2013}
{Quanz}, S.~P., {Avenhaus}, H., {Buenzli}, E., {et~al.} 2013, \apjl, 766, L2, \dodoi{10.1088/2041-8205/766/1/L2}

\bibitem[{{Ruane} {et~al.}(2019){Ruane}, {Ngo}, {Mawet}, {Absil}, {Choquet}, {Cook}, {Gomez Gonzalez}, {Huby}, {Matthews}, {Meshkat}, {Reggiani}, {Serabyn}, {Wallack}, \& {Xuan}}]{Ruane2019}
{Ruane}, G., {Ngo}, H., {Mawet}, D., {et~al.} 2019, \aj, 157, 118, \dodoi{10.3847/1538-3881/aafee2}

\bibitem[{{Rustamkulov} {et~al.}(2023){Rustamkulov}, {Sing}, {Mukherjee}, {May}, {Kirk}, {Schlawin}, {Line}, {Piaulet}, {Carter}, {Batalha}, {Goyal}, {L{\'o}pez-Morales}, {Lothringer}, {MacDonald}, {Moran}, {Stevenson}, {Wakeford}, {Espinoza}, {Bean}, {Batalha}, {Benneke}, {Berta-Thompson}, {Crossfield}, {Gao}, {Kreidberg}, {Powell}, {Cubillos}, {Gibson}, {Leconte}, {Molaverdikhani}, {Nikolov}, {Parmentier}, {Roy}, {Taylor}, {Turner}, {Wheatley}, {Aggarwal}, {Ahrer}, {Alam}, {Alderson}, {Allen}, {Banerjee}, {Barat}, {Barrado}, {Barstow}, {Bell}, {Blecic}, {Brande}, {Casewell}, {Changeat}, {Chubb}, {Crouzet}, {Daylan}, {Decin}, {D{\'e}sert}, {Mikal-Evans}, {Feinstein}, {Flagg}, {Fortney}, {Harrington}, {Heng}, {Hong}, {Hu}, {Iro}, {Kataria}, {Kempton}, {Krick}, {Lendl}, {Lillo-Box}, {Louca}, {Lustig-Yaeger}, {Mancini}, {Mansfield}, {Mayne}, {Miguel}, {Morello}, {Ohno}, {Palle}, {Petit dit de la Roche}, {Rackham}, {Radica}, {Ramos-Rosado}, {Redfield}, {Rogers}, {Shkolnik}, {Southworth}, {Teske}, {Tremblin},
  {Tucker}, {Venot}, {Waalkes}, {Welbanks}, {Zhang}, \& {Zieba}}]{Rustamkulov2023}
{Rustamkulov}, Z., {Sing}, D.~K., {Mukherjee}, S., {et~al.} 2023, \nat, 614, 659, \dodoi{10.1038/s41586-022-05677-y}

\bibitem[{Sanghi {et~al.}(2024)Sanghi, Xuan, Wang, Mawet, Bowler, Ngo, Bryan, Ruane, Absil, \& Huby}]{Sanghi2024}
Sanghi, A., Xuan, J.~W., Wang, J.~J., {et~al.} 2024, The Astronomical Journal, 168, 215, \dodoi{10.3847/1538-3881/ad769f}

\bibitem[{{Schlafly} {et~al.}(2010){Schlafly}, {Finkbeiner}, {Schlegel}, {Juri{\'c}}, {Ivezi{\'c}}, {Gibson}, {Knapp}, \& {Weaver}}]{Schlafly2010}
{Schlafly}, E.~F., {Finkbeiner}, D.~P., {Schlegel}, D.~J., {et~al.} 2010, \apj, 725, 1175, \dodoi{10.1088/0004-637X/725/1/1175}

\bibitem[{{Schlafly} {et~al.}(2014){Schlafly}, {Green}, {Finkbeiner}, {Rix}, {Bell}, {Burgett}, {Chambers}, {Draper}, {Hodapp}, {Kaiser}, {Magnier}, {Martin}, {Metcalfe}, {Price}, \& {Tonry}}]{Schlafly2014}
{Schlafly}, E.~F., {Green}, G., {Finkbeiner}, D.~P., {et~al.} 2014, \apj, 786, 29, \dodoi{10.1088/0004-637X/786/1/29}

\bibitem[{{Schlegel} {et~al.}(1998){Schlegel}, {Finkbeiner}, \& {Davis}}]{Schlegel1998}
{Schlegel}, D.~J., {Finkbeiner}, D.~P., \& {Davis}, M. 1998, \apj, 500, 525, \dodoi{10.1086/305772}

\bibitem[{{Shakura} \& {Sunyaev}(1976)}]{Shakura1976}
{Shakura}, N.~I., \& {Sunyaev}, R.~A. 1976, \mnras, 175, 613, \dodoi{10.1093/mnras/175.3.613}

\bibitem[{{Sigurdsson} {et~al.}(2003){Sigurdsson}, {Richer}, {Hansen}, {Stairs}, \& {Thorsett}}]{Sigurdsson2003}
{Sigurdsson}, S., {Richer}, H.~B., {Hansen}, B.~M., {Stairs}, I.~H., \& {Thorsett}, S.~E. 2003, Science, 301, 193, \dodoi{10.1126/science.1086326}

\bibitem[{{Smith} \& {Terrile}(1984)}]{Smith1984}
{Smith}, B.~A., \& {Terrile}, R.~J. 1984, Science, 226, 1421, \dodoi{10.1126/science.226.4681.1421}

\bibitem[{{Somers} {et~al.}(2020){Somers}, {Cao}, \& {Pinsonneault}}]{Somers2020}
{Somers}, G., {Cao}, L., \& {Pinsonneault}, M.~H. 2020, \apj, 891, 29, \dodoi{10.3847/1538-4357/ab722e}

\bibitem[{{Soummer} {et~al.}(2012){Soummer}, {Pueyo}, \& {Larkin}}]{Soummer2012}
{Soummer}, R., {Pueyo}, L., \& {Larkin}, J. 2012, \apjl, 755, L28, \dodoi{10.1088/2041-8205/755/2/L28}

\bibitem[{{Stapper} \& {Ginski}(2022)}]{Stapper2022}
{Stapper}, L.~M., \& {Ginski}, C. 2022, \aap, 668, A50, \dodoi{10.1051/0004-6361/202142820}

\bibitem[{{Stassun} {et~al.}(2019){Stassun}, {Oelkers}, {Paegert}, {Torres}, {Pepper}, {De Lee}, {Collins}, {Latham}, {Muirhead}, {Chittidi}, {Rojas-Ayala}, {Fleming}, {Rose}, {Tenenbaum}, {Ting}, {Kane}, {Barclay}, {Bean}, {Brassuer}, {Charbonneau}, {Ge}, {Lissauer}, {Mann}, {McLean}, {Mullally}, {Narita}, {Plavchan}, {Ricker}, {Sasselov}, {Seager}, {Sharma}, {Shiao}, {Sozzetti}, {Stello}, {Vanderspek}, {Wallace}, \& {Winn}}]{Stassun2019}
{Stassun}, K.~G., {Oelkers}, R.~J., {Paegert}, M., {et~al.} 2019, \aj, 158, 138, \dodoi{10.3847/1538-3881/ab3467}

\bibitem[{{Stolker} {et~al.}(2019){Stolker}, {Bonse}, {Quanz}, {Amara}, {Cugno}, {Bohn}, \& {Boehle}}]{Stolker2019}
{Stolker}, T., {Bonse}, M.~J., {Quanz}, S.~P., {et~al.} 2019, \aap, 621, A59, \dodoi{10.1051/0004-6361/201834136}

\bibitem[{{Stolker} {et~al.}(2020{\natexlab{a}}){Stolker}, {Marleau}, {Cugno}, {Molli{\`e}re}, {Quanz}, {Todorov}, \& {K{\"u}hn}}]{Stolker2020}
{Stolker}, T., {Marleau}, G.~D., {Cugno}, G., {et~al.} 2020{\natexlab{a}}, \aap, 644, A13, \dodoi{10.1051/0004-6361/202038878}

\bibitem[{{Stolker} {et~al.}(2020{\natexlab{b}}){Stolker}, {Quanz}, {Todorov}, {K{\"u}hn}, {Molli{\`e}re}, {Meyer}, {Currie}, {Daemgen}, \& {Lavie}}]{Stolker2020b}
{Stolker}, T., {Quanz}, S.~P., {Todorov}, K.~O., {et~al.} 2020{\natexlab{b}}, \aap, 635, A182, \dodoi{10.1051/0004-6361/201937159}

\bibitem[{{Teague} {et~al.}(2018){Teague}, {Bae}, {Bergin}, {Birnstiel}, \& {Foreman-Mackey}}]{Teague2018}
{Teague}, R., {Bae}, J., {Bergin}, E.~A., {Birnstiel}, T., \& {Foreman-Mackey}, D. 2018, \apjl, 860, L12, \dodoi{10.3847/2041-8213/aac6d7}

\bibitem[{Teague {et~al.}(2025)Teague, Benisty, Facchini, Fukagawa, Pinte, Andrews, Bae, Barraza-Alfaro, Cataldi, Cuello, Curone, Czekala, Fasano, Flock, Galloway-Sprietsma, Garg, Hall, Hammond, Hilder, Huang, Ilee, Izquierdo, Kanagawa, Lesur, Lodato, Longarini, Loomis, Masset, Menard, Orihara, Price, Rosotti, Stadler, Testi, Yen, Wafflard-Fernandez, Wilner, Winter, Wölfer, Yoshida, \& Zawadzki}]{Teague2025}
Teague, R., Benisty, M., Facchini, S., {et~al.} 2025, The Astrophysical Journal Letters, 984, L6, \dodoi{10.3847/2041-8213/adc43b}

\bibitem[{{Torres} {et~al.}(2006){Torres}, {Quast}, {da Silva}, {de La Reza}, {Melo}, \& {Sterzik}}]{Torres2006}
{Torres}, C.~A.~O., {Quast}, G.~R., {da Silva}, L., {et~al.} 2006, \aap, 460, 695, \dodoi{10.1051/0004-6361:20065602}

\bibitem[{{Valeg{\r{a}}rd} {et~al.}(2024){Valeg{\r{a}}rd}, {Ginski}, {Derkink}, {Garufi}, {Dominik}, {Ribas}, {Williams}, {Benisty}, {Birnstiel}, {Facchini}, {Columba}, {Hogerheijde}, {van Holstein}, {Huang}, {Kenworthy}, {Manara}, {Pinilla}, {Rab}, {Sulaiman}, \& {Zurlo}}]{Valegard2024}
{Valeg{\r{a}}rd}, P.~G., {Ginski}, C., {Derkink}, A., {et~al.} 2024, \aap, 685, A54, \dodoi{10.1051/0004-6361/202347452}

\bibitem[{{van Boekel} {et~al.}(2017){van Boekel}, {Henning}, {Menu}, {de Boer}, {Langlois}, {M{\"u}ller}, {Avenhaus}, {Boccaletti}, {Schmid}, {Thalmann}, {Benisty}, {Dominik}, {Ginski}, {Girard}, {Gisler}, {Lobo Gomes}, {Menard}, {Min}, {Pavlov}, {Pohl}, {Quanz}, {Rabou}, {Roelfsema}, {Sauvage}, {Teague}, {Wildi}, \& {Zurlo}}]{vanBoekel2017}
{van Boekel}, R., {Henning}, T., {Menu}, J., {et~al.} 2017, \apj, 837, 132, \dodoi{10.3847/1538-4357/aa5d68}

\bibitem[{{van Holstein} {et~al.}(2020){van Holstein}, {Girard}, {de Boer}, {Snik}, {Milli}, {Stam}, {Ginski}, {Mouillet}, {Wahhaj}, {Schmid}, {Keller}, {Langlois}, {Dohlen}, {Vigan}, {Pohl}, {Carbillet}, {Fantinel}, {Maurel}, {Orign{\'e}}, {Petit}, {Ramos}, {Rigal}, {Sevin}, {Boccaletti}, {Le Coroller}, {Dominik}, {Henning}, {Lagadec}, {M{\'e}nard}, {Turatto}, {Udry}, {Chauvin}, {Feldt}, \& {Beuzit}}]{vanHolstein2020}
{van Holstein}, R.~G., {Girard}, J.~H., {de Boer}, J., {et~al.} 2020, \aap, 633, A64, \dodoi{10.1051/0004-6361/201834996}

\bibitem[{{Virtanen} {et~al.}(2020){Virtanen}, {Gommers}, {Oliphant}, {Haberland}, {Reddy}, {Cournapeau}, {Burovski}, {Peterson}, {Weckesser}, {Bright}, {van der Walt}, {Brett}, {Wilson}, {Jarrod Millman}, {Mayorov}, {Nelson}, {Jones}, {Kern}, {Larson}, {Carey}, {Polat}, {Feng}, {Moore}, {Vand erPlas}, {Laxalde}, {Perktold}, {Cimrman}, {Henriksen}, {Quintero}, {Harris}, {Archibald}, {Ribeiro}, {Pedregosa}, {van Mulbregt}, \& {Contributors}}]{2020SciPy-NMeth}
{Virtanen}, P., {Gommers}, R., {Oliphant}, T.~E., {et~al.} 2020, Nature Methods, 17, 261, \dodoi{https://doi.org/10.1038/s41592-019-0686-2}

\bibitem[{{{\v{Z}}erjal} {et~al.}(2021){{\v{Z}}erjal}, {Rains}, {Ireland}, {Zhou}, {Kammerer}, {Wallace}, {Orenstein}, {Nordlander}, {Abbot}, \& {Chang}}]{Zerjal2021}
{{\v{Z}}erjal}, M., {Rains}, A.~D., {Ireland}, M.~J., {et~al.} 2021, \mnras, 503, 938, \dodoi{10.1093/mnras/stab513}

\bibitem[{{Wahhaj} {et~al.}(2021){Wahhaj}, {Milli}, {Romero}, {Cieza}, {Zurlo}, {Vigan}, {Pe{\~n}a}, {Valdes}, {Cantalloube}, {Girard}, \& {Pantoja}}]{Wahhaj2021}
{Wahhaj}, Z., {Milli}, J., {Romero}, C., {et~al.} 2021, \aap, 648, A26, \dodoi{10.1051/0004-6361/202038794}

\bibitem[{{Wahhaj, Z.} {et~al.}(2024){Wahhaj, Z.}, {Benisty, M.}, {Ginski, C.}, {Swastik, C.}, {Arora, S.}, {van Holstein, R. G.}, {De Rosa, R.}, {Yang, B.}, {Bae, J.}, \& {Ren, B.}}]{Wahhaj2024}
{Wahhaj, Z.}, {Benisty, M.}, {Ginski, C.}, {et~al.} 2024, \AA, 687, A257, \dodoi{10.1051/0004-6361/202349018}

\bibitem[{{Wang} {et~al.}(2021){Wang}, {Vigan}, {Lacour}, {Nowak}, {Stolker}, {De Rosa}, {Ginzburg}, {Gao}, {Abuter}, {Amorim}, {Asensio-Torres}, {Baub{\"o}ck}, {Benisty}, {Berger}, {Beust}, {Beuzit}, {Blunt}, {Boccaletti}, {Bohn}, {Bonnefoy}, {Bonnet}, {Brandner}, {Cantalloube}, {Caselli}, {Charnay}, {Chauvin}, {Choquet}, {Christiaens}, {Cl{\'e}net}, {Coud{\'e} Du Foresto}, {Cridland}, {de Zeeuw}, {Dembet}, {Dexter}, {Drescher}, {Duvert}, {Eckart}, {Eisenhauer}, {Facchini}, {Gao}, {Garcia}, {Garcia Lopez}, {Gardner}, {Gendron}, {Genzel}, {Gillessen}, {Girard}, {Haubois}, {Hei{\ss}el}, {Henning}, {Hinkley}, {Hippler}, {Horrobin}, {Houll{\'e}}, {Hubert}, {Jim{\'e}nez-Rosales}, {Jocou}, {Kammerer}, {Keppler}, {Kervella}, {Meyer}, {Kreidberg}, {Lagrange}, {Lapeyr{\`e}re}, {Le Bouquin}, {L{\'e}na}, {Lutz}, {Maire}, {M{\'e}nard}, {M{\'e}rand}, {Molli{\`e}re}, {Monnier}, {Mouillet}, {M{\"u}ller}, {Nasedkin}, {Ott}, {Otten}, {Paladini}, {Paumard}, {Perraut}, {Perrin}, {Pfuhl}, {Pueyo}, {Rameau}, {Rodet},
  {Rodr{\'\i}guez-Coira}, {Rousset}, {Scheithauer}, {Shangguan}, {Shimizu}, {Stadler}, {Straub}, {Straubmeier}, {Sturm}, {Tacconi}, {van Dishoeck}, {Vincent}, {von Fellenberg}, {Ward-Duong}, {Widmann}, {Wieprecht}, {Wiezorrek}, {Woillez}, \& {Gravity Collaboration}}]{Wang2021}
{Wang}, J.~J., {Vigan}, A., {Lacour}, S., {et~al.} 2021, \aj, 161, 148, \dodoi{10.3847/1538-3881/abdb2d}

\bibitem[{{Watson} {et~al.}(2006){Watson}, {Henden}, \& {Price}}]{Watson2006}
{Watson}, C.~L., {Henden}, A.~A., \& {Price}, A. 2006, Society for Astronomical Sciences Annual Symposium, 25, 47

\bibitem[{{Weaver} {et~al.}(2024){Weaver}, {Jeffries}, \& {Jackson}}]{Weaver2024}
{Weaver}, G., {Jeffries}, R.~D., \& {Jackson}, R.~J. 2024, \mnras, 534, 2014, \dodoi{10.1093/mnras/stae2133}

\bibitem[{{Xie, Chen} {et~al.}(2022){Xie, Chen}, {Choquet, Elodie}, {Vigan, Arthur}, {Cantalloube, Faustine}, {Benisty, Myriam}, {Boccaletti, Anthony}, {Bonnefoy, Mickael}, {Desgrange, Celia}, {Garufi, Antonio}, {Girard, Julien}, {Hagelberg, Janis}, {Janson, Markus}, {Kenworthy, Matthew}, {Lagrange, Anne-Marie}, {Langlois, Maud}, {Menard, François}, \& {Zurlo, Alice}}]{Xie2022}
{Xie, Chen}, {Choquet, Elodie}, {Vigan, Arthur}, {et~al.} 2022, A\&A, 666, A32, \dodoi{10.1051/0004-6361/202243379}

\bibitem[{{Zari} {et~al.}(2018){Zari}, {Hashemi}, {Brown}, {Jardine}, \& {de Zeeuw}}]{Zari2018}
{Zari}, E., {Hashemi}, H., {Brown}, A.~G.~A., {Jardine}, K., \& {de Zeeuw}, P.~T. 2018, \aap, 620, A172, \dodoi{10.1051/0004-6361/201834150}

\bibitem[{Zhang {et~al.}(2018)Zhang, Zhu, Huang, Guzmán, Andrews, Birnstiel, Dullemond, Carpenter, Isella, Pérez, Benisty, Wilner, Baruteau, Bai, \& Ricci}]{Zhang2018}
Zhang, S., Zhu, Z., Huang, J., {et~al.} 2018, The Astrophysical Journal Letters, 869, L47, \dodoi{10.3847/2041-8213/aaf744}

\bibitem[{{Zhang} {et~al.}(2021{\natexlab{a}}){Zhang}, {Snellen}, {Bohn}, {Molli{\`e}re}, {Ginski}, {Hoeijmakers}, {Kenworthy}, {Mamajek}, {Meshkat}, {Reggiani}, \& {Snik}}]{Zhang21b}
{Zhang}, Y., {Snellen}, I. A.~G., {Bohn}, A.~J., {et~al.} 2021{\natexlab{a}}, \nat, 595, 370, \dodoi{10.1038/s41586-021-03616-x}

\bibitem[{{Zhang} {et~al.}(2024){Zhang}, {Gonz{\'a}lez Picos}, {de Regt}, {Snellen}, {Gandhi}, {Ginski}, {Kesseli}, {Landman}, {Molli{\`e}re}, {Nasedkin}, {S{\'a}nchez-L{\'o}pez}, {Stolker}, {Inglis}, {Knutson}, {Mawet}, {Wallack}, \& {Xuan}}]{Zhang24}
{Zhang}, Y., {Gonz{\'a}lez Picos}, D., {de Regt}, S., {et~al.} 2024, \aj, 168, 246, \dodoi{10.3847/1538-3881/ad7ea9}

\bibitem[{{Zhang} {et~al.}(2021{\natexlab{b}}){Zhang}, {Liu}, {Claytor}, {Best}, {Dupuy}, \& {Siverd}}]{Zhang2021}
{Zhang}, Z., {Liu}, M.~C., {Claytor}, Z.~R., {et~al.} 2021{\natexlab{b}}, \apjl, 916, L11, \dodoi{10.3847/2041-8213/ac1123}

\end{thebibliography}
